\documentclass[onecolumn,amsfonts,showpacs,superscriptaddress]{revtex4-1}

\usepackage[dvipdfmx]{graphicx}
\usepackage{amsmath,amssymb,amsthm,booktabs,mathtools}
\usepackage{color}
\usepackage[colorlinks,bookmarks=false,citecolor=blue,linkcolor=red,urlcolor=blue]{hyperref}
\usepackage{tikz}
\usepackage{pgfplots}
\usepackage{natbib}

\setcitestyle{authoryear,round}

\begin{document}

\title{Generalized Hydrodynamics in the 1D Bose gas: theory and experiments}

\author{Isabelle Bouchoule}
\affiliation{Laboratoire Charles Fabry, Institut d’Optique, CNRS, Universit\'e Paris Sud 11, 2 Avenue Augustin Fresnel, 91127 Palaiseau Cedex, France}
\author{J\'er\^ome Dubail}
\affiliation{Universit\'e de Lorraine, CNRS, LPCT, F-54000 Nancy, France}

\begin{abstract}
	We review the recent theoretical and experimental progress regarding the Generalized Hydrodynamics (GHD) behavior of the one-dimensional Bose gas with contact repulsive interactions, also known as the Lieb-Liniger gas. 
	In the first section, we review the theory of the Lieb-Liniger gas, introducing the key notions of the rapidities and of the rapidity distribution. The latter characterizes the Lieb-Liniger gas after relaxation and is at the heart of GHD. We also present the asymptotic regimes of the Lieb-Liniger gas with their dedicated approximate descriptions. In the second section we enter the core of the subject and review the theoretical results on GHD in 1D Bose gases. The third and fourth sections are dedicated to experimental results obtained in cold atoms experiments: the experimental realization of the Lieb-Liniger model is presented in section 3, with a selection of key results for systems at equilibrium, and section 4 presents the experimental tests of the GHD theory.
	In section 5 we review the effects of atom losses, which, assuming slow loss processes, can be described within the GHD framework. We conclude with a few open questions. 
\end{abstract}

\maketitle

\section*{Introduction}

Physical systems of many identical particles behave very differently depending on the distance and time scales at which they are probed. In a very dilute gas, on time scales not larger than the typical time between collisions, the particles are essentially non-interacting. Then two clouds of fluid can collide and simply pass through each other; one example of such phenomenon, familiar from astrophysics, is the one of clouds of stars in colliding galaxies. In contrast, on time scales much longer than the collision time, particles typically undergo a very large number of collisions, so that the fluid has time to locally relax to an equilibrium state. This local relaxation gives rise to hydrodynamic behavior, which is typically much more complex, non-linear, than simple free propagation. For example, one can think of two droplets of water that collide: those will not simply pass through each other. More likely their motion will be more complex, for instance they will coalesce~\citep{brazier1972interaction}.

\vspace{0.5cm}

Fluid dynamics at short times is captured by an evolution equation for the phase-space density of particles $\rho(x,p,t)$ which takes the form of a free transport equation, or collisionless Boltzmann equation. Typically,
\begin{equation}
	\label{eq:liouville}
	\frac{\partial}{\partial t} \rho(x,p,t)  + v(p)  \frac{\partial}{\partial x} \rho(x,p,t) -   \frac{\partial  V(x)}{\partial x}  \frac{\partial}{\partial p} \rho(x,p,t)   \, = \, 0.
\end{equation}
Here we write the equation in one spatial dimension; the extension to higher dimensions is straightforward. In Eq. (\ref{eq:liouville}), $v(p)$ is usually the group velocity $ \partial \varepsilon(p)/\partial p$ of a particle with momentum $p$ and kinetic energy $\varepsilon(p)$, and $V(x)$ is an external potential. Eq.~(\ref{eq:liouville}) is obtained, for instance, for $N$ classical particles described by the non-interacting Hamiltonian $\mathcal{H} = \sum_{j=1}^N [ \varepsilon(p_j) + V(x_j) ]$. Then the evolution of the phase-space density $\rho(x,p,t) = \sum_{j=1}^N  \delta(x - x_j(t) ) \delta(p-p_j(t))$ follows from the evaluation of the Poisson bracket $\partial \rho / \partial t = \{ \mathcal{H}, \rho  \}$. Equations similar to Eq.~(\ref{eq:liouville}) appear in the description of fluids made of both classical particles and quantum particles; we come back to this below.

\vspace{0.5cm}

On time scales much longer than the relaxation time, equation (\ref{eq:liouville}) is superseded by a system of hydrodynamic equations. At that scale, the fluid is locally relaxed to an equilibrium state at any time. Local equilibrium states are parametrized by the conserved quantities in the system, whose time evolution is given by continuity equations. A good example is the one of a Galilean fluid with conserved particle number, conserved momentum and conserved energy. Then a coarse-grained hydrodynamic description, valid at large distance and time scales, is obtained by writing three continuity equations for the mass density $q_M$, the momentum density $q_P$, and the energy density $q_E$,
\begin{equation}
	\label{eq:continuity3}
	\left\{  \begin{array}{ccc}
		\frac{\partial}{\partial t} q_M (x,t)+ \frac{\partial}{\partial x}   j_{M} (x,t)  &=& 0 \\ 
		\frac{\partial}{\partial t}  q_P (x,t) + \frac{\partial}{\partial x}  j_{P} (x,t) &=& - \frac{1}{m}  \frac{\partial V(x)}{\partial x} q_M (x,t )  \\ 
		\frac{\partial}{\partial t}  q_E (x,t) +\frac{\partial}{\partial x}  j_{E} (x,t)   &=& 0  ,
	\end{array} \right.
\end{equation}
where $j_M$, $j_P$ and $j_E$ are the three associated currents. Here the second line is not quite a continuity equation, unless $\partial V/\partial x = 0$. This is simply because momentum is not conserved in the presence of an external force: the right hand side in this evolution equation for $q_P$ is given by Newton second law.

Because of local equilibration, the currents depend on $x$ and $t$ only through their dependence on the charge densities. In general, a current $j$ is a function of all charge densities $q$ and of their spatial derivatives $\partial_x q$, $\partial_x^2 q$, etc. However, for density variations of very long wavelengths, the dependence on the derivatives can be neglected, and 
$j_M$, $j_P$ and $j_E$ are functions of $q_M$, $q_P$ and $q_E$ only. The zeroth-order hydrodynamic equations obtained in this way are usually called `Euler scale' hydrodynamics or `the Euler hydrodynamic limit'. At the Euler scale, the three continuity equations above reduce to the standard Euler equations for a Galilean fluid,
\begin{equation}
	\label{eq:euler}
	\left\{  \begin{array}{ccc}
		\frac{\partial}{\partial t} n + \frac{\partial}{\partial x}  (nu) &=& 0 \\ 
		\frac{\partial}{\partial t} u + u \frac{\partial}{\partial x} u + \frac{1}{m n} \partial_x \mathcal{P} &=& - \frac{1}{m} \frac{\partial V}{\partial x}   \\ 
		\frac{\partial}{\partial t}  e + u \frac{\partial}{\partial x}  e + \frac{1}{n} \mathcal{P} \partial_x u  &=& 0  .
	\end{array} \right.
\end{equation}
Here $m$ is the particles' mass, $n = q_M/m$ is the particle density, $u = q_P/q_M$ is the mean fluid velocity, and $e = (q_E - q_P^2/(2 q_M) - n V(x) ) /n$ is the internal energy per particle. To go from the conservation equations (\ref{eq:continuity3}) to the system (\ref{eq:euler}), one uses the fact that $j_M = q_P$ because of Galilean invariance. Moreover, at the Euler scale, $j_P = q_P^2/q_M + \mathcal{P}$ and $j_E =  (q_E + \mathcal{P}) q_P/q_M$, where $\mathcal{P} = \mathcal{P}(n,e)$ is the equilibrium pressure.

To close the system of equations (\ref{eq:euler}), one needs to know the equilibrium pressure $\mathcal{P}(n,e)$, which is a function of $n$ and $e$ that depends on the microscopic details of the system. In some simple models such as the ideal gas, a simple analytic expression for the pressure is available, but usually there is none. For the one-dimensional Bose gas with contact repulsion, which is at the center of this review article, $\mathcal{P}(n,e)$ can be tabulated numerically (see Subsection \ref{subsec:yangyang}).

To conclude this brief discussion of hydrodynamic equations, we mention that it is of course possible to go `beyond the Euler scale', and to do first-order hydrodynamics by keeping the dependence of the currents on gradients of charge densities. This results in Navier-Stokes-like hydrodynamic equations, which include dissipative terms. In this review article we mostly focus on Euler scale (zeroth order) hydrodynamics.

\vspace{0.5cm}

This review article is about the peculiar fluid-like behavior that emerges in the quantum one-dimensional Bose gas. It is peculiar in the sense that it is simultaneously of the form (\ref{eq:continuity3},\ref{eq:euler}) and of the form (\ref{eq:liouville}), on time scales much longer than the inverse collision rate. The same peculiar behavior is common to all one-dimensional classical and quantum integrable systems, and it has become known as `Generalized Hydrodynamics' or `GHD' since 2016~\citep{castro2016emergent,bertini2016transport}. Here the word `Generalized' is used in the same way as it is in `Generalized Gibbs Ensemble' \citep{rigol2007relaxation,rigol2008thermalization}: it designates the extension of a concept (`Gibbs Ensemble' or `Hydrodynamics') from the case with a small, finite, number of conserved quantities to the case with infinitely many of them.

\vspace{0.5cm}

\begin{figure}[ht]
    \centering
    \begin{tikzpicture}
    \draw (0,0.4) node{\includegraphics[width=0.35\textwidth]{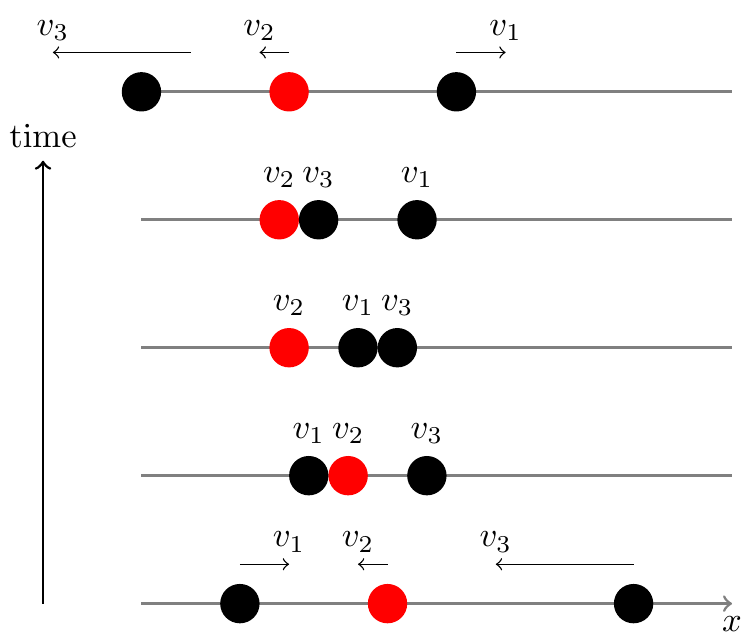}};
    \draw (7.7,0) node{\includegraphics[width=0.6\textwidth]{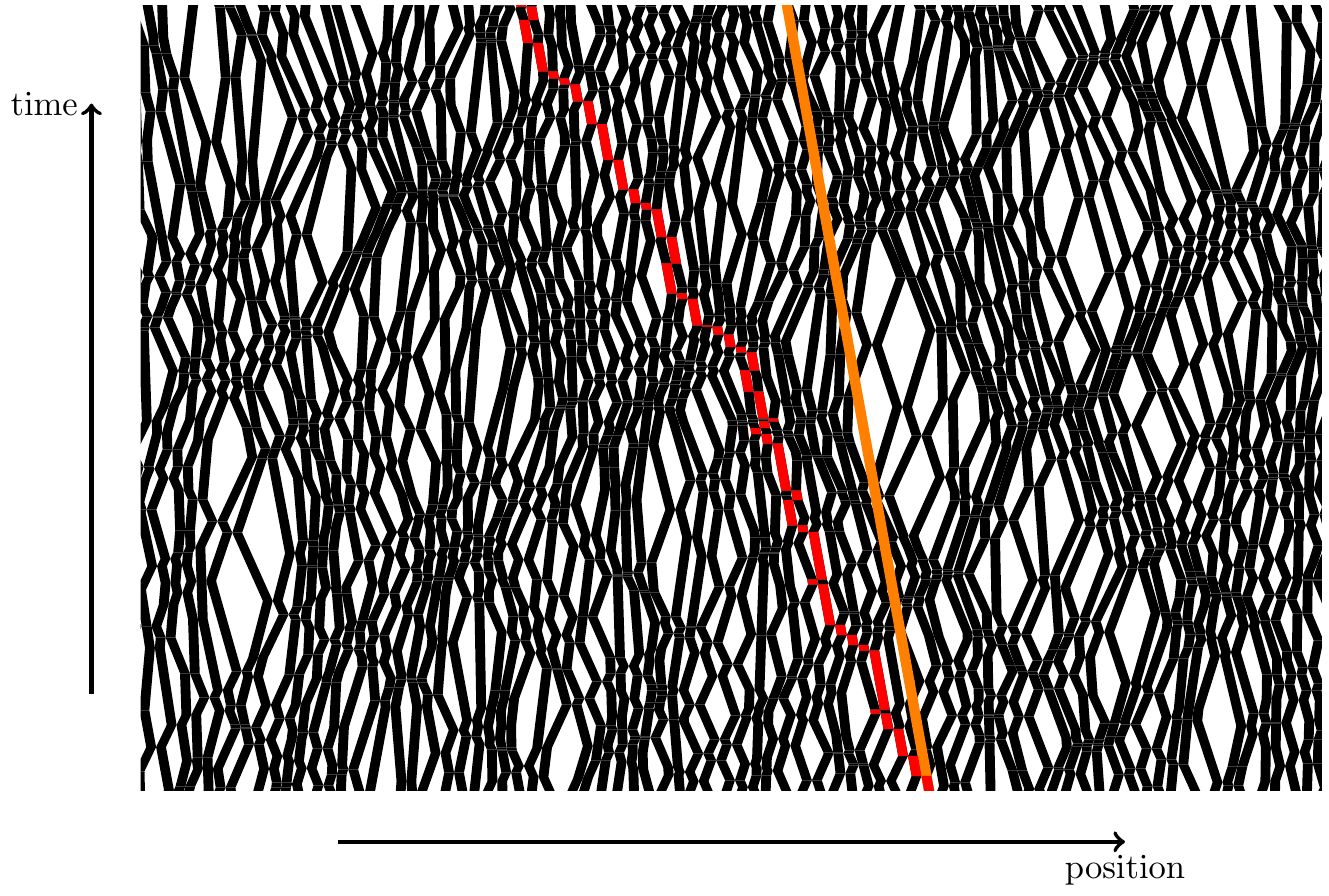}};
    \end{tikzpicture}
    \caption{The simplest system that exhibits `Generalized Hydrodynamics' is arguably the classical hard rod gas, i.e. identical billiard balls whose motion is restricted to a line. Left: the balls collide elastically and exchange their velocities. One can re-index the balls after each collision so that the `bare' velocity $v_j$ is constant (here the red ball is the one with velocity $v_2$ at any time). Right: at large distance and time scales, the effective velocity of the red ball $v^{\rm eff}$ (red trajectory) is different from its `bare' velocity $v$ (orange trajectory). The `Generalized Hydrodynamics' description of the hard rod gas (\ref{eq:GHDintro}) is an Euler hydrodynamic limit where the interactions between the particles enter through the effective velocity.}
    \label{fig:hardrod}
\end{figure}

To illustrate the emergence of `Generalized Hydrodynamics' in a system with infinitely many conserved quantities, it is instructive to think about $N$ identical billiard balls of diameter $| \Delta|$ whose motion is restricted to a one-dimensional line, see Fig.~\ref{fig:hardrod}. Here we take $\Delta < 0$. [This funny convention ensures that the hydrodynamic equations for the hard core gas (\ref{eq:GHDintro}) are almost the same as the ones for the Lieb-Liniger gas, see Eq.~(\ref{eq:ghd}). $\Delta$ is positive in the repulsive one-dimensional Bose gas, see Subsection~\ref{subsec:Wigner}.] This model for a classical one-dimensional gas is known as the `hard rod gas' in the statistical physics literature, see e.g.~\citep{percus1976equilibrium,lebowitz1967kinetic,aizenman1975ergodic,boldrighini1983one,spohn2012large,boldrighini1997one,doyon2017dynamics,cao2018incomplete}. The balls are at position $x_j$ and move at velocity $v_j$, $j=1,\dots , N$. When two balls collide elastically, they exchange their velocities, so the set of velocities is conserved at any time. Thus, this many-particle system has infinitely many conserved quantities that are independent in the thermodynamic limit $N\rightarrow \infty$. Indeed, for any function $f$ of the velocity, the charge $Q[f] := \sum_{j=1}^N f(v_j)$ is conserved.

One can introduce a coarse-grained phase-space density of balls $\rho(x,v) = \sum_{j=1}^N \delta_\ell (x-x_j) \delta_\sigma (v-v_j)$, where $\delta_\ell$ and $\delta_\sigma$ are smooth distributions with weight one peaked around the origin, for instance two Gaussians of width $\ell$ and $\sigma$. When $\ell$ and $\sigma$ are large enough so that the phase space volume $[x,x+ \ell] \times [v,v+\sigma]$ contains a very large number of balls, but small enough so that the density stays constant through the volume, the coarse-grained density evolves according to the two equations
\begin{equation}
    \label{eq:GHDintro}
    \left\{ \begin{array}{c}
       \displaystyle \partial_t \rho(x,v,t) + \partial_x \left( v^{\rm eff}[\rho](v) \, \rho(x,v,t) \right) - \frac{1}{m}\frac{\partial V(x)}{\partial x} \partial_v \rho(x,v,t) \, = \, 0, \\
       \displaystyle  v^{\rm eff}[\rho](v) = v - \Delta  \int_{-\infty}^\infty   \left( v^{\rm eff}[\rho](v) - v^{\rm eff}[\rho](w) \right)  \rho(w) dw ,
    \end{array} \right.
\end{equation}
where we have included an external potential $V(x)$. These are the Generalized Hydrodynamics equations for the hard rod gas, initially derived by~\cite{percus1976equilibrium}, and proved by \cite{boldrighini1983one} for $V(x)=0$. The inclusion of the trapping potential, and its breaking of the conservation laws, was investigated more recently by~\cite{cao2018incomplete}.

The first equation (\ref{eq:GHDintro}) is similar to the transport equation (\ref{eq:liouville}), although two important differences need to be stressed. The first difference lies in the range of applicability of Eq.~(\ref{eq:GHDintro}): it is a  coarse-grained description of the hard rod gas based on local relaxation, which is valid only at the Euler scale. The free transport equation~(\ref{eq:liouville}), on the other hand, does not rely on hydrodynamic assumptions. The second difference is that, instead of the single-particle group velocity, Eq.~(\ref{eq:GHDintro}) involves an `effective velocity'. That effective velocity is a functional of the density $\rho(v)$ at a given position $x$ and time $t$, defined by the second equation (\ref{eq:GHDintro}). It has a simple interpretation, see Fig.~\ref{fig:hardrod}. At each collision, the labels of the colliding balls can be switched, so that each velocity $v_j$ stays constant, but the position $x_j$ changes instantaneously by $\pm |\Delta|$ (the diameter of the balls). For a finite density of balls, these jumps result in a modification of the propagation velocity of the ball with velocity $v_j$ through the gas, $v_j \rightarrow v^{\rm eff}[\rho](v_j)$.

The Generalized Hydrodynamics equations~(\ref{eq:GHDintro}) are also analogous to the Euler hydrodynamic equations (\ref{eq:continuity3})-(\ref{eq:euler}), but for infinitely many charges. All the charges are conserved in the absence of an external potential ($V(x) = 0$), while for $V(x) \neq 0$  only the conservation of mass and energy are expected to survive, generically. To see this, consider the aforementioned charges $Q[f] = \sum_{j=1}^N f(v_j)$, and their associated charge densities $q[f](x,t) = \int_{-\infty}^\infty f(v) \rho(x,v,t) dv$. Those charge densities evolve according to
\begin{equation}
    \label{eq:introGHD_continuity}
    \begin{array}{l}
   \displaystyle     \frac{\partial}{\partial t} q[f](x,t) + \frac{\partial}{\partial_x} j[f](x,t) \, = \, -\frac{1}{m} \frac{\partial V(x)}{\partial x} \int_{-\infty}^\infty  f'(v) \rho(x,v,t)  dv, \\
   \displaystyle {\rm with}  \qquad j[f](x,t) = v^{\rm eff}[\rho](v) q[f](x,t) .
    \end{array}
\end{equation}
When $V(x) = 0$, the first equation is a continuity equation that expresses the conservation of $Q[f]$. The second line gives the expectation value of the current as a function of the charge densities under the hydrodynamic assumptions.

Thus, as claimed above, Generalized Hydrodynamics captures a peculiar fluid-like behavior which resembles both a fluid obeying the free transport equation~(\ref{eq:liouville}), and one obeying the Euler hydrodynamic equations~ (\ref{eq:continuity3},\ref{eq:euler}).

\vspace{0.5cm}

Remarkably, the Generalized Hydrodynamics equations (\ref{eq:GHDintro}) have reappeared in 2016, in the context of quantum integrable one-dimensional systems~\citep{castro2016emergent,bertini2016transport}. In the decade that preceded this 2016 breakthrough, tremendous progress had been made on out-of-equilibrium quantum dynamics, largely driven by advances in cold atom experiments. To name but one example, the 2006 Quantum Newton Cradle experiment of~\cite{kinoshita2006quantum}, where two one-dimensional  clouds of interacting atoms in a harmonic potential $V(x)$ undergo thousands of collisions, seemingly escaping convergence towards thermal equilibrium, had become an important source of inspiration and a challenge for quantum many-body theorists. Many important conceptual advances on the thermalization (or absence thereof) of isolated quantum systems, in particular the developments around the notion of Generalized Gibbs Ensemble, occurred between 2006 and 2016. Yet, a quantitatively reliable modeling of the Quantum Newton Cradle setup, with experimentally realistic parameters, had remained completely out of reach. As usual with quantum many-body systems, the exponential growth of the Hilbert space with the number of atoms $N$ seemingly prevented direct numerical simulations of the dynamics.

The 2016 breakthrough of Generalized Hydrodynamics has completely changed this state of affairs. Realizing that the dynamics of one-dimensional ultracold quantum gases in experiments such as the Quantum Newton Cradle is captured by Generalized Hydrodynamics equations of the form (\ref{eq:GHDintro}) has ushered in a new era for their theory description.

\vspace{1cm}

{\bf Goal of this review and organization.} Our purpose is to 
give a pedagogical overview of the developments that have occurred on the 1D Bose gas since the 2016 discovery of Generalized Hydrodynamics in quantum integrable systems \citep{castro2016emergent,bertini2016transport}. Because the topic is of interest to both cold atom physicists and quantum statistical physicists, we have attempted to write this review article in a way that makes it accessible to all.

The article is organized as follows. In Sec.~\ref{sec:LiebLiniger} we review the basic facts about the theory of the 1D Bose gas that are useful to understand the development of Generalized Hydrodynamics. We provide an introduction to the repulsive Lieb-Liniger model, with a strong emphasis on the key concept of the rapidities. We also review the asymptotic regimes of the 1D Bose gas (quasicondensate, ideal Bose gas, and hard-core regimes),  which are often important in the description of experiments.
In Sec.~\ref{sec:GHDtheory} we present the Generalized Hydrodynamics description of the 1D Bose gas, and review the theory results that have been obtained since 2016 with this approach. In Sec.~\ref{sec:experiments_beforeGHD} we briefly review the experimental setups that have been used to realize 1D Bose gases, and the main experimental results obtained in connection with integrability. In that section we focus mostly on results obtained prior to the advent of Generalized Hydrodynamics. Then, in Sec.~\ref{sec:GHDexperiment}, we present the experimental tests of GHD, and recent experiments whose descriptions have relied on GHD. In  Sec.~\ref{sec:losses}, we briefly discuss the recent theory developments that aim at describing the effect of atom losses. We discuss some perspectives and open questions in the Conclusion.

\tableofcontents

\newpage

\section{The Lieb-Liniger model and the rapidities}
\label{sec:LiebLiniger}

In the absence of an external potential, the Hamiltonian of one-dimensional bosons with delta repulsion is, in second quantized form, 
\begin{equation}
	\label{eq:hamLL}
	H = \int \Psi^\dagger(x) \left[ -\frac{\hbar^2 \partial_x^2}{2m} - \mu + \frac{g}{2} \Psi^\dagger(x) \Psi(x) \right] \Psi(x) \, dx .
\end{equation}
Here $ \Psi^\dagger(x)$ and $ \Psi(x)$ are the boson creation/annihilation operators that satisfy the canonical commutation relations $\left[ \Psi(x) , \Psi^\dagger (x') \right] \, = \, \delta (x-x')$, $m$ is the mass of the bosons, $g>0$ is the 1D repulsion strength, and $\mu$ is the chemical potential. The total number of particles in the system is $N = \int \left< \Psi^\dagger(x) \Psi (x)\right> dx$. In the literature, it is customary to define the parameter $c = mg/\hbar^2$, homogeneous to an inverse length. In a box of length $L$, the ratio of $c$ to the particle density $n=N/L$ gives the dimensionless repulsion strength, or Lieb parameter,
\begin{equation}
	\label{eq:gamma}
	\gamma = \frac{c}{n} = \frac{m g}{\hbar^2 n} .
\end{equation}
In the rest of this section we review some basic facts about the exact solution of the model (\ref{eq:hamLL}) of \cite{lieb1963exact}, see~\citep{korepin1997quantum,gaudin2014bethe} for introductions. In particular, we emphasize the crucial concept of the  {\it rapidities}, and we review a number of results that have proved useful in the recent developments of Generalized Hydrodynamics. We set $\hbar = m =1$.

\subsection{The scattering shift (or Wigner time delay)}
\label{subsec:Wigner}

It is instructive to start with the case of $N=2$ particles on an infinite line. In first quantization, using center-of-mass and relative coordinates $X = (x_1+x_2)/2$ and $Y = x_1-x_2$, the Hamiltonian (\ref{eq:hamLL}) splits into a sum of two independent one-body problems,
\begin{eqnarray}
	\label{eq:ham_2p}
	H &=& -\frac{1}{2} \partial_{x_1}^2 - \frac{1}{2} \partial_{x_2}^2 + c \, \delta(x_1-x_2)  \, = \,  -\frac{1}{4}  \partial_X^2 -  \partial_Y^2  + c \, \delta(Y)  .
\end{eqnarray}
The eigenstates of the center-of-mass Hamiltonian $-\frac{1}{4} \partial_X^2$ are plane waves, and the Hamiltonian for the relative coordinate $Y$ is the one of a particle of mass $1/2$ in the presence of a delta potential at $Y=0$. Because of that delta potential, the first derivative of the wavefunction $\varphi(Y)$ must have a discontinuity at $Y=0$: $ \varphi'(0^+) - \varphi'(0^-)   \, - \, c \, \varphi(0)  = 0$. Coming back to the original coordinates, one sees that the two-body wavefunction $\varphi(x_1,x_2) = \left<  0\right| \Psi(x_1) \Psi(x_2) \left| \varphi \right>$ satisfies
\begin{equation}
	\label{eq:singularity}
	\lim_{x_2 \rightarrow x_1^+} \left[  \partial_{x_2} \varphi(x_1,x_2) - \partial_{x_1} \varphi(x_1,x_2)  \, - \, c \,  \varphi(x_1,x_2) \right]  = 0 .
\end{equation}
The same condition holds for $x_1$ exchanged with $x_2$, since the wavefunction is symmetric. Thus the eigenstates of (\ref{eq:ham_2p}) are
\begin{equation}
	\label{eq:2bd_wf}
	\varphi (x_1,x_2) \, \propto \, \left\{\begin{array}{ccl}
		(\theta_2 - \theta_1 - i c)  e^{i x_1 \theta_1 + i x_2 \theta_2} - (\theta_1 - \theta_2 - i c)  e^{i  x_1 \theta_2 + i x_2 \theta_1} &{\rm if} & x_1 < x_2  \\
		(x_1 \leftrightarrow x_2)  &{\rm if} & x_1 > x_2 ,
	\end{array} \right.
\end{equation}
corresponding to the eigenvalues $(\theta_1^2+\theta_2^2)/2$. For $\theta_1 > \theta_2$, the two terms $e^{i x_1 \theta_1 + i x_2 \theta_2}$ and $e^{i x_1 \theta_2 + i x_2 \theta_1}$ correspond to the in-coming and out-coming pairs of particles in a two-body scattering process. The ratio of their amplitudes is the two-body scattering phase,
\begin{equation}
	\label{eq:2bd_scattering_phase}
	e^{i \phi (\theta_1 - \theta_2)} := \frac{\theta_1 - \theta_2 - i c}{\theta_2 - \theta_1 - i c} .
\end{equation}
An equivalent expression for that phase, often used in the literature and which we also use below, is $\phi (\theta) = 2 \, {\rm arctan} (\theta/c) \in [-\pi,\pi]$.

\begin{figure}
    \centering
    \begin{tikzpicture}
        \begin{scope}
        \draw[thick,->] (-1.5,-1.5) -- (-0.8,-0.8);
        \draw[thick,] (-1,-1) -- (0,0);
        \draw[thick,->] (1.5,-1.5) -- (0.8,-0.8);
        \draw[thick,] (1,-1) -- (0,0);
        \draw[thick,dashed] (0,0) -- (1.8,1.8);
        \draw[thick,dashed] (0,0) -- (-1.8,1.8);

        \draw[thick,->] (-0.4,0) -- ++(1,1);
        \draw[thick] (-0.4,0) -- ++(1.8,1.8);
        \draw[thick,->] (0.4,0) -- ++(-1,1);
        \draw[thick] (0.4,0) -- ++(-1.8,1.8);

        \fill[blue,draw=black] (0,0.1) circle (0.5cm);
        \draw[<->] (1.4,1.9) -- (1.8,1.9);
        \draw (1.6,2.2) node{$\Delta$};
        \draw[<->] (-1.8,1.9) -- (-1.4,1.9);
        \draw (-1.6,2.2) node{$\Delta$};
        
        \draw[->] (-2.5,-2.5) -- (2,-2.5);
        \draw (2,-2.5) node[above]{$x$};
        \draw[->] (-2.5,-2.5) -- ++(0,4.5) node[left]{$t$};

        \draw (-1.6,-1.9) node{$e^{i x_1 \theta_1}$};
        \draw (1.6,-1.9) node{$e^{i x_2 \theta_2}$};
        \draw (-0.8,1.9) node{$e^{i x_1 \theta_2}$};
        \draw (0.9,1.9) node{$e^{i x_2 \theta_1}$};
        \end{scope}
        
        \begin{scope}[xshift=8cm,yshift=2.5cm]
        \draw[thick,] (-1.5,-2) -- (1.4,1);
        \draw[thick,] (-0.7,-2) -- (1,1);
        \draw[thick,] (0,-2) -- (0.3,1);
        \draw[thick,] (0.4,-2) -- (-1.1,1);
        \draw[thick,] (1.5,-2) -- (-2,1);
 
        \fill[blue,draw=black] (-0.1,-0.45) circle (1cm);

        \draw[->] (-2.5,-2.5) -- (3,-2.5) node[right]{$x$};
        \draw[->] (-2.5,-2.5) -- ++(0,3.5) node[left]{$t$};

        \draw (-1.6,-2.2) node{$e^{i x_1 \theta_1}$};
        \draw (-0.7,-2.2) node{$e^{i x_2 \theta_2}$};
        \draw (1.7,-2.2) node{$e^{i x_N \theta_N}$};

        \draw (-1.8,1.3) node{$e^{i x_1 \theta_N}$};
        \draw (-0.6,1.3) node{$e^{i x_2 \theta_{N-1}}$};
        \draw (1.8,1.3) node{$e^{i x_N \theta_1}$};
        \end{scope}
        
        \begin{scope}[xshift=8cm,yshift=-2.5cm]
        \draw[thick,] (-1.5,-2) -- (1.4,1);
        \draw[thick,] (-0.7,-2) -- (1,1);
        \draw[thick,] (0,-2) -- (0.3,1);
        \draw[thick,] (0.4,-2) -- (-1.1,1);
        \draw[thick,] (1.5,-2) -- (-2,1);
 
        \fill[blue,draw=black] (-0.4,-0.35) circle (1.2mm);
        \fill[blue,draw=black] (-0.15,-0.54) circle (1.1mm);
        \fill[blue,draw=black] (-0.3,-0.72) circle (1.2mm);
        \fill[blue,draw=black] (-0.1,-1) circle (1.2mm);
        \fill[blue,draw=black] (0.03,-0.7) circle (0.8mm);
        \fill[blue,draw=black] (0.05,-1.3) circle (1mm);
        \fill[blue,draw=black] (0.1,-0.85) circle (0.8mm);
        \fill[blue,draw=black] (0.15,-0.5) circle (0.8mm);
        \fill[blue,draw=black] (0.17,-0.22) circle (1mm);
        \fill[blue,draw=black] (0.42,0) circle (1.2mm);

        \draw[->] (-2.5,-2.5) -- (3,-2.5) node[right]{$x$};
        \draw[->] (-2.5,-2.5) -- ++(0,3.5) node[left]{$t$};

        \draw (-1.6,-2.2) node{$e^{i x_1 \theta_1}$};
        \draw (-0.7,-2.2) node{$e^{i x_2 \theta_2}$};
        \draw (1.7,-2.2) node{$e^{i x_N \theta_N}$};

        \draw (-1.8,1.3) node{$e^{i x_1 \theta_N}$};
        \draw (-0.6,1.3) node{$e^{i x_2 \theta_{N-1}}$};
        \draw (1.8,1.3) node{$e^{i x_N \theta_1}$};

        \end{scope}
        
        \draw[->] (5,1.8) arc (110:250:1cm and 2.5cm) node[left]{factorization};
        
    \end{tikzpicture}
    \caption{Left: the wavefunction (\ref{eq:2bd_wf}) on the infinite line corresponds to a two-body scattering process. Semiclassically, the scattering phase in that two-body process is reflected in the scattering shift (\ref{eq:scattering_phase}): after the collision, the position of the particle has been shifted by a distance  $\Delta(\theta_1-\theta_2)$. Right: the Bethe wavefunction (\ref{eq:bethewf}) on the infinite line corresponds to an $N$-body scattering process which factorizes into two-body processes (the scattering shift $\Delta$ is also present here, but it is not drawn in the cartoon). In that $N$-body process, the rapidities $\theta_j$ are the asymptotic momenta of the bosons.}
    \label{fig:scattering}
\end{figure}
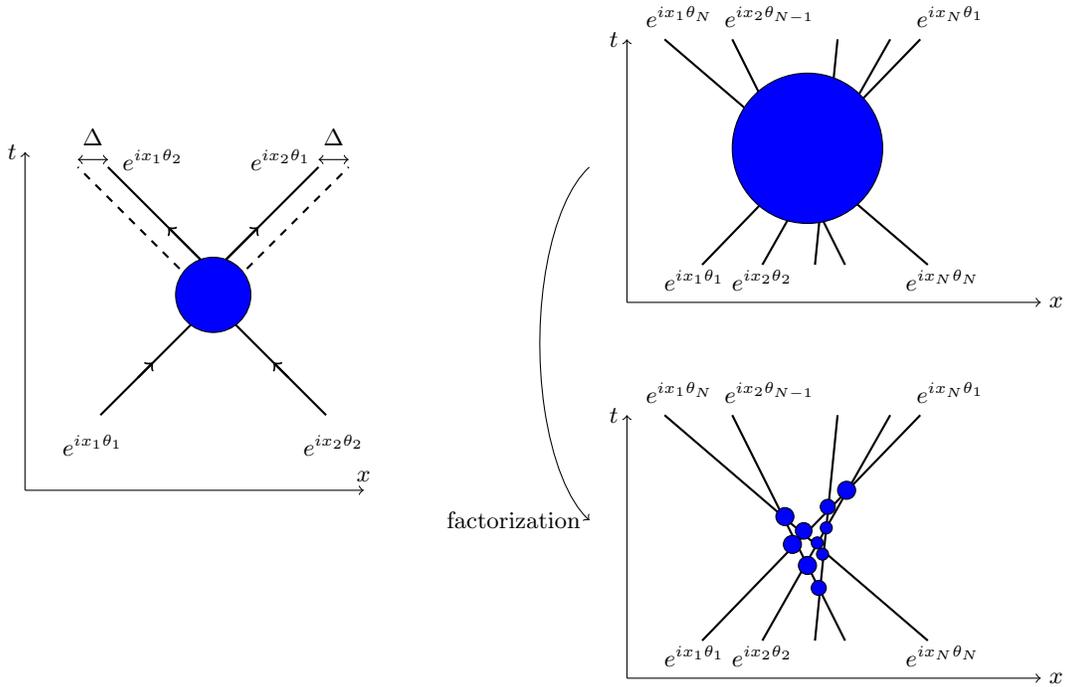

It was pointed out by~\cite{eisenbud1948formal} and by~\cite{wigner1955lower} that the scattering phase may be viewed semiclassically as a `time delay'. Let us briefly sketch the argument of~\cite{wigner1955lower}. First, we note that, for a single particle, a simple substitute for a wavepacket is a superposition of two plane waves with momenta $\theta$ and $\theta + \delta \theta$,
\begin{equation}
	e^{i x \theta   }  + e^{i x (\theta + \delta \theta)  } .
\end{equation}
Such a superposition evolves in time as $e^{i x \theta  - i  t \varepsilon(\theta )  }  + e^{i x (\theta + \delta \theta)  - i  t \varepsilon(\theta + \delta \theta)}$, 
where $\varepsilon(\theta) = \theta^2/2$ is the energy. The center of this `wave packet' is at the position where the phases of the two terms coincide, namely the point where $x \delta \theta  - t [\varepsilon(\theta+\delta \theta) - \varepsilon(\theta)]  = 0$, which gives $x \simeq v t$ with the group velocity $v = d \varepsilon/d \theta = \theta$. So this is indeed a `wave packet' moving at speed $\theta$. Next, consider two incoming particles in a state such that the center of mass $(x_1+x_2)/2$ has momentum $\theta_1+\theta_2$, while the relative coordinate $x_1 - x_2$ is in a `wave packet' moving at velocity $(\theta_1 - \theta_2)/2$,
\begin{eqnarray}
	\label{eq:incoming2}
\nonumber	\psi_{{\rm inc.}} (x_1,x_2) & \propto & e^{i \frac{x_1+x_2}{2} (\theta_1 + \theta_2) } \left( e^{i (x_1 - x_2) \frac{\theta_1 - \theta_2}{2} } + e^{i (x_1 - x_2) ( \frac{\theta_1 - \theta_2}{2} + \delta \theta) } \right)  \\
	& = & e^{i x_1 \theta_1 + i x_2 \theta_2}  + e^{i x_1 (\theta_1 + \delta \theta) + i x_2 (\theta_2 - \delta \theta ) } . 
\end{eqnarray}
According to Eqs.~(\ref{eq:2bd_wf})-(\ref{eq:2bd_scattering_phase}), the corresponding out-coming state would be
\begin{eqnarray}
\nonumber	\psi_{{\rm outc.}} (x_1,x_2) & \propto &  - e^{i \phi(\theta_1 - \theta_2)}  e^{i x_1 \theta_2 + i x_2 \theta_1}    - e^{i \phi(\theta_1  - \theta_2 +2 \delta \theta )}  e^{i x_1 (\theta_2 - \delta \theta) + i x_2 (\theta_1 + \delta \theta ) }  \\
		& = & e^{i \frac{x_1+x_2}{2} (\theta_1 + \theta_2) } \left( - e^{i \phi (\theta_1 -\theta_2)} e^{i (x_2 - x_1) \frac{\theta_1 - \theta_2}{2} } - e^{i \phi (\theta_1 -\theta_2 + 2 \delta \theta)}  e^{i (x_2 - x_1) ( \frac{\theta_1 - \theta_2}{2} + \delta \theta) } \right)  . \qquad
\end{eqnarray}
Then, repeating the previous argument of phase stationarity, one finds that the relative coordinate is at position $x_1 - x_2 \simeq  \frac{\theta_1-\theta_2}{2} t - 2 d\phi/d\theta$ at time $t$. Since the center of mass is not affected by the collision and moves at the group velocity $(\theta_1+\theta_2)/2$, we see that the position of the two semiclassical particles after the collision will be
\begin{equation}
	x_1 \simeq \theta_1 t - \Delta (\theta_2- \theta_1) ,  \qquad x_2 \simeq \theta_2 t + \Delta (\theta_2- \theta_1)  , 
\end{equation}
where the scattering shift $\Delta (\theta)$ is given by the derivative of the scattering phase,
\begin{equation}
	\label{eq:scattering_phase}
	\Delta (\theta) \, := \, \frac{d\phi (\theta)}{d\theta}  \, = \, \frac{2c}{c^2 + \theta^2} .
\end{equation}
The two particles are delayed: their position after the collision is the same as if they were late by a time $\delta t_1 = \Delta(\theta_2- \theta_1)/v_1$ and $\delta t_2 = \Delta(\theta_2- \theta_1)/v_2$ respectively.

\subsection{The Bethe wavefunction, and the rapidities as asymptotic momenta}
\label{subsec:asymptotic_momenta}

For more particles, the eigenstates of the Hamiltonian (\ref{eq:hamLL}) on the infinite line are Bethe states  $\left| \{\theta_a\} \right>$ labeled by a set of $N$ numbers $\{ \theta_a\}_{1 \leq a \leq N}$, called the rapidities. In the domain $x_1<x_2< \dots < x_N$, the wavefunction is~\citep{lieb1963exact,korepin1997quantum,gaudin2014bethe}
\begin{eqnarray}
	\label{eq:bethewf}
	\varphi_{\{ \theta_a \} }(x_1,\dots, x_N) & = & \left< 0 \right| \Psi(x_1) \dots \Psi(x_N) \left| \{\theta_a\} \right> \\
\nonumber	&\propto &  \sum_\sigma (-1)^{|\sigma |}  \left(\prod_{1 \leq a<b\leq N} (\theta_{\sigma (b)} - \theta_{\sigma (a)} - i c )\right) e^{i \sum_{j} x_j  \theta_{\sigma (j)}  } ,
\end{eqnarray}
and it is extended to other domains by symmetry $x_i \leftrightarrow x_j$. Here the sum runs over all permutations $\sigma$ of $N$ elements (so there are $N!$ terms) and $(-1)^{|\sigma|}$ is the signature of the permutation. The momentum and energy of the eigenstate (\ref{eq:bethewf}) are
\begin{equation}
    \label{eq:thetaPE}
    P = \sum_{a=1}^N \theta_a , \qquad  E = \sum_{a=1}^N \frac{\theta_a^2}{2} .
\end{equation}
The rapidities $\theta_a$ are  conveniently thought of as the  asymptotic momenta in an $N$-body scattering process. For $\theta_1 > \theta_2 > \dots > \theta_N$, the combination of two terms
\begin{equation}
    e^{i x_1 \theta_1 + \dots + i x_N \theta_N} + \left( \prod_{1 \leq a < b \leq N} -e^{i \phi(\theta_a - \theta_b) } \right) e^{i x_1 \theta_N + \dots + i x_N \theta_1} 
\end{equation}
that appears in (\ref{eq:bethewf}) can be viewed as the sum of in-coming ($e^{i x_1 \theta_1 + \dots + i x_N \theta_N}$) and out-coming states ($e^{i x_1 \theta_N + \dots + i x_N \theta_1}$) in an $N$-body scattering process, see Fig.~\ref{fig:scattering}. Their respective amplitude $\prod_{1 \leq a < b \leq N} -e^{i \phi(\theta_a - \theta_b) }$ is a many-body phase, which depends on all in-coming rapidities. Crucially, this many-body phase factorizes into a product of two-body scattering phases (\ref{eq:2bd_scattering_phase}): this is a central property of all quantum integrable systems.

The fact that the rapidities are the asymptotic momenta in a scattering process implies that they can be measured by letting the bosons expand freely along the infinite line \citep{rigol2005fermionization,minguzzi2005exact,buljan2008fermi,jukic2008free,del2008fermionization,bolech2012long,campbell_sudden_2015,mei2016unveiling,caux2019hydrodynamics,wilson_observation_2020,malvania2020generalized}. Here we follow the argument of \citep{campbell_sudden_2015}. 

Consider a state $\left| \psi_t \right>$ of $N$ bosons confined to some interval around the origin at time $t$. This could be, for instance, the ground state in a trapping potential $V(x)$, or some out-of-equilibrium state produced by some quench protocol, also in a trapping potential to ensure that the bosons are initially confined. This many-body state can be expanded in the basis of Bethe states,
\begin{eqnarray}
    \label{eq:decomp_psit}
\nonumber    \left| \psi_t \right> & = &  \int_{\theta_1> \theta_2 > \dots > \theta_N} d\theta_1 d\theta_2 \dots d\theta_N \, \left< \{ \theta_a\} \left| \psi_t \right> \right.  \, \left| \{ \theta_a\} \right> \\
    & = &  \frac{1}{N!} \int  d\theta_1 d\theta_2 \dots d\theta_N \, \left< \{ \theta_a\} \left| \psi_t \right> \right.  \, \left| \{ \theta_a\} \right>,
\end{eqnarray}
where the Bethe states on the infinite line are normalized such that $\left< \{ \theta_a\} \left| \{ \theta'_a\} \right> \right. = \prod_{a=1}^N \delta(\theta_a - \theta'_a) $ (assuming that both sets of rapidities are ordered, $\theta_1 > \dots> \theta_N$ and $\theta_1' > \dots >\theta_N'$). The integral is restricted to the domain $\theta_1> \theta_2 > \dots > \theta_N$ in the first line to avoid double-counting. Notice that, with the definition (\ref{eq:bethewf}), the Bethe states are anti-symmetric under exchange of two rapidities $\theta_a \leftrightarrow \theta_b$. Then, plugging (\ref{eq:bethewf}) into (\ref{eq:decomp_psit}), and using this antisymmetry, one obtains~\citep{campbell_sudden_2015}
\begin{equation}
    \left< 0 \right| \Psi(x_1) \dots \Psi(x_N) \left| \psi_t \right> \, \propto \,  \int d\theta_1 \dots d\theta_N  \, \left< \{ \theta_a\} \left| \psi_t \right>\right. e^{i \sum_{a<b} \phi( \theta_a - \theta_b) }  e^{i \sum_a x_a \theta_{N+1-a} } ,
\end{equation}
for $x_1 < x_2 < \dots < x_N$. This expression is particularly convenient to analyze the expansion. When the trapping potential $V(x)$ is switched off at time $t$, and the bosons are let to evolve freely along the infinite line, the probability to find them at positions $x_1, x_2, \dots , x_N$ after an expansion time $t_{\rm exp}$ is
\begin{eqnarray}
    \label{eq:1D_expansion}
\nonumber    && P_{\rm exp}(x_1, \dots, x_N) \, =\, \left| \left< 0 \right| \Psi(x_1,t_{\rm exp}) \dots \Psi(x_N,t_{\rm exp}) \left| \psi_t \right>  \right|^2 \\
\nonumber    && \qquad \quad  \propto \,  \left| \int d\theta_1 \dots d\theta_N  \, \left< \{ \theta_a\} \left| \psi_t \right>\right. e^{i \sum_{a<b} \phi( \theta_a - \theta_b) }  e^{i \sum_a  ( x_a \theta_{N+1-a} - i \frac{t}{2} \theta_{N+1-a}^2) } \right|^2 \\
    &&  \qquad  \underset{t_{\rm exp. \rightarrow \infty}}{=}  \quad  \frac{1}{t_{\rm exp}^N} \left| \left< \{ x_N/t_{\rm exp}, \dots , x_1/t_{\rm exp} \} \left| \psi_t \right>\right. \right|^2  .
\end{eqnarray}
From the second to the third line, we have used the stationary phase approximation, taking the limit $t_{\rm exp} \rightarrow \infty$ while keeping the ratios $x_1/t_{\rm exp}, \dots , x_N/t_{\rm exp}$ fixed. The proportionality factor is fixed by imposing that $\int_{x_1<\dots<x_N} P(x_1,\dots , x_N) dx_1 \dots dx_N = 1$.

In conclusion, we see from (\ref{eq:1D_expansion}) that the joint probability distribution of the positions of the atoms after a large 1D expansion time directly reflects the distribution of rapidities in the state $\left| \psi_t \right>$ just before the expansion. This is very important because it means that the rapidities can be measured experimentally, by performing such 1D expansions. This has been done experimentally for the first time by \cite{wilson_observation_2020}. This experiment is discussed in Section~\ref{sec:experiments_beforeGHD} below.

\subsection{Finite density and the Bethe equations}

In the two previous subsections, we have focused on a finite number of bosons on the infinite line, corresponding to a vanishing density of particles. But, to understand the thermodynamic properties of the model, one needs to work with a finite density $N/L$. This can be done by imposing periodic boundary conditions, identifying the points $x=0$ and $x=L$ in the system. Imposing periodic boundary conditions on the Bethe wavefunction (\ref{eq:bethewf}), i.e. $\varphi_{ \{ {\bf \theta}_a \} } (x_1, \dots, x_{N-1}, L) \, = \, \varphi_{ \{ {\bf \theta}_a \} } (0, x_1, \dots, x_{N-1})$, leads to the Bethe equations 
\begin{equation}
	 e^{i \theta_a L} \prod_{b \neq a} e^{i \phi( \theta_a - \theta_b) } = (-1)^{N-1} , \qquad a=1,\dots, N,
\end{equation}
where the two-body scattering phase $\phi(\theta_a - \theta_b)$ is defined in Eq.~(\ref{eq:2bd_scattering_phase}). 
Taking the logarithm on both sides, one gets the following system of $N$ coupled non-linear equations
\begin{equation}
    \label{eq:logbethe}
	\theta_a + \frac{1}{L} \sum_{b \neq a} 2~{\rm arctan} \left( \frac{\theta_a - \theta_b}{c} \right)  \, = \, p_a ,  \qquad {\rm where} \quad \left\{ \begin{array}{l}
        p_a \in \frac{2\pi}{L} \mathbb{Z}  \quad {\rm for } \quad N \quad {\rm odd} \\	
        p_a \in \frac{2\pi}{L} (\mathbb{Z}  +\frac{1}{2}) \quad {\rm for } \quad N \quad {\rm even}.
	\end{array} \right.
\end{equation}
It is convenient to think of the numbers $p_a$ as the momenta of $N$ non-interacting fermions (with periodic or anti-periodic boundary conditions, depending on the parity of $N$). These fermion momenta have the following interpretation. For fixed $N$ and $L$, one can adiabatically follow each eigenstate $\left| \{ \theta_a\}_{1\leq a \leq N} \right>$ as one varies the repulsion strength $c$. In the infinite repulsion limit $c \rightarrow +\infty$, the bosonic wavefunction (\ref{eq:bethewf}) is, up to multiplication by a sign $\prod_{a<b }{{\rm sign}(x_b-x_a)}$, equal to the Slater determinant of $N$ non-interacting fermions, i.e. ${\rm det} \left[ e^{i x_a \theta_b} \right]_{1 \leq a,b \leq N}$~\citep{girardeau1960relationship}. In that limit, the rapidities are nothing but the momenta of these non-interacting fermions: $\theta_a = p_a$. Importantly, the fermions in the $c\rightarrow + \infty$ limit must obey the Pauli exclusion principle, so all momenta should be different: $p_a \neq p_b$ if $a\neq b$. In the following we order both the rapidities and the fermion momenta as
\begin{equation}
    \label{eq:orderrapidities}
    \theta_1 > \theta_2  > \dots > \theta_N , \qquad  \quad p_1 > p_2  > \dots > p_N .
\end{equation}

\begin{figure}
    \centering
    \includegraphics[width=0.65\textwidth]{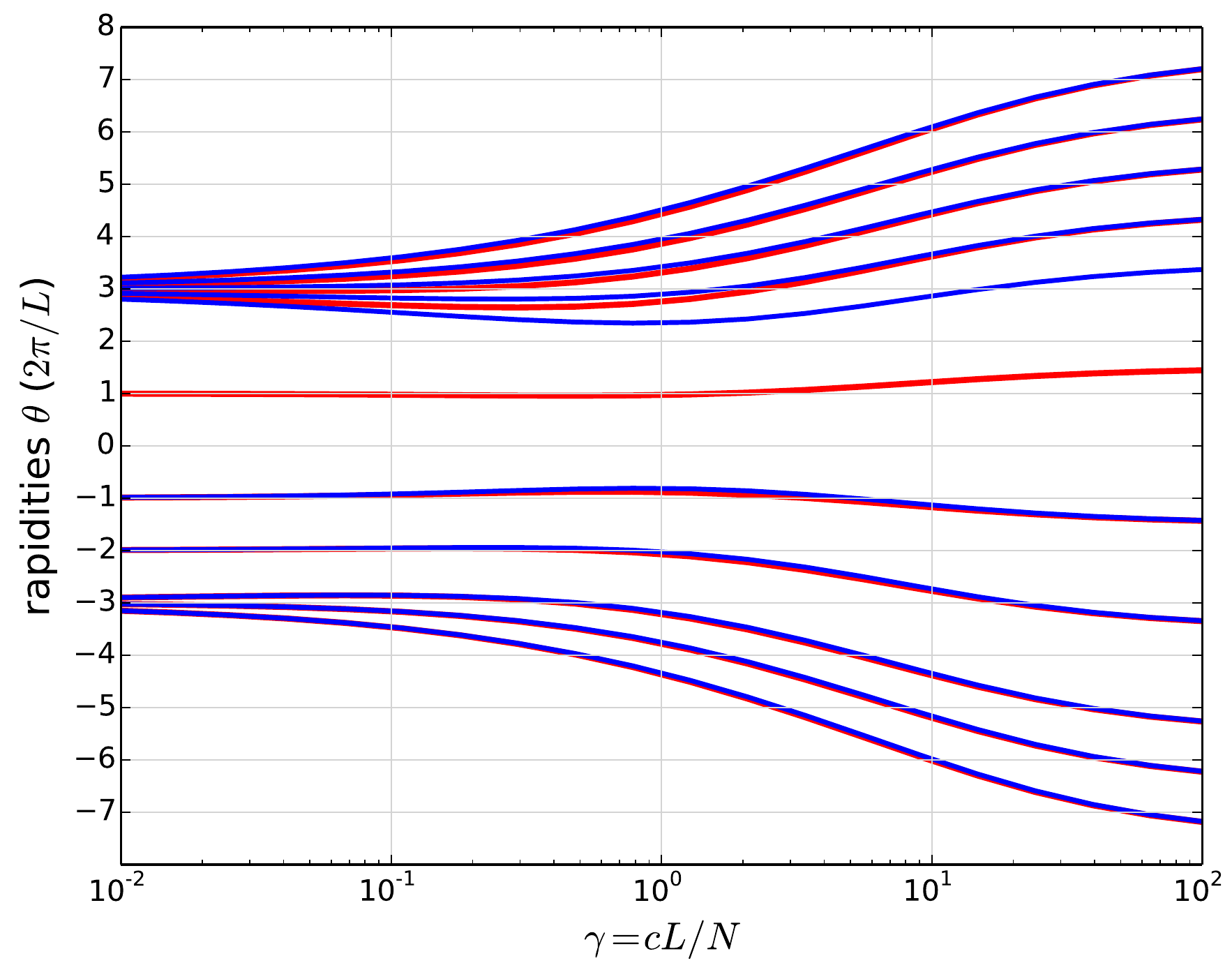}
    \caption{Blue curves: rapidities $\theta_a$ obtained by solving the Bethe equations (\ref{eq:logbethe}) with $N=10$ and $(p_a)_{1 \leq a\leq 10} = \frac{2\pi}{L}  (7.5,6.5,5.5,4.5,3.5,-1.5,-3.5, -5.5,-6.5,-7.5)$, or the equivalent form (\ref{eq:logbethe_bosons}) with $(p^{({\rm B})}_a)_{1 \leq a\leq 10} = \frac{2\pi}{L}  (3,3,3,3,3,-1,-2,-3,-3,-3,-2)$. The rapidity $\theta_a$ interpolates between $p_a^{({\rm B})}$ (when $c \rightarrow 0$) and $p_a$ ($c\rightarrow \infty$). Red curves: rapidities obtained after the modification $(p_5 = 3.5) \rightarrow (p_5 = 1.5)$. It shows that the rapidities are all coupled: changing only one of the $p_a$'s results in small shifts of all the other rapidities.}
    \label{fig:bethe_rapidities}
\end{figure}

It is natural to wonder what happens in the opposite limit of non-interacting bosons, $c \rightarrow 0$. This is easily answered by introducing the `boson momenta' $p^{({\rm B})}_a$, related to the fermion momenta $p_a$ as
\begin{equation}
    p^{({\rm B})}_a \, = \, p_a + a - \frac{N+1}{2} \, \in \frac{2\pi}{L} \mathbb{Z} , \qquad a=1,\dots, N.
\end{equation}
Notice that $p^{({\rm B})}_1 \geq p^{({\rm B})}_2  \geq  \dots \geq p^{({\rm B})}_N $; in particular, two or more boson momenta can coincide. Using the fact that ${\rm arctan(u)} = \frac{\pi}{2} \,{\rm sign}(u) - {\rm arctan(1/u)}$, Eq.~(\ref{eq:logbethe}) is equivalent to
\begin{equation}
    \label{eq:logbethe_bosons}
	\theta_a - \frac{1}{L} \sum_{b \neq a} 2~{\rm arctan} \left( \frac{c}{\theta_a - \theta_b} \right)  \, = \, p^{({\rm B})}_a , \qquad a=1,\dots, N.
\end{equation}
For $c>0$, The `momenta' $p_a^{(B)}$ are just another way of parameterizing the solutions of the Bethe equations; they should not be confused with the momenta of the atoms, which would be obtained by computing the momentum distribution $\left< \{ \theta_a \} \right| \Psi^\dagger_p \Psi_p \left| \{\theta_a \} \right>$, where $\Psi^\dagger_p = \frac{1}{\sqrt{L}} \int e^{i p x} \Psi^\dagger(x) dx$ is the Fourier mode of the creation operator $\Psi^\dagger(x)$. However, in the limit of vanishing repulsion $c \rightarrow 0$, the rapidities are nothing but the boson momenta, $\theta_a \rightarrow p^{({\rm B})}_a$. Moreover, in that limit, the Bethe wavefunction (\ref{eq:bethewf}) is nothing but the permanent ${\rm per} [e^{i x_a \theta_b}]_{1 \leq a,b \leq N}$, i.e. the wavefunction of $N$ non-interacting bosons. So, in that limit, the rapidities coincide with the atom momenta.

Away from these two limits, the rapidities $\theta_a$ correspond to an adiabatic interpolation between the non-interacting fermion ($c \rightarrow +\infty$) and boson ($c \rightarrow 0$) momenta, obtained by solving the Bethe equations~(\ref{eq:logbethe}).

In general, the Bethe equations~(\ref{eq:logbethe}) cannot be solved analytically, but they can easily be solved numerically. One efficient way of doing this is to use the  Newton-Raphson method.

\subsection{Conserved charges and currents}
\label{subsec:chargescurrents}
The eigenstates of the Lieb-Liniger Hamiltonian (\ref{eq:hamLL}) are Bethe states $\left| \{ \theta_a\}_{1 \leq a\leq N} \right>$ labeled by their sets of rapidities. This allows to define a family of charge operators $Q[f]$, diagonal in the eigenbasis and parameterized by functions $f : \mathbb{R} \rightarrow \mathbb{R}$, such that
\begin{equation}
    \label{eq:chargeQf}
    Q[f]  \left| \{ \theta_a \} \right> \, = \, \left(\sum_{b=1}^N f(\theta_b) \right) \left| \{ \theta_a \} \right> .
\end{equation}
Both the momentum operator and the Hamiltonian are of that form, with $f(\theta) = \theta$ and $f(\theta) = \theta^2/2$ respectively, see Eq.~(\ref{eq:thetaPE}). It is the integrability of the model, reflected in the structure of the eigenstates (\ref{eq:bethewf}), which allows us to consider the more general conserved charges (\ref{eq:chargeQf}). By construction, all these operators commute: $[Q[f_1], Q[f_2] ] = 0$. In general, an explicit expression for $Q[f]$ in second-quantized form (like $Q[\theta^2/2]$ given by Eq.~(\ref{eq:hamLL})) is not known, and typically regularization issues appear when one tries to write it~\citep{davies1990higher,davies2011higher}. Nevertheless, even in the absence of such direct expressions for the charges, the conserved charges $Q[f]$ defined formally by Eq.~(\ref{eq:chargeQf}) prove to be very useful. There are other ways to do calculations with these charges, that do not require to know their explicit second-quantized form, in particular the algebraic Bethe Ansatz, see e.g. \citep{korepin1997quantum} for an introduction.

From their definition (\ref{eq:chargeQf}), one expects the $Q[f]$ charges to be extensive with $N$, and to be the integral of a charge density
\begin{equation}
    Q[f] \, = \, \int_0^L q[f] (x) dx .
\end{equation}
For $f$ sufficiently regular, the charge density $q[f] (x)$ is sufficiently local, meaning that it acts as the identity far away from the point $x$. [In the hard core limit $g \rightarrow +\infty$, this is a consequence of the Paley-Wiener theorem. At finite repulsion strength $g$, and more generally in interacting integrable models, the locality properties of charge densities are an advanced topic that is  beyond the scope of this review; see e.g. \citep{ilievski2016quasilocal,doyon2017thermalization,palmai2018quasilocal} for introductions.] By definition, the expectation value of the charge density in a Bethe state (normalized as $\left< \{\theta_a \} \left|  \{ \theta_a \} \right> \right. =1$) is
\begin{equation}
    \label{eq:expectationqf}
 \left< \{\theta_a \} \right|  q [f](x)  \left| \{ \theta_a\} \right> = \frac{1}{L} \sum_{b=1}^N f(\theta_b) .
\end{equation}
It is independent of $x$ because the Bethe state is translation invariant.

To the charge density $q[f]$, one associates a current operator $j[f]$ through the continuity equation,
\begin{eqnarray}
    \label{eq:jfx}
    \frac{\partial}{\partial t}  q[f](x) + \frac{\partial}{\partial x} j[f](x) = i \left[ H, q[f](x) \right]  + \frac{\partial}{\partial x} j[f](x) = 0.
\end{eqnarray}
As sketched in the introduction, continuity equations are the basic ingredient of hydrodynamics. To write useful hydrodynamic equations, however, one must be able to evaluate the currents in given stationary states. Until very recently, it was not known how to evaluate the expectation values of the current $j[f]$. However, thanks to developments in integrability, in particular in form factor techniques and algebraic Bethe Ansatz, a remarkable exact formula has just  been discovered by~\cite{borsi2020current} for the expectation value of $j [f]$ in a Bethe state (see also \citep{pozsgay2020algebraic,pozsgay2020current} for further developments in the context of spin chains, as well as the review article by Pozsgay, Borsi and Pristy\'ak in this Volume),
\begin{equation}
    \label{eq:pozsgay}
 \left< \{ \theta_a \} \right|  j [f](x)  \left| \{ \theta_a \} \right> = \frac{1}{L} \sum_{a,b} \varepsilon' (\theta_a)  [\mathcal{G}^{-1}]_{ab}  f (\theta_b) .
\end{equation}
Here $\varepsilon' (\theta) = \theta$ is the derivative of $\varepsilon (\theta) = \theta^2/2$, and $\mathcal{G}$ is the Jacobian matrix of the transformation from the $p_a$'s to the $\theta_b$'s defined by Eq.~(\ref{eq:logbethe}), known as the Gaudin matrix,
\begin{equation}
	\label{eq:gaudin}
	\mathcal{G}_{ab} \, = \, \frac{\partial p_a}{\partial \theta_b} 
\end{equation}
(where the $p_a$'s in Eq.~(\ref{eq:logbethe}) are no longer restricted to be in $\frac{2\pi}{L} \mathbb{Z}$). 
The Gaudin matrix is symmetric, $\mathcal{G}^T = \mathcal{G}$, as a consequence of the fact that the scattering phase $\phi$ depends on the rapidities $\theta_a$ and $\theta_b$ only through the difference $\theta_a-\theta_b$, see Eq.~(\ref{eq:scattering_phase}).

Let us mention that the remarkable formula (\ref{eq:gaudin}) is a particular case of a more general result, also obtained in~\citep{borsi2020current,pozsgay2020algebraic,pozsgay2020current}. One can define generalized currents $j[h,f](x)$ through the generalization of the continuity equation~(\ref{eq:jfx}):
\begin{equation}
    \label{eq:jgfx}
    i \left[ Q[h] , q[f](x) \right]  + \frac{\partial}{\partial x} j[h,f](x) = 0 .
\end{equation}
Then the general formula for the expectation value reads
\begin{equation}
    \label{eq:gpozsgay}
 \left< \{ \theta_a \} \right|  j [h,f](x)  \left| \{ \theta_a \} \right> = \frac{1}{L} \sum_{a,b} h' (\theta_a)  [\mathcal{G}^{-1}]_{ab}  f (\theta_b) ,
\end{equation}
and the above physical current is the special case $h(\theta) = \varepsilon(\theta)= \theta^2/2$. We note that such generalized currents had also been considered in \citep{castro2016emergent} in the thermodynamic limit.

The discovery and proof of formula (\ref{eq:jgfx}) or its generalization (\ref{eq:gpozsgay}) required advanced techniques~\citep{borsi2020current,pozsgay2020algebraic,pozsgay2020current}, however the result is simple and its physical interpretation is quite clear. The $b^{\rm th}$ boson, with rapidity $\theta_b$, carries an amount of charge density $\frac{1}{L} f(\theta_b)$. In the absence of other particles, it would travel at the single-particle group velocity $v_b = \partial \varepsilon(\theta_b)/ \partial p (\theta_b) = \partial \varepsilon(\theta_b)/ \partial \theta_b$ (or its generalization $v_b = \partial h(\theta_b)/ \partial p (\theta_b)) = \partial h(\theta_b)/ \partial \theta_b)$, resulting in the current $j[h,f] = \frac{1}{L} v_b f(\theta_b) $.

In the presence of other particles, the group velocity of the $b^{\rm th}$ boson is modified. To compute it, one can consider a small variation of the fermion momentum $p_b \rightarrow p_b + \delta p_b$ in Eq.~(\ref{eq:logbethe}). This results in a small change of the total momentum $\delta P = \delta p_b$, and of the total energy $\delta E = \delta \left( \sum_{a=1}^N \varepsilon(\theta_a) \right) = \sum_{a} \frac{\partial \varepsilon(\theta_a) }{\partial p_b} \delta p_b = \sum_{a} \varepsilon'(\theta_a) [\mathcal{G}^{-1}]_{ab} \delta p_b$ (more generally, of the total charge $\delta Q[h] =  \sum_{a} \frac{\partial h(\theta_a) }{\partial p_b} \delta p_b = \sum_{a} h'(\theta_a) [\mathcal{G}^{-1}]_{ab} \delta p_b)$). Thus, the modified group velocity is $\delta E/\delta p_b = \sum_{a} \varepsilon'(\theta_a) [\mathcal{G}^{-1}]_{ab}$ resulting in formula (\ref{eq:pozsgay}), or more generally $\delta Q[h]/\delta p_b = \sum_{a} h'(\theta_a) [\mathcal{G}^{-1}]_{ab}$ resulting in (\ref{eq:gpozsgay}).

For further discussions of the physical interpretation of Eqs.~(\ref{eq:pozsgay},\ref{eq:gpozsgay}), see~\citep{borsi2020current}, and also~\citep{bonnes2014light,bertini2016transport,castro2016emergent,doyon2018soliton,doyon2019lecture}, where similar discussions had been given previously for the thermodynamic version of these formulas (see Eq.~(\ref{eq:currents_ghd}) below). See also the two reviews by Borsi, Pozsgay and Pristy\'ak and by Cubero, Yoshimura and Spohn in this Volume.

\subsection{Thermodynamic limit}
\label{subsec:thermodynamic_limit}

So far we have focused on a finite number of bosons $N$, first on an infinite line, and then in a periodic box of length $L$. To do hydrodynamics, one needs first to understand the thermodynamic properties of the system. In this subsection, we briefly review the techniques for taking the thermodynamic limit $N,L \rightarrow \infty$, keeping the density of bosons $n=N/L$ fixed.

The key idea is to focus on an infinite sequence of eigenstates $\left( \left| \{ \theta_a \}_{1 \leq a \leq N} \right> \right)_{N \in \mathbb{N}}$ of the Lieb-Liniger Hamiltonian (\ref{eq:hamLL}), with $L = N/n$, such that the limit of the distribution of rapidities
\begin{equation}
    \label{eq:rapidity_dist}
    \rho(\theta) \, := \, \lim_{N \rightarrow \infty} \frac{1}{L} \sum_{a=1}^N \delta (\theta-\theta_a)
\end{equation}
is well defined and is a (piecewise) smooth function of $\theta$. The thermodynamic properties of the system (such as its energy density, pressure, etc.) then become particular functionals of that rapidity density $\rho(\theta)$, and the goal is to find these functionals and to evaluate them. In what follows, we write `$\lim_{\rm therm.}$' for this limiting procedure.

For example, consider the expectation values of the charge densities (\ref{eq:expectationqf}): in the thermodynamic limit, these become
\begin{equation}
    {\rm lim}_{\rm therm.} \left< \{ \theta_a \} \right| q[f] \left|\{ \theta_a \} \right> \, = \, \int_{-\infty}^\infty f(\theta) \rho(\theta) d\theta .
\end{equation}
In particular, the density of particles, the momentum density, and the energy density are, respectively, $n= \left< q[1]\right> = \int \rho(\theta)d\theta$, $\left< q[\theta] \right> = \int \theta \rho(\theta)d\theta$ and $ \left< q[\theta^2/2] \right> = \int \frac{\theta^2}{2} \rho(\theta)d\theta$.

\subsubsection{Thermodynamic form of the Bethe equations}

Crucially, since all the states in the infinite sequence $\left( \left| \{ \theta_a \}_{1 \leq a \leq N} \right> \right)_{N \in \mathbb{N}}$ are Bethe states, each set of rapidities $\{ \theta_a \}_{1 \leq a \leq N}$ must satisfy the Bethe equations (\ref{eq:logbethe}). To implement that constraint, it is customary to consider the set of fermion momenta $\{ p_a \}_{1 \leq a \leq N}$ associated to the set of rapidities $\{ \theta_a \}_{1 \leq a \leq N}$, both of them ordered as in~(\ref{eq:orderrapidities}), and to define the density of states $\rho_{\rm s} (\theta)$ as
\begin{equation}
    \label{eq:def:rhos}
    2 \pi \rho_{\rm s} (\theta) := {\rm lim}_{\rm therm.}  \frac{| p_a - p_{a+1} |}{ | \theta_a - \theta_{a+1} |} ,
\end{equation}
where the sequence of indices $a$ in the r.h.s is chosen so that ${\rm lim}_{\rm therm.}  \theta_a = \theta$.
\begin{figure}[h]
    \centering
    \includegraphics[width=0.6\textwidth]{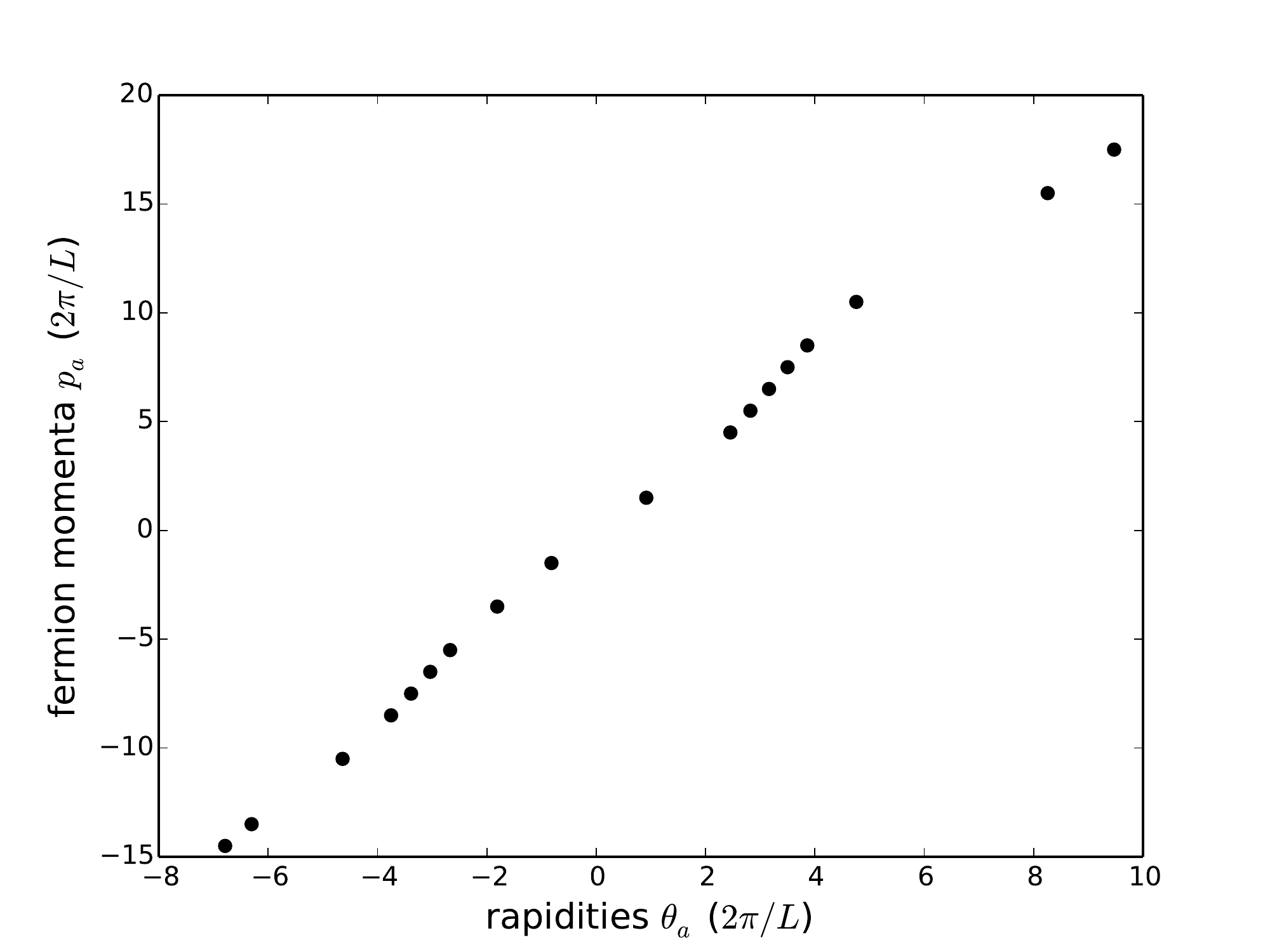}
    \caption{Fermion momenta $p_a$ plotted against the rapidities $\theta_a$ for the solution of the Bethe equations (\ref{eq:logbethe}) defined by $p_a = \frac{2\pi}{L} (-14.5,-13.5,-10.5,-8.5,-7.5,-6.5,-5.5,-3.5,-1.5,1.5,4.5,5.5,6.5,7.5,8.5,10.5,15.5,17.5)$ with $\gamma=0.2$. As the density of momenta $p_a$ and of rapidities $\theta_a$ increases, this becomes a smooth curve, whose slope is $ 2\pi$ times the density of states $\rho_{\rm s}(\theta)$, see Eq.~(\ref{eq:def:rhos}).}
    \label{fig:bethe_rho_s}
\end{figure}
Because the fermion momenta $p_a$ must satisfy the Pauli exclusion principle (they must all be different), it is clear that $| p_a - p_{a+1}| \geq \frac{2\pi}{L}$. Also, notice that, by definition, ${\rm lim}_{\rm therm.} \frac{1}{L |\theta_a - \theta_{a+1} |} = \rho(\theta)$. Consequently, the Fermi occupation ratio
\begin{equation}
    \nu (\theta) := \frac{\rho(\theta)}{\rho_{\rm s}(\theta)}
\end{equation}
must always satisfy
\begin{equation}
    0 \leq \nu (\theta) \leq 1 .
\end{equation}
Moreover, the rapidity density $\rho(\theta)$ and the density of states $\rho_{\rm s}(\theta)$ are related by the thermodynamic version of the Bethe equation (\ref{eq:logbethe}). Plugging Eq.~(\ref{eq:logbethe}) into the definition (\ref{eq:def:rhos}) leads to the constitutive equation
\begin{eqnarray}
    \label{eq:constitutive}
 \nonumber   2\pi \rho_{\rm s}(\theta) &=&  {\rm lim}_{\rm therm.}  \frac{1}{\theta_a- \theta_{a+1}} \left[
    \left( \theta_a + \frac{1}{L} \sum_{b} 2 \, \arctan \left( \frac{\theta_a - \theta_b}{2} \right) \right) \right. \\
\nonumber   && \left. \qquad \qquad \quad - \left( \theta_{a+1} + \frac{1}{L} \sum_{b} 2 \, \arctan \left( \frac{\theta_{a+1} - \theta_b}{2} \right) \right) \right] \\
    &=& 1 +\int_{-\infty}^\infty \Delta(\theta - \theta') \rho(\theta') d\theta' ,
\end{eqnarray}
where $\Delta(\theta - \theta')$ is the differential two-body scattering shift (\ref{eq:scattering_phase}).

In practice, to construct interesting thermodynamic states, one can specify the Fermi occupation ratio $\nu(\theta)$, and then use the constitutive equation (\ref{eq:constitutive}) to reconstruct the rapidity density $\rho(\theta)$ and the density of states $\rho_{\rm s}(\theta)$. One important example of this is the ground state of the Lieb-Liniger Hamiltonian, which corresponds to an occupation ratio which is a rectangular function: $\nu(\theta) = 1$ for $\theta \in [-\theta_{\rm F} , \theta_{\rm F}]$, and $\nu(\theta) = 0$ otherwise. Here $\theta_{\rm F}$ is the Fermi rapidity, which is a function of the density of particles $n$. In that case, the constitutive equation becomes the Lieb equation \citep{lieb1963exact} (also known as the Love equation \citep{love1949electrostatic}; for studies of this particular equation see e.g.~\cite{takahashi1975validity,popov1977theory,lang2017ground,prolhac2017ground,marino2019exact}). Another important example is the one of a thermal equilibrium distribution $\rho(\theta)$ obtained by solving the Yang-Yang equation (Eq.~(\ref{eq:yangyang}) below).

In general, the constitutive equation cannot be solved analytically, however, since it is linear, it is easily solved numerically by discretizing the integral.

\subsubsection{The dressing}

In thermodynamic manipulations, it turns out that the following operation is ubiquitous: to a function $f(\theta)$, one has to associate its `dressed' counterpart $f^{\rm dr} (\theta)$, defined by the integral equation
\begin{equation}
    \label{eq:dressing}
	 f^{\rm dr} (\theta) = f(\theta) + \int \frac{d\theta'}{2\pi} \Delta(\theta-\theta') \nu(\theta') f^{\rm dr}(\theta')  .
\end{equation}
Although it is not explicit in the notation, $f^{\rm dr} (\theta)$ is always a functional of the rapidity distribution, through its dependence on the Fermi occupation ratio. For instance, with this definition, the constitutive equation (\ref{eq:constitutive}) is recast as
\begin{equation}
    2\pi \rho_{\rm s}(\theta) = 1^{\rm dr} (\theta),
\end{equation}
where $1 (\theta) = 1$ is the constant function.

Another example where the dressing (\ref{eq:dressing}) pops out is in manipulations that involve the Gaudin matrix. This is important for this review article, because to establish hydrodynamic equations one needs the thermodynamic limit of the expectation value of the current, see Eq.~(\ref{eq:pozsgay}). The following identity holds:
\begin{equation}
    \label{eq:gaudin_dressing}
	\frac{f^{\rm dr} (\theta)}{2\pi \rho_s(\theta)} \, = \,    {\rm lim}_{\rm therm.} \, \sum_{b=1}^N [\mathcal{G}^{-1}]_{ab} f(\theta_b) ,
\end{equation}
where, again, the relation between $\theta$ in the l.h.s and the rapidity $\theta_a$ in the r.h.s is $\theta = {\rm lim}_{\rm therm.} \theta_a$. This identity is easily derived as follows. Using the definition (\ref{eq:gaudin}), 
\begin{eqnarray}
\nonumber	\sum_b \mathcal{G}_{ab} h (\theta_b) && \; = \;  \left( 1 + \frac{1}{L} \sum_{b \neq a} \Delta(\theta_a - \theta_b)  \right)  h(\theta_a) -  \frac{1}{L} \sum_{b \neq a} \Delta ( \theta_b - \theta_a ) h(\theta_b) \\
\nonumber	&& \underset{{\rm therm. \,lim.}}{\longrightarrow} 2\pi \rho_s(\theta) \, h (\theta)  -  \int \frac{d\theta'}{2\pi} \Delta ( \theta - \theta' ) n(\theta')  2\pi \rho_s(\theta')  h(\theta') \\
&& \; = \; \, [2\pi \rho_s \, h ]^{\rm undr} (\theta) ,
\end{eqnarray}
where the `undressing' is the inverse of the dressing, i.e. $(f^{{\rm undr }})^{{\rm dr}} (\theta) = f(\theta)$. Inverting this formula gives Eq.~(\ref{eq:gaudin_dressing}).

We will see a few more examples of physical quantities whose computation involves the dressing operation below. References where this operation is used extensively include e.g. the original derivation of the GHD equations in integrable quantum field theories \citep{castro2016emergent}, the calculation of Drude weights and other two-point correlations of charge and currents in the Lieb-Liniger model~\citep{doyon2017drude}, the inclusion of force fields~\citep{doyon2017note} or adiabatically varying interactions~\citep{bastianello2019generalized} or diffusive corrections~\citep{de2018hydrodynamic,gopalakrishnan2018hydrodynamics,de2019diffusion} into the GHD equations.

\subsubsection{Expectation values of the currents in the thermodynamic limit}
\label{subsec:expectation_thermodynamic}

We now present the central ingredient of Generalized Hydrodynamics. The thermodynamic expectation value of the current $j[f]$ (see subsection~\ref{subsec:chargescurrents}) is
\begin{equation}
    \label{eq:currents_ghd}
	{\rm lim}_{\rm therm.}  \left< \{\theta_a \} \right|  j [f]  \left| \{ \theta_a \} \right>  \, = \, \int  v^{\rm eff}[\rho] (\theta) f(\theta) \rho(\theta) d \theta,
\end{equation}
where the `effective velocity' is a functional of the rapidity distribution defined by
\begin{equation}
    \label{eq:veff}
	v^{\rm eff}[\rho] (\theta) \, := \, \frac{ (\varepsilon')^{\rm dr} (\theta) }{2 \pi \rho_{\rm s}(\theta)} \, = \, \frac{ {\rm id}^{\rm dr} (\theta) }{1^{\rm dr}(\theta)}  ,
\end{equation}
with $\varepsilon'(\theta) = {\rm id}(\theta) = \theta$ and $1(\theta) = 1$. The remarkable result (\ref{eq:currents_ghd}) was first obtained by \citep{castro2016emergent,bertini2016transport}, and it was the key observation that triggered all the later developments of GHD in quantum integrable systems. \cite{bertini2016transport} relied partially on \citep{bonnes2014light}, where the formula for the effective velocity (\ref{eq:veff}) had first appeared in the context of a quantum integrable system. In retrospect, the thermodynamic result (\ref{eq:currents_ghd}) can be viewed as a consequence of the finite-size formula (\ref{eq:pozsgay}) of \citep{borsi2020current,pozsgay2020algebraic,pozsgay2020current}, using the fact that the dressing is the thermodynamic limit of the Gaudin matrix, see Eq.~(\ref{eq:gaudin_dressing}). Historically though, the thermodynamic result was discovered before its finite-size counterpart. Since 2016, several works have aimed at  establishing the validity of the thermodynamic formula~(\ref{eq:currents_ghd}) in various models, by relying on various approaches. Let us mention the form factor approaches of \citep{vu2019equations,cubero2019generalized,cubero2020generalized} for quantum field theories, arguments based on the symmetry of the charge-current correlations \citep{yoshimura2020collision}, or exact results in the classical integrable model of the Toda chain~\citep{bulchandani2019kinetic,cao2019gge,doyon2019generalized,spohn2020collision}. We also refer to the two reviews on this topic by Cubero, Yoshimura and Spohn and by Pozsgay, Borsi and Pristy\'ak in this Volume.

The effective velocity (\ref{eq:veff}) solves the equation
\begin{equation}
    \label{eq:veff_quantum}
\displaystyle v^{\rm eff}[\rho] (\theta) = \theta - \int_{-\infty}^\infty \Delta(\theta- \theta')  \left(  v^{\rm eff}[\rho] (\theta)  - v^{\rm eff}[\rho] (\theta') \right) \rho(\theta') d\theta' .
\end{equation}
This is analogous to Eq.~(\ref{eq:GHDintro}) in the introduction, which defines the effective velocity in the hard rod gas. The main difference is that the scattering shift $\Delta(\theta-\theta')$ is now rapidity-dependent, while in the hard rod gas $\Delta$ is a constant equal to minus the diameter of the balls. The physical interpretation of Eq.~(\ref{eq:veff_quantum}) is analogous to the one in Fig.~\ref{fig:hardrod}: a `tracer' quasiparticle with rapidity (asymptotic momentum) $\theta$, which would normally travel at constant speed $\theta$ in the vacuum, finds its velocity modified by the presence of a finite density $\rho(\theta')$ of other quasiparticles. From time $t$ to $t+ \delta t$, the tracer typically scatters against a number $\delta t \times | v^{\rm eff}[\rho] (\theta)  - v^{\rm eff}[\rho] (\theta') | \rho(\theta')$ of quasiparticles with rapidity $\theta'$. At each collision, the tracer is shifted backwards by an amount $\Delta (\theta-\theta')$: this is the physical effect that is encoded by formula (\ref{eq:veff_quantum}).

To check that the effective velocity (\ref{eq:veff}) solves Eq.~(\ref{eq:veff_quantum}) as claimed, one can use the definition of the dressing and the constitutive relation:
\begin{eqnarray*}
    && \int_{-\infty}^\infty \Delta(\theta- \theta')  \left(  v^{\rm eff}[\rho] (\theta)  - v^{\rm eff}[\rho] (\theta') \right) \rho(\theta') d\theta' \\
    &&= \,  v^{\rm eff}[\rho] (\theta)  \int_{-\infty}^\infty \Delta(\theta- \theta')  \rho(\theta') d\theta' - \int_{-\infty}^\infty \frac{d\theta'}{2\pi} \Delta(\theta- \theta')  \nu(\theta') {\rm id}^{\rm dr} (\theta')   \\
    &&= \,  v^{\rm eff}[\rho] (\theta) \, (2\pi \rho_{\rm s}(\theta) - 1 ) - ( {\rm id}^{\rm dr} (\theta) - {\rm id} (\theta)  )  = \theta -  v^{\rm eff}[\rho] (\theta) .
\end{eqnarray*}

\subsection{Entropy maximization: the Yang-Yang equation}
\label{subsec:yangyang}
In the previous Subsection we illustrated how physical observables, such as the expectation values of charges and currents, become functionals of the rapidity distribution $\rho(\theta)$ in the thermodynamic limit. We did no explain how to construct physically meaningful rapidity distributions though (except for the ground state of the Lieb-Liniger Hamiltonian, for which $\nu(\theta)$ is a rectangular function, see Subsection~\ref{subsec:thermodynamic_limit}).

For instance, what is the rapidity distribution corresponding to a thermal equilibrium state at non-zero temperature? This question was answered in the pioneering work of  \cite{yang1969thermodynamics}, which we now briefly review.

First, we observe that there are many different choices of sequences of eigenstates $(\{ \theta_a \}_{1 \leq a \leq N} )_{N \in \mathbb{Z}}$ that lead to the same thermodynamic rapidity distribution (\ref{eq:rapidity_dist}). The description of the system in terms of a rapidity distribution $\rho(\theta)$ is only a coarse-grained description: one should think of the rapidity distribution $\rho(\theta)$ as characterizing a macrostate of the system, corresponding to a very large number of possible microstates $\left| \{ \theta_a\} \right>$. To do thermodynamics, one needs to estimate the number of such microstates.

To estimate that number, one focuses on a small rapidity cell $[\theta, \theta + \delta \theta]$, which contains $L \rho(\theta) \delta \theta$ rapidities. The Bethe equations (\ref{eq:logbethe}) relate these rapidities to fermion momenta $p_a$ in a momentum cell $[p , p + \delta p]$, where $\delta p/ \delta \theta \simeq 2\pi \rho_{\rm s}(\theta)$, see Eq.~(\ref{eq:def:rhos}). Importantly, the fermion momenta $p_a$ satisfy the Pauli exclusion principle. Then the number of microstates is evaluated by counting how many configurations of mutually distinct $L \rho(\theta) \delta \theta$ fermion momenta can fit into the box $[p , p + \delta p]$. Since the minimal spacing between two momenta is $\frac{2\pi}{L}$, the answer is
\begin{equation}
    \label{eq:SYY}
   \# {\rm conf.} \, \simeq \, \frac{[L \rho_{\rm s}(\theta) \delta \theta] !}{[L \rho(\theta) \delta \theta] ! [L (\rho_{\rm s}(\theta)- \rho(\theta)) \delta \theta] !} .
\end{equation}
The total number of microstates is the product of all such configurations over all the rapidity cells $[\theta, \theta+\delta \theta]$. Taking the logarithm, and replacing the sum by an integral over $d\theta$, we obtain the Yang-Yang entropy
\begin{eqnarray}
\nonumber  && \log \left( \, \# {\rm microstates} \, \right) \, \simeq \, L \, S_{\rm YY}[\rho] , \\
  &&   S_{\rm YY}[\rho] \, := \, \int_{-\infty}^\infty \left( \rho_{\rm s} \log \rho_{\rm s} - \rho \log \rho - ( \rho_{\rm s}-\rho) \log ( \rho_{\rm s} - \rho) \right) d\theta  .
\end{eqnarray}
The notation indicates that the Yang-Yang entropy is a functional of $\rho$ only, and not of $\rho_{\rm s}$; this is because $\rho_{\rm s}$ must always be obtained from $\rho$ by the constitutive equation (\ref{eq:constitutive}).

Now let us consider the thermal equilibrium density matrix at temperature $T$,
\begin{equation}
    \label{eq:thermalrho}
    \hat{\rho}_{\rm thermal} \, \propto \, e^{-  H/T} \, = \, \sum_{\left| \theta_a \right> } e^{-  \sum_{a} (\varepsilon(\theta_a) -\mu)/T } \left|  \{ \theta_a \} \right> \left<  \{ \theta_a \} \right|  ,
\end{equation}
where the sum runs over all eigenstates. In fact, a straightforward generalization consists in considering the  Generalized Gibbs Ensemble \citep{rigol2007relaxation,rigol2008thermalization} density matrix
\begin{equation}
    \label{eq:GGErho}
    \hat{\rho}_{\rm GGE}[f] \, \propto \, e^{- Q[f]} \, = \, \sum_{\left| \theta_a \right> } e^{- \sum_{a} f(\theta_a) } \left|  \{ \theta_a \} \right> \left<  \{ \theta_a \} \right| ,
\end{equation}
for some function $f$. We would like to compute expectation values w.r.t this density matrix, e.g.
\begin{equation}
    \left< \mathcal{O} \right>_{\rm GGE} \, := \, \frac{{\rm tr} [\mathcal{O} e^{-Q[f]} ]}{ {\rm tr}[e^{-Q[f]}] } \, = \, \frac{    \sum_{\left| \theta_a \right> } \left<  \{ \theta_a \} \right| \mathcal{O} \left|  \{ \theta_a \} \right> e^{- \sum_{a} f(\theta_a) }  }{ \sum_{\left| \theta_a \right> } e^{- \sum_{a} f(\theta_a) }  }
\end{equation}
for some observable $\mathcal{O}$. When the observable $\mathcal{O}$ is sufficiently local, it is believed that the expectation value $\left<  \{ \theta_a \} \right| \mathcal{O} \left|  \{ \theta_a \} \right>$ does not depend on the specific microstate of the system, so that it becomes a functional of $\rho$ in the thermodynamic limit,
\begin{equation}
\label{eq:generalized-ETH}
    {\rm lim}_{\rm therm.} \left<  \{ \theta_a \} \right| \mathcal{O} \left|  \{ \theta_a \} \right> \, = \, \left< \mathcal{O}\right>_{[\rho]} .
\end{equation}
This assumption is related to a `Generalized Eigenstate Thermalization Hypothesis', see e.g.~\citep{cassidy2011generalized,pozsgay2011mean,he2013single,pozsgay2014failure,vidmar2016generalized,dymarsky2019generalized}. Under that assumption, one can replace the above sum over all eigenstates by a functional integral over the coarse-grained rapidity distribution $\rho$,
\begin{equation}
    \label{eq:thermO_GGE}
    {\rm lim}_{\rm therm.}  \left< \mathcal{O} \right>_{\rm GGE} \, = \, \frac{  \int \mathcal{D} \rho \, \left< \mathcal{O} \right>_{[\rho]}  e^{L ( S_{\rm YY}[\rho] - \int f(\theta) \rho(\theta) d\theta ) }  }{  \int \mathcal{D} \rho \, e^{L ( S_{\rm YY}[\rho] - \int f(\theta) \rho(\theta) d\theta ) }  } .
\end{equation}
The functional integral is then dominated by the root distribution which minimizes a (generalized) free energy functional:
\begin{equation}
    \frac{\delta}{\delta \rho} \left[ \int f(\theta) \rho(\theta) d\theta  - S_{\rm YY}[\rho] \right] \, = \, 0.
\end{equation}
Using the definition of the Yang-Yang entropy and the constitutive equation (\ref{eq:logbethe}), one obtains the following relation between the function $f(\theta)$ defining the diagonal density matrix (\ref{eq:GGErho}) and the rapidity distribution $\rho(\theta)$ dominating the functional integral (\ref{eq:thermO_GGE}):
\begin{equation}
    \label{eq:yangyang}
   f(\theta) \, =  \, \log \left( \frac{\rho_{\rm s}(\theta)}{\rho(\theta)} -1 \right) - \int \frac{d\theta'}{2\pi} \Delta (\theta- \theta')   \log \left( 1 - \frac{\rho(\theta')}{\rho_{\rm s} (\theta')} \right).
\end{equation}
This equation is known as the (Generalized) Yang-Yang equation, or (Generalized) Thermodynamic Bethe Ansatz equation. Again, the term `Generalized' refers to the replacement of the thermal equilibrium density matrix by a Generalized Gibbs Ensemble, (\ref{eq:thermalrho})$\rightarrow$(\ref{eq:GGErho}), see e.g.~\citep{caux2012constructing,wouters2014quenching}. Like most of the equations encountered so far, in general the Yang-Yang equation cannot be solved analytically, but it can be efficiently solved numerically, by iteration. In particular, this allows to compute the rapidity distribution $\rho(\theta)$ at thermal equilibrium.

This is particularly useful in applications discussed later in this review, because in experiments, one often assumes that the system is (at least initially) at thermal equilibrium.

For instance, using the Yang-Yang equation, it is possible to tabulate the equilibrium pressure $\mathcal{P}(n,e)$ as a function of the particle density $n$ and the energy per particle $e$. To do this, one needs to first solve numerically Eq.~(\ref{eq:yangyang}) with $f(\theta) =  (\varepsilon(\theta)-\mu )/T$, and Eq.~(\ref{eq:constitutive}), to get $\rho(\theta)$ and $\rho_{\rm s}(\theta)$. Then the equilibrium pressure is given by \citep{yang1969thermodynamics,korepin1997quantum}
\begin{equation}
    \label{eq:pressure}
        \mathcal{P} = - \left( \frac{\partial F}{\partial L} \right)_T = - T \int \frac{d \theta}{2\pi} \log \left( 1 - \frac{\rho(\theta)}{\rho_{\rm s}(\theta)} \right) ,
\end{equation}
where the free energy is $F/L =  \int (\varepsilon(\theta) - \mu) \rho(\theta) d \theta - T S_{\rm YY} [\rho] $. This gives the thermodynamic equilibrium pressure at density $n = \int \rho(\theta) d\theta$ and energy per particle $e = \int \frac{\theta^2}{2} \rho(\theta) d\theta/n$. Alternatively, the pressure can be identified with the momentum current $j_P = j[{\rm id}]$ (with ${\rm id}(\theta) = \theta$) as we did in the introduction, see Eq.~(\ref{eq:euler}). Thus, according to Eq.~(\ref{eq:currents_ghd}), we must also have
\begin{equation}
    \label{eq:pressure2}
        \mathcal{P} = {\rm lim}_{\rm therm.} \left< j[{\rm id}] \right> = \int  v^{\rm eff}(\theta)  \,\theta \, \rho(\theta) d\theta .
\end{equation}
The equivalence between the two formulas (\ref{eq:pressure}) and (\ref{eq:pressure2})  follows from manipulations of the dressing operation (\ref{eq:dressing}), which we leave as an exercise to the interested reader. [Hint: with the definition of the effective velocity, formula (\ref{eq:pressure2}) is equivalent to $\mathcal{P} = \int \theta \nu(\theta) {\rm id}^{\rm dr}(\theta) \frac{d\theta}{2\pi}$, while differentiating (\ref{eq:yangyang}) w.r.t. $\theta$ and using the definition of the dressing operation leads to ${\rm id}^{\rm dr}(\theta) =  T \frac{1}{\nu(\theta)} \frac{d}{d\theta} \log(1-\nu(\theta))$.]

\subsection{Relaxation in the Lieb-Liniger model}
\label{subsec:GGE}
 It is not {\it a prior} clear wether
Generalized Gibbs Ensembles are relevant
in the context of an isolated Lieb-Liniger gas. 
This question is linked to 
the notion of relaxation in isolated many-body quantum systems, which 
was at the heart of many studies in the last decades~\citep{polkovnikov2011colloquium}.

It is now well established that Generalized Gibbs Ensembles are relevant to describe 
locally an isolated Lieb-Liniger gas after relaxation, see for instance the review articles~\citep{vidmar2016generalized,caux2016quench,essler2016quench}.

Since this point is essential in the Generalized Hydrodynamics theory, we briefly recall the underlying physics. 
Let us consider a Lieb-Liniger gas, confined in a box-like potential, and let us assume it is initially in an out-of-equilibrium state that is a pure quantum state. This quantum state expands onto many Bethe-Ansatz eigenstates:
\begin{equation}
    |\psi\rangle = \sum_{\{\theta_a\}} c_{\{\theta_a\}} |\{\theta_a\}\rangle.
\end{equation}
Typically, the rapidity distributions of the Bethe-Ansatz states 
involved in this expansion gather around a given averaged rapidity distribution. During the time-evolution, the different Bethe-Ansatz states, which evolve each with its own energy, will dephase. Because of this dephasing, the contribution of cross terms will vanish at long time when computing the mean value of an observable. Thus mean values of observable will undergo a relaxation and take the asymptotic value
\begin{equation}
\langle {\cal O} \rangle \underset{t\rightarrow \infty} = 
\sum_{\{\theta_a\}} |c_{\{\theta_a\}}|^2 \langle \{\theta_a\}|{\cal O} |\{\theta_a\} \rangle
\end{equation}
 We then invoke the generalized eigenstate thermalisation hypothesis which states that expectation values of a local observable does not depend on the specific Bethe-Ansatz state, but is a smooth functionnal of the rapidity distribution.     
Thus expectation values of local observables are identical for all 
diagonal ensembles, provided they are peaked onto a given rapidity
distribution. One can choose the Generalized Gibbs Ensemble corresponding to the correct rapidity distribution, as done in Eq.~\eqref{eq:generalized-ETH}.  
One then find that 
\begin{equation}
\langle {\cal O} \rangle \underset{t\rightarrow \infty} = 
 \langle {\cal O} \rangle_{[\rho]}.
\end{equation}

While, to compute local observables after relaxation, one can represent the whole isolated system of length $L$ by any diagonal ensemble peaked around the the correct rapidity distribution, the GGE plays a special role to describe a subsystem of length $l$. If $l$ is both much smaller than $L$ and much larger than microscopic correlation lengths, then the subsystem is correctly described by a GGE, the GGE accounting properly for the fluctuations in the subsystem. This property was used when analysing local fluctuation measurements~\citep{armijo_probing_2010,jacqmin_sub-poissonian_2011}. The physical picture supporting the relevance of the GGE to describe the small system is that the large system acts as a reservoir of rapidities for the subsystem. 

In order to show that the GGE indeed describe a system in contact with reservoirs of rapidities, we propose the following picture. 
Let us first discretized the rapidity space: we split it in intervals $[\theta_i,\theta_i+\delta \theta]$,
where $i\in Z$, $\theta_i=i\delta \theta$ and $\delta \theta$ is much smaller than the scale of variation of $\rho$.
We assume now that the system is, for each integer $i$, in contact with a reservoir of rapidities lying in the interval $[\theta_i,\theta_i+\delta \theta]$, and we label the reservoir with  the integer $i$. 
Then, statistical mechanics tell us that the density matrix of the system is 
\begin{equation}
    \hat{\rho}\propto
    \sum_{|\{\theta_a\}\rangle}
     e^{-\sum_i f_i N_i} |\{\theta_a\}\rangle \langle \{\theta_a\}|
     \label{eq:rho_reservoirs}
\end{equation}
where $N_i$ is the number of rapidities
of the state $|\{\theta_a\}\rangle$ lying in the interval 
$[\theta_i,\theta_i+\delta \theta]$, and $f_i$ is the temperature parameter associated to the reservoir
number $i$. Eq.~\ref{eq:rho_reservoirs} is nothing else than the GGE ensemble given in Eq.~\ref{eq:GGErho}, with a discretized function $f$: in each segment $[\theta_i,\theta_i+\delta \theta]$,
 $f$ takes the constant value $f_i$.

\subsection{Asymptotic regimes of the Lieb-Liniger gas and approximate descriptions}
\label{sec:regimes}

So far we have seen that the thermodynamics of the Lieb-Liniger model is determined by the rapidity distribution $\rho(\theta)$, which parameterizes an infinite family of stationary states. This is in contrast with  
generic chaotic Galilean invariant gases, where the thermodynamic properties would depend only on the atomic density, the momentum density and the energy density. Despite the infinite-dimensional parameter space of stationary states, the Lieb-Liniger gas possesses a small number of asymptotic regimes where its description simplifies.

In this Subsection, we review the three main asymptotic regimes of the Lieb-Liniger gas: the ideal Bose gas, the quasicondensate, and the hard-core regimes. These three regimes arise when one compares the typical energy per atom $e$ to two energy scales: the scattering energy $mg^2/\hbar^2$ and the mean-field interaction energy $gn$ (where $n$ is the atom density):
\begin{itemize}
    \item $e \gg mg^2/\hbar^2 , g n$: ideal Bose gas regime
    \item $e \simeq g n   \gg  mg^2/\hbar^2 $: quasicondensate regime
    \item $mg^2/\hbar^2 \gg e$: hard-core regime.
\end{itemize}
In the following, we discuss the phenomenology of these three regimes.

\subsubsection{Ideal Bose gas regime}

When $e \gg mg^2/\hbar^2 , g n$, the interactions are negligible, and the gas behaves like a gas of non-interacting bosons. The rapidity distribution coincides with the momentum distribution of the bosons, as discussed after  Eq.~(\ref{eq:logbethe_bosons}). Alternatively, this can also be understood as a consequence of the fact that the momentum distribution of the ideal gas is preserved during a 1D expansion, and the fact that the rapidities are the asymptotic momenta after such an expansion, see Subsection~\ref{subsec:asymptotic_momenta}.

The Generalized Gibbs Ensemble that describes the local properties of the gas takes the form of a Gaussian density matrix, $\hat{\rho}_{\rm GGE} \propto \exp \left( \sum_{p} h(p) \Psi^\dagger_p \Psi_p \right)$, for some function $h(p)$.  Here $\Psi_p = \int e^{-ip x} \Psi(x) dx$ is the Fourier mode of the boson annihilation operator $\Psi(x)$. One of the consequences of that general Gaussian form is that, because of Wick's theorem, the two-body zero-distance correlation
\begin{eqnarray}
    \label{eq:g2_IBG}
 \nonumber   g^{(2)}(0) & := & \langle (\Psi^\dagger(0))^2  (\Psi(0))^2 \rangle/n^2 \\
    & = & 2. \qquad \quad ({\rm ideal \; Bose \; gas})
\end{eqnarray}
Thus, the gas exhibits the bosonic bunching phenomenon: whenever a boson is found inside a small interval $[x,x+d x]$, the probability to find another boson in that same interval is enhanced (i.e. it is larger than $n \,d x$).

We stress that the ideal Bose gas regime  is not restricted to the classical, or non-degenarate, limit. The population of some bosonic modes can be highly occupied, and thus the ideal Bose gas can be highly degenerate.

This is exemplified by the case of thermal equilibrium at temperature $T$ and chemical potential $\mu$, with $\mu <0$ for the ideal Bose gas. The distribution of bosons is given by the Bose-Einstein distribution $1/(e^{(\frac{p^2}{2m}-\mu)/(k_{\rm B} T)}-1)$, and there is a crossover between the classical gas (which corresponds to $|\mu|\gg k_{\rm B}T$) and the degenerate ideal gas ($|\mu|\ll k_{\rm B}T$). In the degenerate regime, for momenta $p\ll \sqrt{m k_{\rm B} T}$, the momentum distribution is close to a Lorentzian of half-width at half maximum  $\sqrt{2}m |\mu|$. The atom density is then $n\simeq 
k_{\rm B}T/\sqrt{2|\mu|/m}/\hbar$, so we can estimate that the gas is degenerate as long as $m (k_{\rm B} T)^2/(\hbar^2 n^2) \sim |\mu| \ll k_{\rm B}T$, or, equivalently, as long as
\begin{equation}
    \label{eq:condition_IBG_degenerate}
      k_{\rm B} T \ll \frac{\hbar^2 n^2}{m} .
\end{equation}
On the other hand, the typical energy per particle is $e\simeq |\mu|$. The above condition $e\gg g n$, which ensures that the gas, although it is degenerate, is in the ideal Bose gas regime as opposed to the quasicondensate regime, then reads
\begin{equation}
    \label{eq:condition_IBG_equilibrium}
    \hbar \sqrt{g/m} \, n^{3/2} \ll k_{\rm B} T .
\end{equation}
As long as both conditions (\ref{eq:condition_IBG_degenerate}) and (\ref{eq:condition_IBG_equilibrium}) are fulfilled, the thermal equilibrium gas is not in the quasi-condensate regime, even though it is highly degenerate. We note that the condition (\ref{eq:condition_IBG_equilibrium}) was first established by ~\cite{kheruntsyan2003pair}, who estimated the effects of interactions on $g^{(2)}(0)$ perturbatively, asking that they remain small.

\subsubsection{The quasicondensate regime}
This regime is reached when $\gamma := \frac{m g}{\hbar^2 n} \ll 1$ and 
the typical energy per atom $e$ stays close to its value in the ground state, $e \simeq gn/2$.
It is characterized by very small density fluctuations, with
\begin{equation}
    \label{eq:g2_quasicondensate}
    g^{(2)}(0) \simeq 1 . \qquad \quad ({\rm quasicondensate \; regime})
\end{equation}
Correlations are weak in this regime: the probability to find an atom in a small interval $[x, x+ d x]$ is barely affected by the presence of another atom in this interval.

A good description of the gas in that regime is provided by Bogoliubov theory, or more precisely the extension of Bogoliubov theory to quasicondensates \citep{mora_extension_2003}. This approach assumes a phase-density representation of the bosonic field: one writes the atomic field $\Psi(x)$ as $\sqrt{n + \delta n(x)} e^{i \phi(x)}$ where $\phi$ and $\delta n$ are the phase and density fluctuation fields, which fulfill $[\delta n(x), \phi(x')] = i \delta(x-x')$. This approach is a coarse grained approximation, valid for length scales much larger than the interparticle distance. The Bogoliubov approximation assumes small density fluctuations, $\delta n(x) \ll n$, and small phase gradient, $\partial \phi(x)/\partial x \ll n $. Inserting this phase-density representation into the Hamiltonian (\ref{eq:hamLL}), one finds to second order:
\begin{equation}
    \label{eq:ham_quasicondensate}
  H \, \simeq \, \int \left[ \frac{\hbar^2}{8m n} (\partial_x \delta n)^2  + \frac{g}{2} \delta n^2 + \frac{\hbar^2 n}{2m} (\partial_x \phi)^2  \right] dx .
\end{equation}
This quadratic Hamiltonian allows to grasp quantum fluctuations around the classical profile which solves the Gross-Pitaevski equation, i.e. $n = N/L = \mu/g$ where $\mu>0$ is the chemical potential. It is easily diagonalized by a Bogoliubov transformation. Defining the bosonic mode $B(x) = \frac{1}{2 \sqrt{n}} \delta n(x) + i \sqrt{n} \phi(x)$ such that $[B(x) , B^\dagger(x')] = \delta(x-x')$, and its Fourier transform $B_q = \int e^{-i q x/\hbar} B(x) dx/\sqrt{L}$ with $q\in (2\pi \hbar/L) \mathbb{Z}$, one finds that the quadratic Hamiltonian becomes, up to a constant term,
\begin{equation}
 H \, \simeq \,  \frac{1}{2} \sum_q \left( \begin{array}{c} B_q \\B_{-q}^\dagger \end{array} \right)^\dagger \left( \begin{array}{cc}
    \frac{q^2}{2m} + \mu &  \mu  \\
   \mu & \frac{q^2}{2m} + \mu 
 \end{array} \right)  \left( \begin{array}{c} B_q \\B_{-q}^\dagger \end{array} \right)  ,
\end{equation}
where we have used $\mu = g n$. Then the Bogoliubov transformation
$$
\left( \begin{array}{c}
    B_q \\ B_{-q}^\dagger
\end{array} \right) = \left( \begin{array}{cc}
    \bar{u}_q & \bar{v}^*_{q} \\
    \bar{v}_{-q} & \bar{u}^*_{-q }
\end{array}\right) \left( \begin{array}{c}
    b_q \\ b_{-q}^\dagger
\end{array} \right)
$$
with $\bar{u}_q = \bar{u}_q^* = \cosh (\theta_q/2)$ and $\bar{v}_q = \bar{v}^*_q = -\sinh (\theta_q/2)$, where
$\tanh \theta_q = \mu/(\mu + \frac{q^2}{2m})$, gives
\begin{equation}
    \label{eq:ham_quasicondensate_diagonal}
 H  \, \simeq \, \sum_q \varepsilon_q b_q^\dagger b_q \,+\, {\rm const.},
\end{equation}
with a dispersion relation $\varepsilon_q = \sqrt{\frac{q^2}{2m} \left( \frac{q^2}{2m} + \mu \right)}$.

The Bogoliubov model (\ref{eq:ham_quasicondensate_diagonal}) is obviously an integrable model, since it amounts to a collection of independent harmonic modes, and its integrals of motion are the population in each mode.  Making the link between the Bogoliubov modes and the rapidities in the Bethe-Ansatz solution of the Hamiltonian (\ref{eq:hamLL}) is, however, a difficult task. In his seminal work, \cite{lieb_exact_1963} described this link for states close to the ground state. He identified 
the so-called `Lieb-I excitation' (or `particle excitation') branch to the Bogoliubov modes. In a more recent investigation, \cite{ristivojevic_excitation_2014} found that this holds in fact only for large enough momenta. Thus, to our knowledge, making the connection between rapidities and Bogoliubov modes precise remains an open problem. The difficulty of this problem is related to the difficulty of developing schemes for expansions of the thermodynamic form of the Bethe equations (\ref{eq:constitutive}) at small $\gamma$, see e.g.~\citep{takahashi1975validity,popov1977theory,lang2017ground,prolhac2017ground,marino2019exact}.

\subsubsection{Hard-core regime}
The hard-core regime is reached when the scattering energy $m g^2/\hbar^2$ is much larger than all other intensive energy scales in the system. This is equivalent to taking $g \rightarrow +\infty$. Then, in all two-body scattering processes, the scattering phase factor (\ref{eq:2bd_scattering_phase}) is one, and the scattering shift (\ref{eq:scattering_phase}) vanishes. In that regime, two atoms can never be at the same position, which results
in
\begin{equation}
    \label{eq:g2_hardcore}
    g^{(2)} (0) \, = \, 0. \qquad \quad ({\rm hard-core \; regime}) 
\end{equation}
In that regime, when a boson is found in a small interval $[x,x+dx]$, then the probability to find another one in the same interval is zero. This property reflects the Pauli principle satisfied by 
non-interacting fermions, which are related to the hard-core bosons by the non-local transformation
\begin{equation}
    \label{eq:Jordan_Wigner}
    \Psi_{\rm F}(x) := \exp \left( i \pi \int_{y<x} \Psi^\dagger(y) \Psi(y) dy \right) \Psi(x).
\end{equation}
This transformation, closely related to the Jordan-Wigner transformation between lattice hard-core bosons and spin chains of spin-$1/2$, is defined such that the fermion creation/annihilation operators satisfy the canonical anticommutation relation $\{ \Psi_{\rm F}(x) , \Psi_{\rm F}^\dagger (y) \} = \delta(x-y)$. In terms of these fermions, the Lieb-Liniger Hamiltonian (\ref{eq:hamLL}) with $g \rightarrow +\infty$ becomes
\begin{equation}
   H \, =\, \int \Psi_{\rm F}^\dagger(x) \left[ - \frac{\hbar^2 \partial_x^2}{2m} - \mu \right] \Psi_{\rm F}(x) dx .
\end{equation}
This is the Hamiltonian of a non-interacting Fermi gas. The identification of hard-core bosons with non-interacting fermions remains valid in the presence of an external potential $V(x)$.

The Bethe wavefunction is, up to a sign $\prod_{a<b} {{\rm sign} (x_b-x_a)}$, the Slater determinant of $N$ non-interacting fermions  \citep{girardeau1960relationship},  and the rapidities are simply the momenta of the underlying non-interacting fermions, as discussed below Eq.~(\ref{eq:logbethe}). The GGE, which describe relaxed states and which are characterized by the rapidity distribution, corresponds, for the fermionic gas, to GGE states that are obtained as a product of Gaussian density matrix for each momentum state, $\hat{\rho}_{\rm GGE} \propto \exp \left( \sum_p h(p) \Psi^\dagger_{{\rm F},p} \Psi_{{\rm F},p} \right)$, where $\Psi^\dagger_{{\rm F},p} = \frac{1}{\sqrt{L}} \int e^{i p x/\hbar} \Psi^\dagger_{\rm F}(x)< dx$ is the Fourier mode of the above fermion creation operator. [To be more precise, for finite $L$ the fermions obey either periodic or anti-periodic boundary conditions depending on the parity of the particle number $N$, so the GGE is rather of the form $\hat{\rho}_{\rm GGE} \propto P_{N \,{\rm even}} \cdot \exp \left( \sum_{p \in \frac{2\pi \hbar}{L} (\mathbb{Z} + \frac{1}{2})} h(p) \Psi^\dagger_{{\rm F},p} \Psi_{{\rm F},p} \right) + P_{N \,{\rm odd}} \cdot \exp \left( \sum_{p \in \frac{2\pi \hbar}{L} (\mathbb{Z})} h(p) \Psi^\dagger_{{\rm F},p} \Psi_{{\rm F},p} \right)$, where $P_{N \,{\rm even}}$ and $P_{N \,{\rm odd}}$ are projectors onto the even and odd sectors. This complication can usually be omitted when one is interested in expectation values of local observables. For an example where it cannot be omitted, see e.g. \citep{bouchoule_effect_2020} where the change of boundary conditions plays a key role in determining the effect of particle losses on the resulting GGE.]

The number of works that have exploited the mapping from hard-core bosons to non-interacting fermions is too large to review them here. Here we simply mention a few such works that are  representative in that they illustrate the typical calculations that can be done in that regime. For instance, early studies of the momentum distribution 
\citep{lenard1964momentum,vaidya1979one} revealed the presence of algebraically decaying ground-state correlations, as well as the presence of tails in the momentum distribution of the atoms decaying as $1/p^{4}$~\citep{minguzzi2002high,rigol2004universal}. As is often the case with the 1D Bose gas, these observations remain valid beyond the hard-core regime, see e.g. the review articles~\citep{cazalilla2004bosonizing,cazalilla2011one} for long-range correlations, or~\citep{olshanii2003short} about the tails of the momentum distribution. 
Many advances have been obtained in out-of-equilibrium quantum dynamics thanks to the study of the hard-core limit, for instance the early works on Generalized Gibbs Ensembles~\citep{rigol2007relaxation}, or more recently investigations of trap releases \citep{collura2013equilibration,collura2013quench} and equilibration towards a GGE, or Floquet dynamics in harmonic traps~\citep{scopa2017one,scopa2018exact}. For more references on hard-core bosons, we refer to the bibliographies of these papers.

\subsubsection{The thermal equilibrium phase diagram} In general, because of its integrability, an isolated Lieb-Liniger gas has
no reason to be described by a thermal equilibrium state.
Even after relaxation, the system is expected to be described by a Generalized Gibbs Ensemble (GGE) parameterized by a whole function (the rapidity distribution, see Subsections~\ref{subsec:yangyang}, \ref{subsec:GGE}), rather than by a Gibbs ensemble parameterized by only 2 parameters: the atom density $n$ and the energy density $e$. 
However, in experiments, weak
perturbations violate integrability, for instance the
presence of transversely excited states~\citep{li_relaxation_2020,mazets_breakdown_2008}
or the longitudinal
potential~\citep{bastianello_thermalization_2020}. 
The Lieb-Liniger GGE will then be observable on an intermediate time-scale, long compared to the relaxation time of the Lieb-Liniger model, but short enough so that 
effect of integrability breaking perturbations is still
negligible. The system on this intermediate time-scale 
is called prethermalized~\citep{berges_prethermalization_2004}.
At very long times, the integrability breaking mechanisms will induce relaxation towards a thermal equilibrium state. Here we discuss the thermal equilibrium behavior of the 1D Bose gas, following \citep{petrov_regimes_2000,gangardt2003local} and especially \citep{kheruntsyan2003pair}.

\begin{figure}
  \centerline{ \scalebox{0.8}{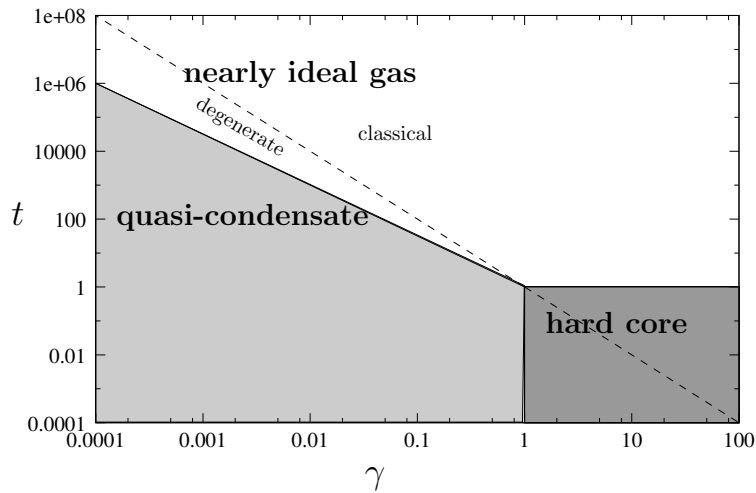}}
  \caption{Phase diagram of the Lieb-Liniger model at thermal equilibrium.
    Different asymptotic regimes are separated by smooth crossovers.
    The crossover between the ideal Bose gas regime and the quasicondensate
    regime occurs for $t\simeq \gamma^{-3/2}$, the crossover
    between the quasicondensate regime and the hard-core regime
    occurs for $\gamma \simeq 1$ and the crossover between the
    hard-core regime and the ideal Bose gas regime occurs
    for $t\simeq 1$. The dashed line represent the quantum
    degeneracy condition, which writes $t\simeq \gamma^{-2}$.
    Note that thermal equilibrium is not granted for the Lieb-Liniger model, because of its integrability.}
    \label{fig:phase_diagram}
\end{figure}

Assuming that the homogeneous 1D Bose gas is at thermal equilibrium, its state is characterized by only two dimensionless parameters: the dimensionless interaction strength $\gamma$, and the dimensionless temperature $t$ (not to be confused with a time in this Subsection)
\begin{equation}
    \gamma = \frac{m g}{\hbar^2 n} , \qquad \quad  t= \frac{k_{\rm B} T}{m g^2/\hbar^2}.
\end{equation}
To identify the different asymptotic regimes discussed above in the phase diagram ($\gamma$,$t$), one can rely on the fact that $g^{(2)}(0)$ allows to distinguish between them~\citep{kheruntsyan2003pair}, see Eqs.~(\ref{eq:g2_IBG},\ref{eq:g2_quasicondensate},\ref{eq:g2_hardcore}). One can compute
$g^{(2)}(0)$ using the Hellmann-Feynman theorem,
\begin{equation}
\label{eq:HelmanFeynam}
    n^2 g^{(2)}(0)=2(\partial F/\partial g)_{n,T},
\end{equation}
where $F$ is the free energy per unit length, which can be computed using Yang-Yang thermodynamics~\citep{yang1969thermodynamics}, see Subsection~\ref{subsec:yangyang}. As imposed by the Hohenberg-Mermin-Wagner theorem, there is no phase transition, however the three aforementioned asymptotic regimes appear in the phase diagram, separated by smooth crossovers, see Fig.~\ref{fig:phase_diagram}.

In Fig.~\ref{fig:phase_diagram}, we represent the crossover between the ideal Bose gas regime and the quasicondensate regime, $t \simeq \gamma^{-3/2}$, see the condition (\ref{eq:condition_IBG_equilibrium}). The dashed line is the quantum degeneracy condition $t\simeq \gamma^{-2}$, see (\ref{eq:condition_IBG_degenerate}).
Above this line, 
the occupation numbers of single particle quantum states are small: quantum effects are small and the gas behaves mainly as a  classical gas. Below this line, the behavior depends on the regime. In the ideal Bose gas regime, low energy single particle states get highly populated. 
In the hard-core regime, the rapidity distribution, which corresponds to the momentum distribution of the equivalent fermi gas, becomes close to the one of a zero-temperature Fermi sea, namely a rectangular function.

\subsubsection{The classical field approximation.}
\label{subsubsec:classicalfield}
We conclude this survey of the regimes of the Lieb-Liniger model by discussing  the 
classical field approach, which, owing to its simplicity and its relevance to the description of the crossover between the quasicondensate and degenerate ideal Bose gas, is a popular technique. In this approach, the quantization of the atomic field, i.e. the discrete nature of atoms, is ignored, and the physics boils down to that of a classical complex field $\psi(x)$. The energy functional of this field is 
\begin{equation}
    E[\psi]=\int dx \left 
    (
    \frac{\hbar^2}{2m}\left | \frac{\partial \psi}{\partial x} \right |^2 + \frac{g}{2} |\psi|^4 - \mu |\psi|^2
    \right ),
\end{equation}
and the Lagrangian is $L=(i\hbar/2)\int dx (\psi^* \partial \psi/\partial t -\partial \psi^*/\partial t\psi)-E[\psi]$, such that the time evolution of $\psi$ is given by the Gross-Pitaevski equation
\begin{equation}
    i\hbar \frac{\partial \psi}{\partial t} = -\frac{\hbar^2}{2m} \frac{\partial^2 \psi}{\partial x^2} +g |\psi|^2\psi -\mu \psi.
\end{equation} 
The effect of an external potential is easily taken into account within this approach, by adding a term $\int dx V(x) |\psi|^2$ (resp. $\int dx V(x) \psi$ ) to the energy functional (resp. to the Gross-Pitaevski equation).
This approach is expected to be meaningful in the degenerate ideal Bose gas regime, as well as in the high temperature quasicondensate regime, where the population of the modes is high. In particular, it captures the crossover between the ideal bose gas regime and the quasicondensate regime. This approach has been used, at thermal equilibrium, to compute correlation functions~\citep{castin_coherence_2000,castin_simple_2004,jacqmin_momentum_2012,bouchoule_two-body_2012} and full counting statistics~\citep{arzamasovs_full_2019}, and to investigate non-equilibrium dynamics~\citep{bouchoule_finite-temperature_2016,thomas_thermalization_2021}. 
In the absence of an external potential, the Gross-Pitaevski equation is the non-linear Shr\"odinger equation, a classical field integrable model. The link between the classical integrals of motion and the rapidity distribution of the quantum model has been discussed in~\citep{vecchio_exact_2020,bettelheim2020whitham}.


The classical field  approach is plagued by an overestimation of the role of high wavevector components of $\psi$: their thermal mean value scales as $m k_{{\rm B}} T/(2\hbar)$ in the classical field approach, instead of the expected Gaussian behavior $e^{-\hbar^2 k^2/(2m k_{{\rm B}} T)}$. 
This can affect strongly some observables. In higher dimensions, this induces  a UV divergence of the density, which leads to the well known black body problem.
In 1D, the density does not diverge within the classical field approximation, but other observables do, like the energy density.  
To cure this problem, refined classical field approaches have been developed, that include a cut-off~\citep{davis_simulations_2001,blakie_dynamics_2008,cockburn_comparison_2011,cockburn_quantitative_2011}.

\newpage

\section{Generalized Hydrodynamics of the 1D Bose gas: theory results}
\label{sec:GHDtheory}

In the previous Section, we reviewed the thermodynamic properties of the homogeneous Lieb-Liniger gas. In particular, we emphasized the key role of the distribution of rapidities $\rho(\theta)$. In the thermodynamic limit, expectation values of physical observables, like charge densities or currents,  become functionals of the rapidity distribution.

\begin{figure}[ht]
    \centering
    \includegraphics[width=0.45\textwidth]{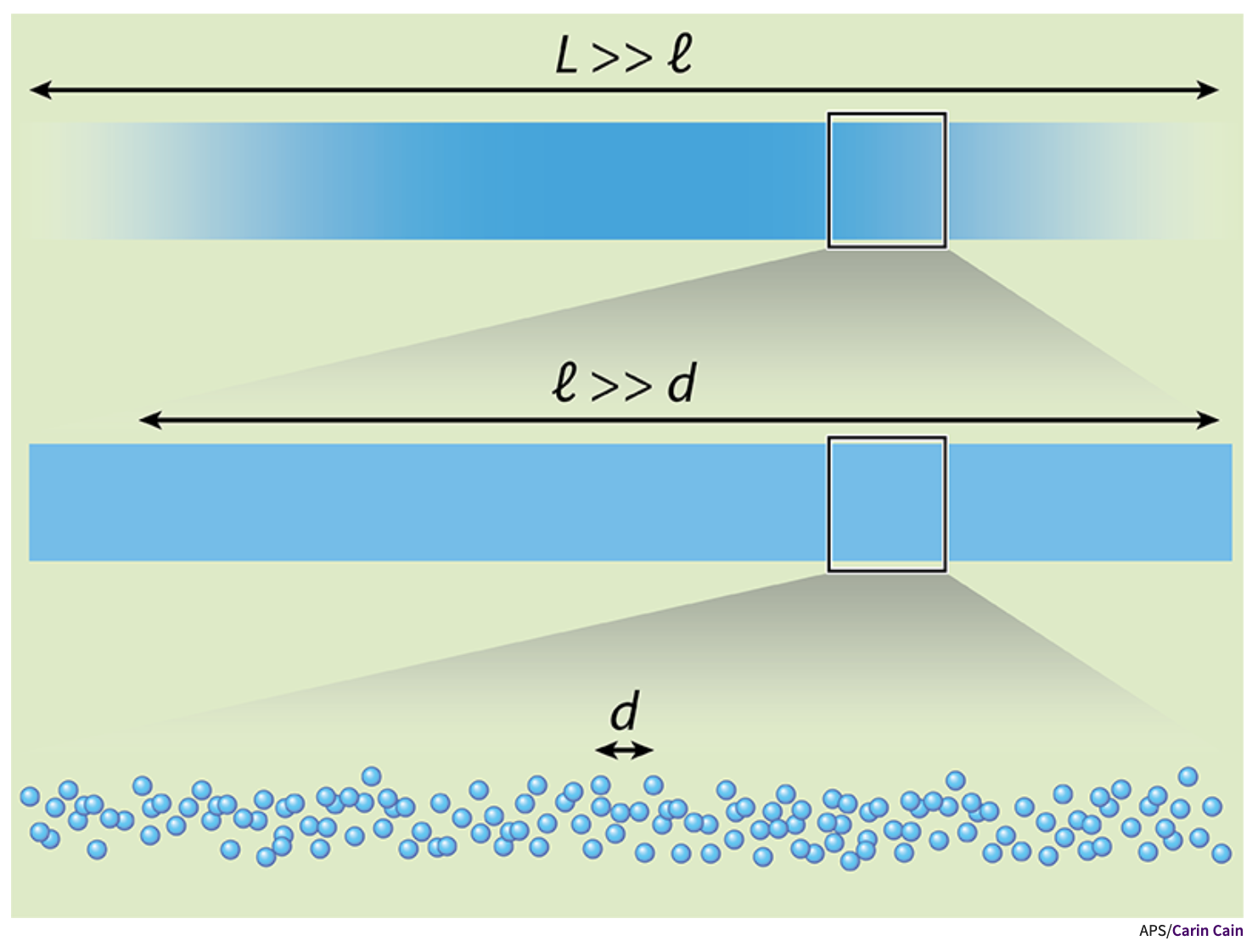}
    \caption{[From \citep{dubail2016more}.] Like any other hydrodynamic approach, Generalized Hydrodynamics applies in the limit where separations of distance and time scales hold. When the characteristic distance $L$ over which densities vary is much larger than the microscopic length scale $d$, one can view the system as a continuum of locally homogeneous fluid cells of size $\ell$ ($d \ll \ell \ll L$) which contain a thermodynamically large number of particle. Similarly, assuming slow variation in time, each fluid cell is locally relaxed to a stationary state. In standard hydrodynamics, this local stationary state is a thermal equilibrium state, while in Generalized Hydrodynamics it is a Generalized Gibbs Ensemble.}
    \label{fig:sos}
\end{figure}

In this Section we turn to the Euler scale hydrodynamic equations that follow from the thermodynamics of the Lieb-Liniger model. As in any hydrodynamic approach, the starting point is the assumption of separation of scales, see Fig.~\ref{fig:sos}. When the charge densities in the gas vary sufficiently slowly in space and in time, one can view the gas as a continuum of fluid cells, each of which contains a thermodynamically large number of particles that have relaxed to a stationary state.

Under the assumption of separation of scales, the gas is described by its distribution of rapidities $\rho(x,\theta,t)$ within each fluid cell $[x,x+dx]$ at time $t$. This time- and position-dependent rapidity density evolves according to the Generalized Hydrodynamic equations,
\begin{equation}
    \label{eq:ghd}
    \begin{array}{c}
       \displaystyle \partial_t \rho(x,\theta,t) + \partial_x \left( v^{\rm eff}[\rho] (\theta) \, \rho(x,\theta,t) \right) - (\partial_x V(x)) \partial_\theta \rho(x,\theta,t)  \, = \, 0 \\
       \displaystyle v^{\rm eff}[\rho] (\theta) = \theta - \int_{-\infty}^\infty \Delta(\theta- \theta')  \left(  v^{\rm eff}[\rho] (\theta)  - v^{\rm eff}[\rho] (\theta') \right) \rho(\theta') d\theta' .
    \end{array}
\end{equation}
These equations were first derived for quantum integrable systems by \cite{castro2016emergent,bertini2016transport} (more precisely, they were derived in the absence of an external potential $V(x)$; the additional term, which corresponds to Newton's second law, was added later by \cite{doyon2017note}). These equations are of the same form as Eqs.~(\ref{eq:GHDintro}) for the classical integrable gas discussed in the introduction, with two main nuances. The first is that it is the density of rapidities, or asymptotic momenta, that enters the equations; not a `bare' velocity as in the hard rod gas. The second is that $\Delta(\theta-\theta') = \frac{2mg/\hbar}{(m g/\hbar)^2 + (\theta - \theta')^2}$ is now the scattering shift (\ref{eq:scattering_phase}), which depends on the rapidities,  while $\Delta$ in the classical gas in the introduction was just a constant equal to minus the diameter of the balls. The effective velocity that solves the second equation (\ref{eq:ghd}) can be written as $v^{\rm eff}[\rho](\theta) = {\rm id}^{\rm dr}(\theta)/1^{\rm dr}(\theta)$, as discussed in Subsection \ref{subsec:expectation_thermodynamic}.

\subsection{Hydrodynamic approaches to the 1D Bose gas that preceded GHD}
\label{subsec:conventional_hydro}

The idea of a hydrodynamic description of the 1D Bose gas does not date back to 2016, it is of course much older. One popular hydrodynamic approach in the atomic gas literature is to start from the Gross-Pitaevskii description of the weakly interacting gas at zero temperature. Writing the wavefunction of the quasicondensate as $\psi=\sqrt{n} e^{i \phi}$, and defining the velocity $u= \frac{\hbar}{m} \partial_x \phi$), one gets the Madelung form of the Gross-Pitaevskii equation \citep{stringari1996collective,stringari1998dynamics,cazalilla2011one},
\begin{equation}
    \label{eq:GPmadelung}
    \left\{\begin{array}{rcl}
            \partial_t n + \partial_x (n u) & = & 0 \\
            \partial_t u + u \partial_x u + \frac{1}{m n} \partial_x \mathcal{P} + \partial_x \left( - \frac{\hbar^2}{2m^2} \frac{\partial_x^2 \sqrt{n}}{\sqrt{n}} \right) &= & - \frac{1}{m} \partial_x V ,
    \end{array}\right.
\end{equation}
where $\mathcal{P}(n) = \frac{1}{2}g n^2$ is the pressure of the gas in the quasicondensate regime. Clearly, these two equations look like the first two Euler hydrodynamic equations (\ref{eq:euler}) in the introduction, up to the so-called quantum pressure term $- \frac{\hbar^2}{2m^2} \frac{\partial_x^2 \sqrt{n}}{\sqrt{n}} $. This term is beyond the Euler scale though: it involves higher order derivatives, so in the Euler limit of a slowly varying density $n(x)$, this term vanishes. The reason why there are only two equations in (\ref{eq:GPmadelung}), instead of the three Euler equations in the introduction, is that the temperature is zero. Indeed, the third Euler equation in (\ref{eq:euler}) can be recast as a conservation law for the entropy of the fluid, which is automatically satisfied at zero temperature because the entropy identically vanishes.

There have been several attempts at extending this description of the gas beyond the weakly interacting regime, see e.g.~\citep{kolomeisky2000low,menotti2002collective,damski2006shock}. One idea that is often used \citep{menotti2002collective,ohberg2002dynamical,pedri2003violation,damski2006shock,peotta2014quantum,sarishvili2016pulse} is to replace the pressure $\mathcal{P}(n)$ of the quasi-condensate regime by the true pressure of the Lieb-Liniger model at zero temperature, calculated from Eq.~(\ref{eq:pressure2}). This gives a closed system of hydrodynamic equations for the gas that can be solved numerically. This approach can also be used at finite temperature~\citep{bouchoule_finite-temperature_2016,doyon2017large,schemmer2019generalized}: in that case one numerically solves the three Euler hydrodynamics equations from the introduction with the equilibrium pressure at finite temperature.

\begin{figure}
    \centering
    \begin{tikzpicture}
        \draw (0,-0.5) node{ \includegraphics[width=0.4\textwidth]{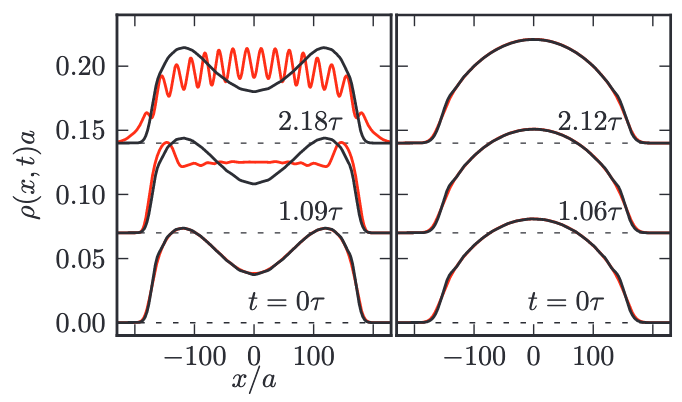}};
       \draw (7.5,-0.2) node{\includegraphics[width=0.4\textwidth]{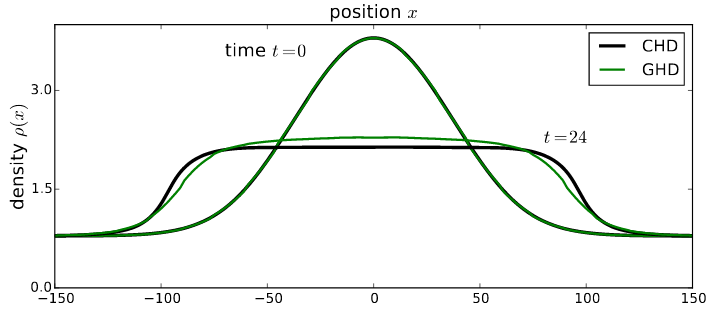}};
    \end{tikzpicture}
    \caption{Left: [from \citep{peotta2014quantum}] evolution of the density profile of the 1D Bose gas after a quench from a double-well to harmonic potential (left column), and after a quench from harmonic to harmonic, with a different frequency (right column). $\tau$ is the oscillation period of the harmonic trap during the evolution. The red curve is obtained by numerically solving the hydrodynamic equation (\ref{eq:GPmadelung}) with the pressure $\mathcal{P}$ of the Lieb-Lininger model; the black curve is a numerically exact t-DMRG simulation. Clearly, the `conventional' hydrodynamics (\ref{eq:GPmadelung}) works well for the second quench (right column), but not for the first one (left column).
    Right: [from \citep{doyon2017large}] evolution of the density profile from an initial thermal equilibrium state at non-zero temperature in an inverted Gaussian potential $V(x) = -5e^{-(x/50)^2}-1$, which creates an initial density bump. At $t>0$ the potential is switched off, $V(x)=0$. The density profile evolves according to   `conventional' Euler hydrodynamics (\ref{eq:euler}) at finite temperature (black curve), or according to GHD (green curve). We see that the predictions of both hydrodynamic theories differ at finite temperature (while they would coincide at zero temperature before the appearance of a shock).}
    \label{fig:peotta_diventra}
\end{figure}

This `conventional' Euler hydrodynamic approach, which assumes local relaxation of the gas to a thermal equilibrium state, has been succesfully applied in states not far from thermal equilibrium, see e.g. \citep{menotti2002collective,ohberg2002dynamical,pedri2003violation}. However, it breaks down away from equilibrium. A good illustration of the problems that are typically encountered can be found in~\citep{peotta2014quantum}, see Fig.~\ref{fig:peotta_diventra}. In that reference, the `conventional' hydrodynamic equations (\ref{eq:GPmadelung}) are solved numerically, and compared to numerically exact t-DMRG calculations for small numbers of bosons. The authors find that the hydrodynamic equations can describe the breathing of the atom cloud very well after a quench of the harmonic trapping frequency. However, those hydrodynamic equations are unable to describe a quench from a double-well to harmonic potential away from the weakly interacting regime.

The reason for the failure of this `conventional' hydrodynamic approach in the latter setup is that it develops a shock after some fraction of the oscillation period. When the atom cloud has initially two well separated density peaks, some atoms from the left peak move to the right with large velocity, while other atoms from the right peak move to the left with large velocity. Then in the center of the cloud, the fluid is similar to a two-component fluid, with one component moving fastly to the right, the other to the left. Locally, this is a state that is very far from a thermal equilibrium state. The above approach, which enforces local thermal equilibrium, fails to capture that situation. Instead, the Euler-scale hydrodynamic equations (\ref{eq:GPmadelung}) develop a shock, which is regulated by higher order derivative terms, like the quantum pressure term. The solution after the shock depends very strongly on the details of the regularization, so that in general it loses its validity after the first shock. The results of \citep{peotta2014quantum} show that the simple insertion of the quantum pressure term into the zero-temperature hydrodynamic equations (\ref{eq:GPmadelung}) does not provide the correct regularization for finite repulsion strength. Only in the limit of weak repulsion $(g \rightarrow 0)$, i.e. when Eq. (\ref{eq:GPmadelung}) is mathematically equivalent to the standard Gross-Pitaevskii equation, does the quantum pressure term provide the right regularization. In that case, the Euler-scale shock is visible in the Gross-Pitaevskii solution through the appearance of strong oscillations of short wavelength in the density profile $n(x,t)$, see e.g. \citep{simmons2020quantum}. These oscillations are beyond the Euler scale, and one can in principle average them over fluid cells to recover the correct Euler-scale description after the shock. \cite{bettelheim2020whitham} has shown that the Euler-scale dynamics obtained from the Gross-Pitaevskii equation in this way exactly coincides with GHD (at zero temperature and in the limit of weak repulsion).

Away from the Gross-Pitaevskii limit, the analysis carried out in \citep{doyon2017large} shows that the above approach is, in fact, well justified only at zero temperature and before the appearance of the first shock. Moreover, in that case, the `conventional' hydrodynamics (\ref{eq:GPmadelung}) at the Euler scale (i.e. neglecting the quantum pressure term) turns out to be exactly equivalent to GHD. However, in any other situation, it is in principle not applicable, and it leads to quantitatively  wrong results. This is illustrated in Fig.~\ref{fig:peotta_diventra}.

\subsection{Modeling the quantum Newton Cradle setup with GHD}
\label{susec:modelingNC}
Contrary to previous hydrodynamic approaches to the 1D Bose gas, GHD is not based on the assumption of local thermal equilibrium, but only on local relaxation to a stationary state which, in general, is a Generalized Gibbs Ensemble. This allows to describe situations that are very far from thermal equilibrium. Arguably, the most paradigmatic such out-of-equilibrium situation is the quantum Newton Cradle setup of~\cite{kinoshita2006quantum}. There, the atoms, which are initially at equilibrium in a harmonic potential, are suddenly given a large momentum $\pm q_{\rm Bragg}$ by a Bragg pulse. Half of the atoms move to the right, the other half move to the left. Because of the harmonic trapping potential, the two packets of atoms oscillate in the trap, colliding twice during each oscillation cycle. In this subsection we review the theory work on this paradigmatic setup. The pioneering experiment of~\cite{kinoshita2006quantum}, which motivated all these theory works, is discussed below in Sec.~\ref{sec:experiments_beforeGHD}.

\begin{figure}
    \centering
    \includegraphics[width=0.95\textwidth]{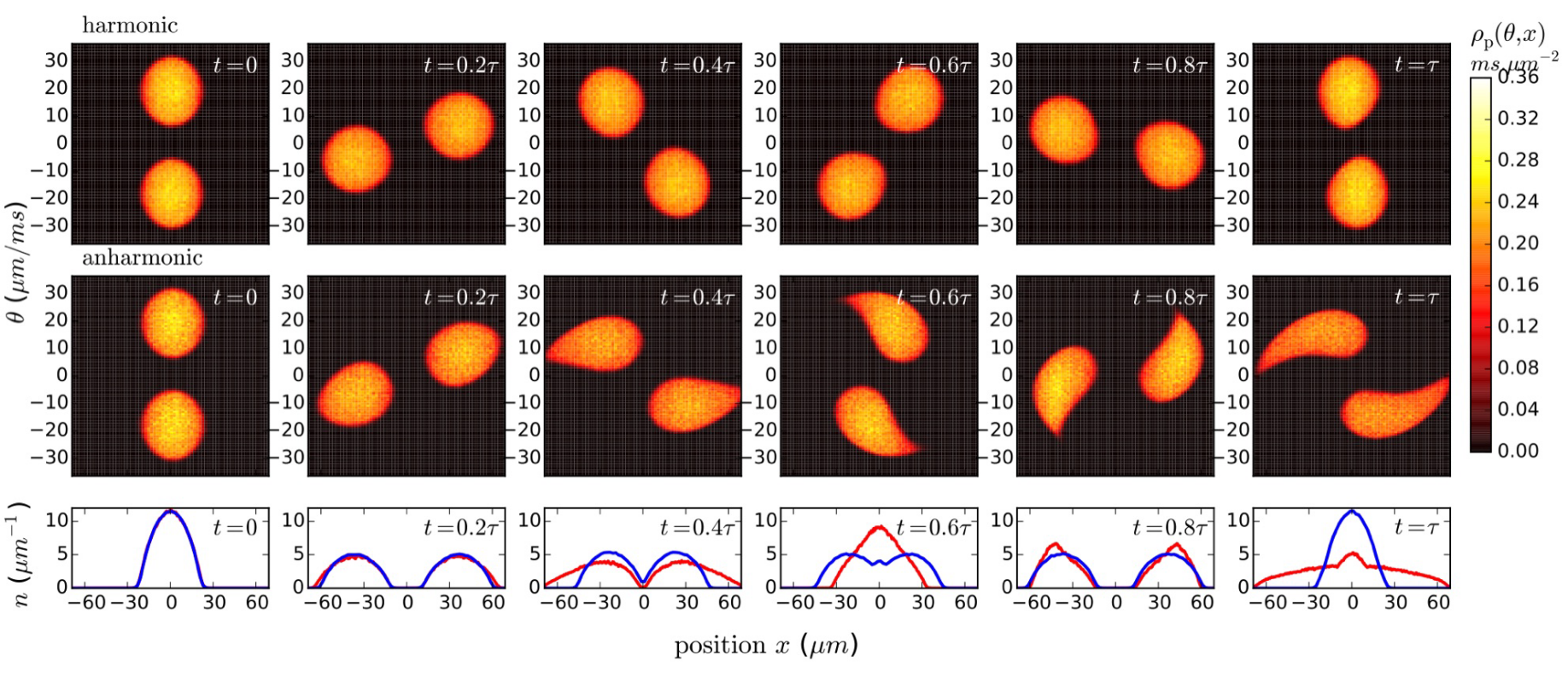}
    \includegraphics[width=0.95\textwidth]{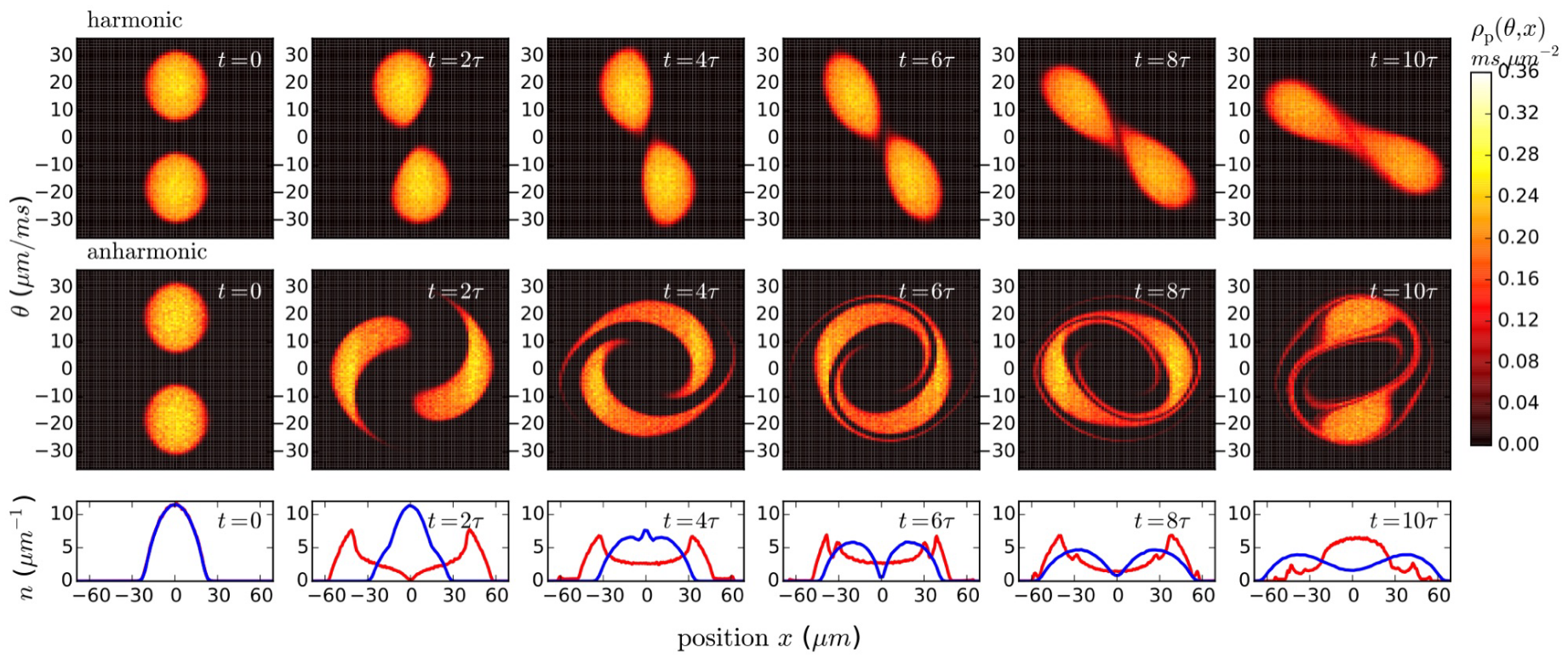}
    \caption{[From \citep{caux2019hydrodynamics}.] GHD simulation of the 1D Bose gas in the Newton Cradle setup. Top figure: the evolution of the phase-space rapidity density $\rho(x,\theta,t)$ during the first oscillation cycle is shown for a harmonic trap with period $\tau$ (first row), and a quasi-harmonic trap with a small anharmonicity (second row). The corresponding density profiles $n(x,t) = \int \rho(x,\theta,t ) d\theta$ are shown in blue and red respectively (third row). Bottom: same as the top one, but on longer time scales.}
    \label{fig:QNCtheory}
\end{figure}

We stress that, before the advent of GHD, a direct simulation of the Quantum Newton Cradle setup with experimentally realistic parameters (in particular, a number of atoms numbers of atoms $N \sim 10^2-10^3$) was completely out of reach. This is of course because of the exponential growth of the Hilbert space in  many-body quantum systems, which makes all direct approaches, such as an exact diagonalization of (a dicretized version of) the Lieb-Liniger Hamiltonian (\ref{eq:hamLL}), numerically untractable for more than a  dozen of atoms. Thanks to the discovery of GHD in 2016, this situation has now completely changed. Nowadays, it is very easy to model the 1D Bose gas in a quantitatively reliable way.

 \cite{bulchandani2017solvable} used GHD to model two packets of atoms colliding against each other on an infinite line. Then, a complete study of the Newton Cradle setup, including the trapping potential that gives rise to oscillations of the packets and therefore multiple collisions, was performed by \cite{caux2019hydrodynamics}, see Fig.~\ref{fig:QNCtheory}. The numerical solution of the GHD equation in that reference was obtained by a classical molecular dynamics simulation of the so-called flea gas model \citep{doyon2018soliton}, which is an extension of the classical hard rod gas that incorparates the scattering shift discussed in Subsection~\ref{subsec:Wigner}. There are other ways of numerically solving the GHD equations, which have been discussed, to some extent, in \citep{bulchandani2017solvable,bulchandani2018bethe,doyon2017large,bastianello2019generalized,moller2020introducing,bastianello_thermalization_2020,moller_extension_2021} (see also the appendix of the review by Bastianello, de Luca and Vasseur in this Volume). Among those works, we advertise in particular the GHD code `ifluid' of \cite{moller2020introducing}, which is publicly available.

In Fig.~\ref{fig:QNCtheory}, one can observe the evolution of the rapidity distribution predicted by GHD in a harmonic potential $V(x) = \frac{1}{2}m \omega^2 x^2$ (with oscillation period $\tau = 2\pi/\omega$), and also in a potential with a small anharmonicity potential $V(x) = \frac{m \omega^2}{\pi^2 \ell^2} (1 - \cos \frac{\pi x}{\ell} )$. The initial state is constructed so as to mimic the effect of the Bragg pulse sequence that imparts their initial momentum to the atoms. Before the sequence, the gas is described in a hydrostatic (or local density) approximation by its local distribution of rapidities $\rho(x,\theta,t<0) = \rho_{\rm thermal}(x,\theta)$, obtained by solving the Yang-Yang equation (\ref{eq:yangyang}) for a thermal equilibrium distribution. Then momentum $\pm q_{\rm Bragg}$ is imparted in a random fashion to all quasiparticles in the system. This results in a distribution of rapidities at $t=0$
\begin{equation}
    \label{eq:initial_Bragg}
    \rho( {x,\theta,t=0}) \, \simeq \,  \frac{1}{2} \rho_{\rm thermal}(x,\theta - q_{\rm Bragg})  + \frac{1}{2} \rho_{\rm thermal}(x,\theta + q_{\rm Bragg}) .
\end{equation}
This simple Ansatz for the initial state can be justified using results of \citep{van2016separation}, which showed that the momentum distribution function of the bosons is affected in this way by a Bragg pulse, and to a good approximation the same holds for the rapidities. The rapidity distribution is evaluated at later times by solving the GHD equation (\ref{eq:ghd}). One observes that the two blobs, initially well separated in momentum space for sufficiently large $q_{\rm Bragg}$, evolve by performing a deformed rotation-like movement around the origin of phase space. In the harmonic case, over the first two or three oscillations cycles, their evolution is not drastically affected by the collisions. However at later times, the two blobs ultimately merge due to inter-cloud interactions. With a small anharmonicity, the two blobs get deformed much more quickly and the distribution $\rho(x,\theta,t)$ gets more and more stirred up after few periods. This dephasing effect would also be present for the single particle in an anharmonic trap, see e.g.~\citep{bastianello2017quenches}. Many-body dephasing is also present: without interactions, the original blobs would disintegrate into long spiraling filaments; instead, here the filaments merge and high-energy (longer-period) tails scatter to lower energies, leading to the reformation of new blobs.

Importantly, the ability to perform GHD modeling of the Newton Cradle setup has opened the possibility to study theoretically one fundamental question raised by the experiment of \cite{kinoshita2006quantum}. If one waits long enough, does the gas in the trap ultimately reach thermal equilibrium?

This question is non-trivial because, although the Lieb-Liniger gas is integrable, its integrability is broken by the trapping potential $V(x)$. Yet, at the Euler scale, the potential $V(x)$ varies very slowly compared to microscopic scales, so the breaking of integrability by the external potential is weak. It is not obvious how much of the original conservation laws should be reflected in the stationary state.

\cite{cao2018incomplete} studied this question for the classical hard rod gas in a trapping potential, and found that the answer is negative: the gas exhibits `incomplete thermalization'. It reaches a stationary state of the GHD equation (\ref{eq:ghd}), namely a rapidity distribution $\rho_{\rm stat.}(x,\theta)$ which satisfies
\begin{equation}
    \label{eq:GHDstationary}
    \partial_x \left( v^{\rm eff}[\rho_{\rm stat.}](\theta) \, \rho_{\rm stat.}(x,\theta) \right) - (\partial_x V) \partial_\theta \rho_{\rm stat.}(x,\theta) = 0, 
\end{equation}
but this distribution does not need to be a thermal equilibrium distribution in the trap. The possibility of  GHD-stationary distributions of the form (\ref{eq:GHDstationary}) that are not thermal has also been discussed in \citep{doyon2017note}. \cite{caux2019hydrodynamics} arrived at the same conclusion for the 1D Bose gas in the Newton Cradle setup, within the framework of the aforementioned flea gas model. In \citep{caux2019hydrodynamics}, the existence of  non-thermal stationary states was argued to be a consequence of the conservation of certain quantities $S[f]$ under evolution generated by the Euler-scale GHD equation (\ref{eq:ghd}), even in the presence of an external potential $V(x)$. The conservation of these quantities is incompatible with convergence towards thermal equilibrium. These quantities can be constructed out of the Fermi occupation ratio $\nu(x,\theta,t)$, and read
\begin{equation}
    \label{eq:Sf_conserved}
    S[f] \, = \, \int dx \, d \theta  \, \rho_{\rm s}(x,\theta,t)  f( \nu(x,\theta,t) ) ,
\end{equation}
for arbitrary functions $f$. We stress that these quantities are different from the standard conserved charges of the form (\ref{eq:chargeQf}). Instead, the quantities $S[f]$ look more like generalizations of the Yang-Yang entropy: the Yang-Yang entropy, integrated over space, corresponds to the specific  choice $f(\nu) = -\nu \log \nu - (1-\nu) \log (1-\nu)$, see Eq.~(\ref{eq:SYY}). The fact that
\begin{equation}
    \frac{d}{dt} S[f] \, = \, 0
\end{equation}
can be checked directly using the GHD equation (\ref{eq:ghd}), see \citep{caux2019hydrodynamics}. We stress that the Yang-Yang entropy, and more generally the quantities $S[f]$, are conserved only at the Euler scale. When higher-order terms are included into the hydrodynamic equations (\ref{eq:ghd}), as discussed in Subsection~\ref{sec:diffusive_corrections} below, the entropy increases with time, and the other quantities $S[f]$ are no longer constant. In particular, it has been argued recently by \citep{bastianello_thermalization_2020} that the inclusion of a Navier-Stokes-like higher order term in (\ref{eq:ghd}) does lead to `complete thermalization', see Subsection~\ref{sec:diffusive_corrections} below.

\subsection{Other setups}
\label{subsec:other_setups}
Let us briefly review some other physically relevant setups that have been investigated with the new toolbox provided by GHD. 

\cite{de2018edge} considered the case of two atom clouds that are prepared at different temperatures $T_1 \neq T_2$, put in the same trapping potential. At the junction between the two clouds, the local state displays an edge singularity in its response function and quasilong-range order.

\begin{figure}
    \centering
    \begin{tikzpicture}
        \draw (-3.2,0) node{\includegraphics[width=0.5\textwidth]{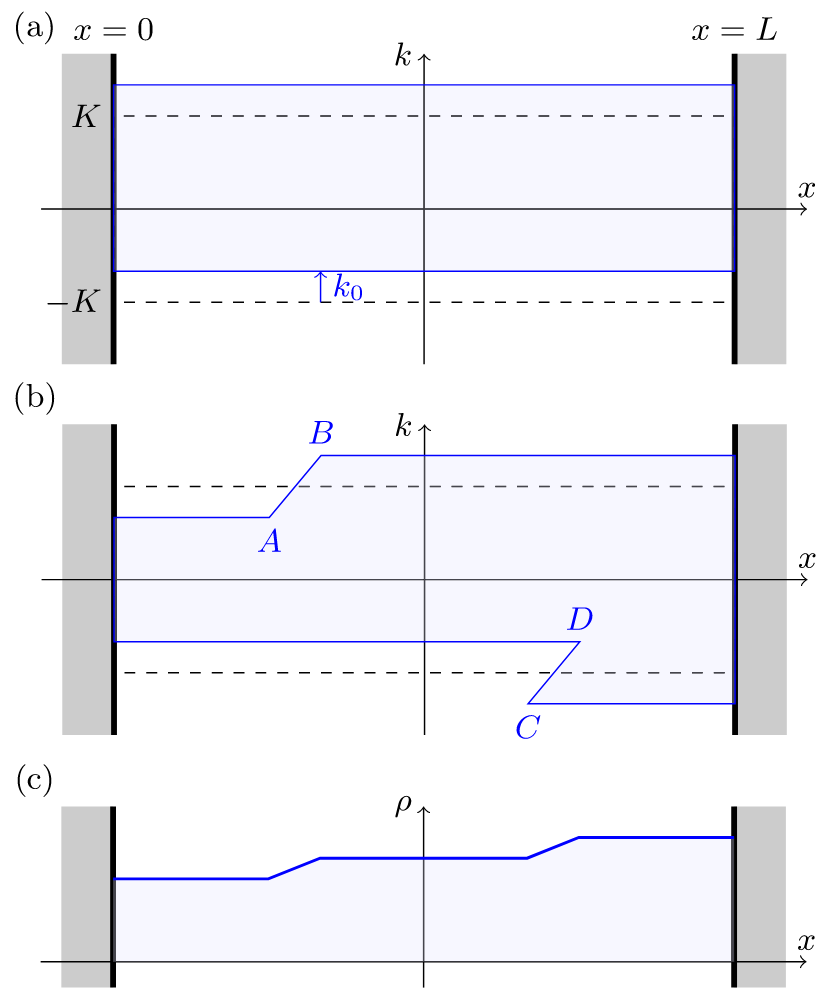}};
        \draw (4,2.2) node{\includegraphics[width=0.4\textwidth]{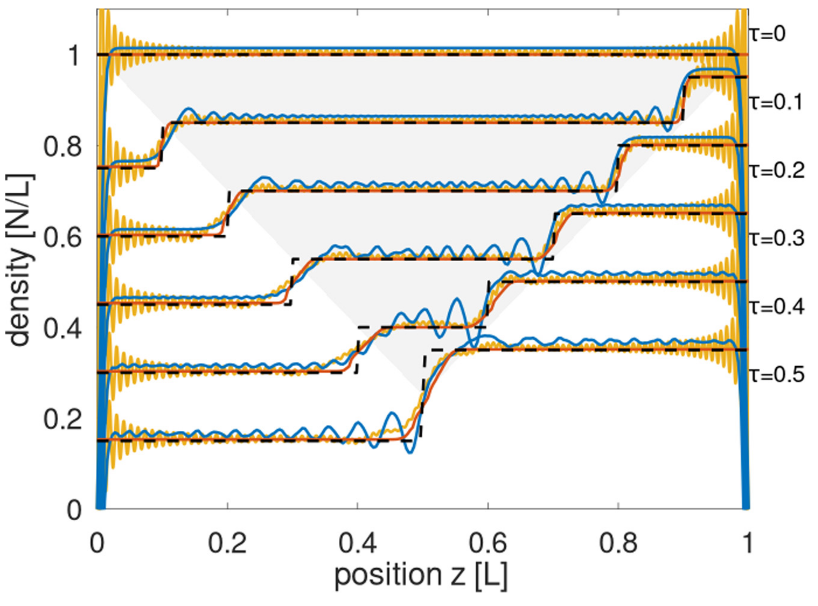}};
        \draw (4,-2.8) node{\includegraphics[width=0.4\textwidth]{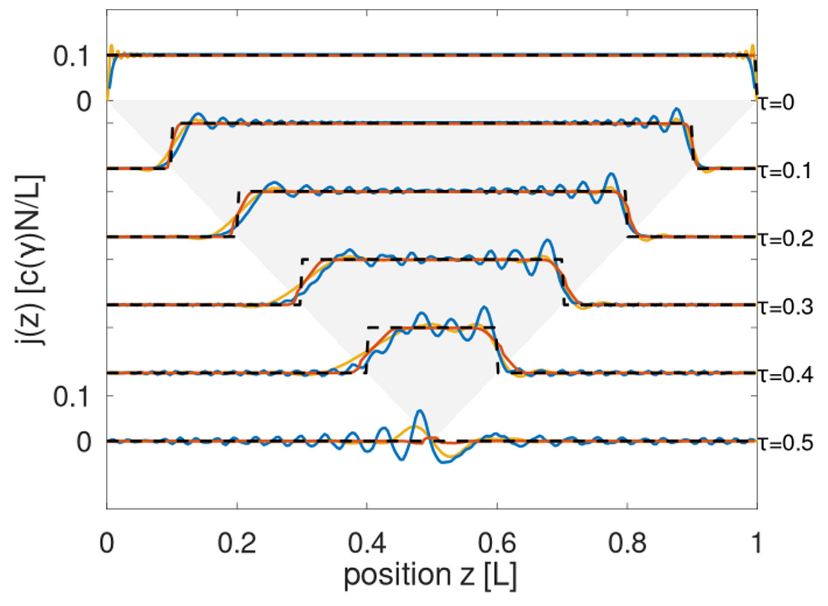}};
    \end{tikzpicture}
    \caption{[From \citep{dubessy2021universal}] Left: GHD simulation of a gas in an infinite box trap. (a) The initial state is the ground state, on which a boost of $+k_0$ is applied at $t=0$. The blue region is the region of phase space where the Fermi occupation ratio $\nu (x,\theta, t) = \rho(x,\theta, t)/ \rho_{\rm s}(x,\theta, t)$ is one. Outside this region, it is zero. (b) After some time the contour of the blue region gets deformed.  (c) The corresponding real-space density profile $n(x,t) = \int \rho(x,\theta, t) d\theta$. Right: evolution of the particle density (top) and current (bottom). Here $\tau = t/(v L)$ where $v$ is the sound velocity. The blue curve is obtained with the Gross-Pitaevskii equation (valid for small $\gamma$), the yellow curve is obtained from a free fermion calculation in the $\gamma \rightarrow \infty$ limit, and the red curve is the GHD simulation at $\gamma=1$. The dashed black line correspond to an effective model, see \citep{dubessy2021universal}.
    }
    \label{fig:dubessy}
\end{figure}

\vspace{0.5cm}

\cite{dubessy2021universal} studied the Lieb-Liniger gas in an infinite flat box potential (Fig.~\ref{fig:dubessy}). Initially the gas is in its ground state. At time $t=0$, it is instantaneously boosted by a momentum $k_0$, a protocol that can be realized experimentally by phase imprinting. Then the particles in the gas start reflecting against the two infinite walls at $x=0$ and $x=L$. This is modeled in GHD by the following boundary condition,
\begin{equation}
    \label{eq:wall_boundary_conditions}
    \rho(x = 0, \theta) \, = \, \rho(x = 0, -\theta) .
\end{equation}
The same boundary condition holds at $x=L$. \cite{dubessy2021universal} used a trick to implement easily  these boundary conditions: the system can be glued together with its mirror image, to give a periodic system of length $2L$. The rapidity distribution in that periodic system of size $2L$ is related to the one in the infinite box potential by
\begin{equation}
    \rho_{\rm periodic}(x,\theta) = \left\{ \begin{array}{rcl} \rho (x, \theta)  &{\rm if}& 0 < x < L \\
    \rho (2L-x, -\theta)  &{\rm if}& L < x < 2L.
    \end{array} \right.
\end{equation}
With this trick, \cite{dubessy2021universal} studied the formation of shock waves that oscillate in the box for several  periods, with a period fixed by the sound velocity in the gas, see Fig.~\ref{fig:dubessy}.

\vspace{0.5cm}

\cite{doyon2017large} pointed out that a drastic simplification of the GHD equations occurs when the gas is initially in its ground state. In an external potential $V(x)$, the ground state is modeled by hydrostatics, or equivalently by the local density approximation, which gives the distribution of rapidities $\rho(x,\theta, t=0)$. In the ground state, the Yang-Yang entropy of the gas vanishes. Then, since entropy is always conserved by Euler-scale hydrodynamic equations, it must vanish at all times. This puts very strong restrictions on the class of local stationary states that are explored by the system under GHD evolution: these local states must be either the ground state itself (up to a Galilean boost), or a `split Fermi sea' \citep{fokkema2014split,eliens2016general,eliens2017quantum}. Within the framework of GHD, this is easily understood by using the so-called convective form of the GHD equation (\ref{eq:ghd}), which gives the evolution of the Fermi occupation ratio $\nu(x,\theta,t)=\rho(x,\theta,t)/\rho_{\rm s}(x,\theta,t)$,
\begin{equation}
    \label{eq:ghd_convective}
    \partial_t \nu +  v^{\rm eff} (\theta) \partial_x \nu - (\partial_x V) \partial_\theta \nu  \, = \,  0 .
\end{equation}
[It is easy to see that this form of the GHD equation is equivalent to (\ref{eq:ghd}). Plugging the constitutive relation (\ref{eq:constitutive}) into  (\ref{eq:ghd}), one finds that the first equation (\ref{eq:ghd}) satisfied by $\rho(x,\theta,t)$ is also satisfied by $\rho_{\rm s}(x,\theta,t)$. This directly leads to (\ref{eq:ghd_convective}) for the ratio $\rho/\rho_{\rm s}$.] 

In the ground state, the Fermi occupation ratio is either zero or one: $\nu(x,\theta) = 1 $ if $\theta \in [-\theta_{\rm F}(x),\theta_{\rm F}(x) ]$, and $\nu(x,\theta) = 0$ otherwise. Here $\theta_{\rm F}(x)$ is a position-dependent Fermi momentum, which depends on the atom density $n(x)$, see Subsection~\ref{subsec:thermodynamic_limit}. This specific form of $\nu(x,\theta,t)$ is preserved under (\ref{eq:ghd_convective}): at any time, $\nu(x,\theta,t)$ is either zero or one. Consequently, the state of the system at time $t$ is parameterized by a contour $\Gamma_t$ in phase space (Fig.~\ref{fig:dubessy}, left, and Fig.~\ref{fig:qghd}, top row), which separates the region where $\nu = 1$ from the one where $\nu = 0$, namely
\begin{equation}
    \nu(x,\theta,t) \, = \, \left\{ \begin{array}{rcl}
        1 &{\rm if}& (x,\theta)  \; {\rm is \, inside } \; \Gamma_t \\
        0 &{\rm if}& (x,\theta)  \; {\rm is \; outside } \; \Gamma_t  \, .
    \end{array} \right. 
\end{equation}
Writing the contour as $\Gamma_t = \{(x_t(s), \theta_t (s)) ; \; s \in [0,2\pi) \}$, and plugging this into Eq.~(\ref{eq:ghd_convective}) one finds that its evolution equation reads
\begin{equation}
    \label{eq:zero_entropy_ghd}
    \frac{d}{dt} \left( \begin{array}{c}
            x_t (s)  \\
        \theta_t (s)  
    \end{array}
    \right) \, = \, \left( \begin{array}{c}
            v^{\rm eff} ( \theta_t (s) )  \\
        - \partial_x V(x_t(s)) 
    \end{array}
    \right) .
\end{equation}
This `zero-entropy GHD' \citep{doyon2017large} is very useful for numerical purposes, because it provides a very efficient way of solving the GHD equations: it is easier to compute the evolution of the contour $\Gamma_t$, rather than to simulate  the evolution of the full distribution $\rho(x,\theta, t)$, even though both formulations are equivalent in the end. In Fig.~\ref{fig:qghd} (top row), one can see the evolution of the contour $\Gamma_t$ when the gas is suddenly released from its ground-state in a double-well potential to a harmonic potential. After a fraction of the period of the harmonic trap frequency, one observes the appearance of local split Fermi seas. Hence, such a setup could not be described by the conventional hydrodynamic approaches of Subsection~\ref{subsec:conventional_hydro}, because the appearance of such multiple Fermi seas would translate into shock formations in those approaches. GHD, on the other hand, does not have shocks \citep{el2005kinetic,bulchandani2017classical,doyon2017large} and remains valid after the appearance of multiple Fermi seas.

\subsection{Extensions of Euler-scale Generalized Hydrodynamics}

So far we have reviewed results obtained on the 1D Bose gas with the Euler-scale GHD equation (\ref{eq:ghd}). Since 2016, the original framework has been extended in several directions that could be relevant to the description of existing or future experiments. We now turn to these developments.

\subsubsection{Adiabatically varying interactions}

The original formulation of GHD of \citep{castro2016emergent,bertini2016transport} was for a time-independent, translation invariant, Hamiltonian acting on a spatially inhomogeneous state. In particular, no external potential term $-(\partial_x V) (\partial_\theta \rho)$ was  included originally. \cite{doyon2017note} considered the addition of generalized potentials to the Hamiltonian $H \rightarrow H+ \int V_f(x) q[f](x) dx$, where $q[f](x)$ is the charge density of the charge operator (\ref{eq:chargeQf}). This includes, in particular, the case of a standard external potential $H \rightarrow H+ \int V(x) \Psi^\dagger(x) \Psi(x) dx$, corresponding to the simplest choice $q[f]=1$, which results in the form (\ref{eq:ghd}) of the GHD equation with the acceleration term $-\partial_x V \partial_\theta \rho$. This is the form of the GHD equation that is most relevant for the description of existing experiments. However we stress that the results of \citep{doyon2017note} are in principle more general.

A further extension of the original GHD equation was obtained by \cite{bastianello2019generalized}, who considered the case of a non-uniform repulsion strength $g(x,t)$ through the gas, see Fig.~\ref{fig:spacetime_interactions}. Under the assumption of slow variation of $g(x,t)$ (or of $c(x,t) = \frac{m g(x,t)}{\hbar^2}$) in position and time, they found the following additional term:
\begin{figure}
    \centering
    \includegraphics[width=0.45\textwidth]{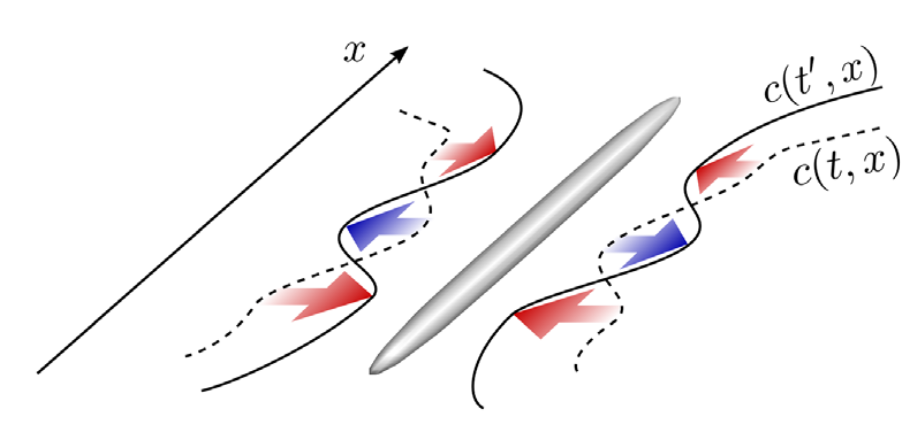}
    \caption{[From \citep{bastianello2019generalized}] The GHD equations (\ref{eq:ghd}) can be extended to include an repulsion strength $c(x,t)$ that varies slowly in time or in space, which would correspond to slowly varying the transverse trapping frequency in an experimental setup.}
    \label{fig:spacetime_interactions}
\end{figure}
\begin{equation}
    \label{eq:inhomogeneous_interaction}
    \partial_t \rho + \partial_x \left( v^{\rm eff} \rho \right) - (\partial_x V) \partial_\theta \rho  + \partial_\theta \left(  \frac{(\partial_t c) F^{\rm dr} + (\partial_x c) \Lambda^{\rm dr} }{1^{\rm dr}}  \rho \right)  \, = \, 0,
\end{equation}
where we have dropped the dependence on $x$, $\theta$ and $t$ for better readability, and where the functions $F$ and $\Lambda$ are
$$
  F(\theta) := \int  \frac{\partial  \phi(\theta-\theta')}{\partial c} \rho(\theta') d\theta' , \qquad   \Lambda(\theta) := \int  \frac{\partial  \phi(\theta-\theta')}{\partial c}  v^{\rm eff}(\theta') \rho(\theta') d \theta' .
$$
For a derivation of this result, see the original paper \citep{bastianello2019generalized}.

\subsubsection{Boltzmann kinetic equation for Bose gas at the 1D-3D crossover}

Taking inspiration from the developments of GHD,  \cite{moller_extension_2021} studied a 3D Bose gas at the crossover to the 1D regime, and introduced a phenomenological description of the gas dynamics at that crossover. The idea is the following. The atoms lie in the 3D potential $V(x) + V_{\perp} (r_\perp)$, where $r_\perp = \sqrt{y^2+z^2}$ is the distance from the axis at $y=z=0$. The transverse potential is harmonic, $V_{\perp} (y,z) = m \omega_\perp^2 r_\perp^2/2$, so that the wavefunction of each atom $\psi(x,r_\perp)$ can be expanded on the eigenstates $\psi_n(r_\perp)$ of the transverse harmonic oscillator, see Eq.~ (\ref{eq:expansion_harmonic_transv}) below. Strictly in the 1D regime (i.e. when $\hbar \omega_\perp$ is much larger than all other energy scales), the transverse ground state ($n=0$) is  the only state that is occupied. But if the transverse confinement is not strong enough, transverse excited states will also be occupied. \cite{moller_extension_2021} consider the first three transverse states ($n=0,1,2$) (their degeneracy is neglected), and regard the resulting system as a three-component 1D Bose gas, with a Hamiltonian of the form 
\begin{eqnarray}
    \label{eq:ham_multicomponent}
\nonumber    H &=& H_0 \, + \, {\rm excitation \; terms}  , \\
    H_0 & = & \sum_{a=0,1,2} \int dx\,  \Psi_a^\dagger \left( - \frac{\partial_x^2}{2m} + V(x) + \frac{g_a}{2} \Psi^\dagger_a \Psi_a \right)  \Psi_a \, + \, \sum_{0 \leq a<b \leq 2} \int dx \, g_{ab} \Psi^\dagger_a \Psi_a \Psi^\dagger_b \Psi_b .
\end{eqnarray}
Here the excitation terms are of the form $\int dx \left( (\Psi^\dagger_a (x))^2 \Psi_b^2(x) + {\rm h.c.} \right)$ and correspond, for instance, to a two-body collision where two atoms in the transverse ground state get excited to the first excited state. The coupling constants $g_a$ and $g_{ab}$ ($a,b = 0,1,2$) are effective 1D coupling constants resulting from the 3D interaction. For a 3D scattering length much smaller than $\hbar/\sqrt{m\omega_\perp}$, 
they can be calculated from the shape of the transverse orbitals, see the discussion in Subsection (\ref{subsec:olshanii}).
\cite{moller_extension_2021} simply assume that they are equal: $g_a = g_{ab} = g$. Under that assumption, the multi-component Bose gas with $V(x)=0$ is integrable~\citep{yang1967some,gaudin2014bethe,caux2009polarization,klauser2011equilibrium,klumper2011efficient,pactu2015thermodynamics,robinson2016exact,robinson2017excitations}, however the description of  its thermodynamic limit is considerably more complicated than the one of the Lieb-Liniger model reviewed in Section~\ref{sec:LiebLiniger}. In particular, while the macrostates of the Lieb-Liniger model in the thermodynamic limit are characterized by their distribution of rapidities $\rho(\theta)$, the ones of the multi-component Bose gas are characterized not only by $\rho(\theta)$, but by infinitely many rapidity distributions for pseudo-spin bound states. \cite{moller_extension_2021} assume that the population of such pseudo-spin bound states is small and can be neglected, and keep only the distribution of rapidities $\rho(\theta)$, whose evolution under the Hamiltonian (\ref{eq:ham_multicomponent}) then coincides with the one of the Lieb-Liniger gas, and is therefore nothing but the GHD equation (\ref{eq:ghd}).

\begin{figure}[tb]
    \centering
    \begin{tikzpicture}
    \draw (0,0) node{\includegraphics[width=0.7\textwidth]{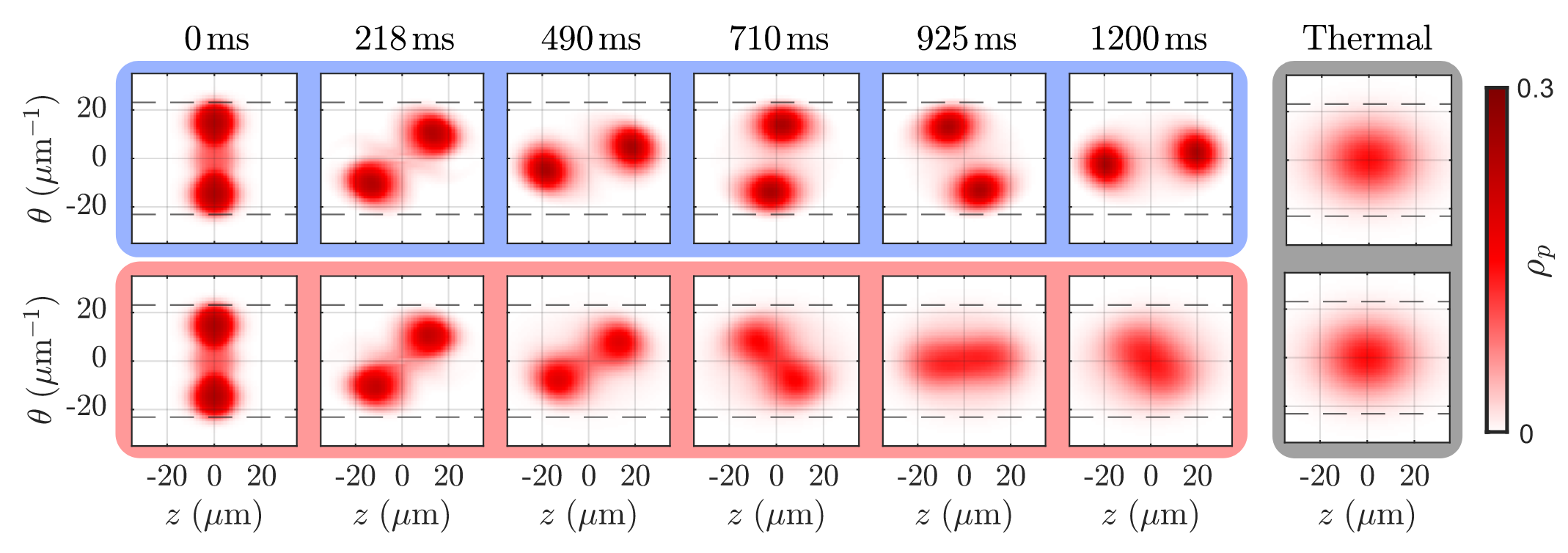}};
    \draw (9,-0.1) node{\includegraphics[width=0.3\textwidth]{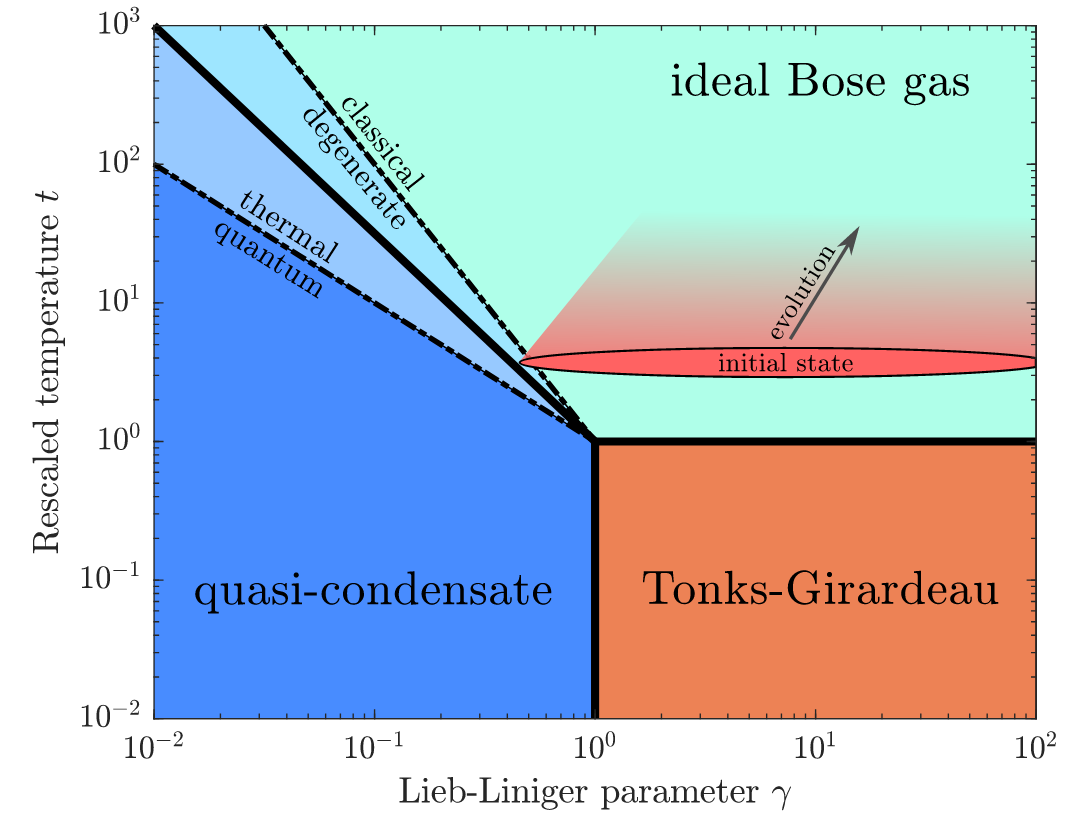}};
    \end{tikzpicture}
    \caption{ [From \citep{moller_extension_2021}] Left: evolution of the rapidity distribution $\rho(\theta)$ under GHD in the quantum Newton Cradle setup (top), compared to the Eq.~(\ref{eq:Boltzmann-Moller}) that includes a phenomenological Boltzmann collision integral modeling excitations to transverse modes when the gas is at the 1D-3D crossover. Right: the cloud is initially in the ideal Bose gas regime, and it stays in that regime under time evolution with Eq.~(\ref{eq:Boltzmann-Moller}).}
    \label{fig:newton_cradle_Dcrossover}
\end{figure}

Finally, the effects of the excitation terms in the Hamiltonian (\ref{eq:ham_multicomponent}) are introduced in the form of a Boltzmann collision integral:
\begin{equation}
    \label{eq:Boltzmann-Moller}
    \partial_t \rho + \partial_x [v^{\rm eff} \rho] - (\partial_x V) \partial_\theta \rho \, = \, \mathcal{I} .
\end{equation}
The collision integral $\mathcal{I}$ is constructed phenomenologically, by considering a simple model for two-body collisions. This phenomenological approach does not incorporate interactions with the other particles, contrary to what usually happens in exact Bethe Ansatz calculations (where the interaction effects appear through the dressing of the various quantities that enter all the formulas). Within the framework of that very simple model, the probability for two colliding atoms with momenta $p_1 = \hbar k_1$, $p_2 =\hbar k_2$ initially in the same transverse state to get excited to a different transverse state is estimated as $P_{\updownarrow} (k,q) \simeq 4 c^2 k q /[k^2 q^2 + c^2 (k+q)^2 ]$ where $k = |k_1-k_2|$ and $q= \sqrt{ |k_1-k_2|^2  - 8 m \omega_\perp/\hbar }$. Then the authors assume that the atom momenta may be replaced by the rapidities, and arrive at an expression of the form
$\mathcal{I} \, \propto \, \sum_{n=1,2} [ \mathcal{I}^-_h \nu_n^{\beta_n} - \mathcal{I}^-_p - \mathcal{I}^+_p \nu_n^{\beta_n} + \mathcal{I}_h^+ ]$, 
where $\nu_n$ is the probability that an atom is in the $n^{\rm th}$ transverse excited state (assumed to be $\ll 1$ for $n=1,2$), $\beta_n$ is the number of atoms changing state in a collision ($\beta_2 = 1$ and $\beta_1 = 2$ in the model of \citep{moller_extension_2021}), and
\begin{equation}
\label{eq:collisionIntegralMoeller}
    \mathcal{I}^+_p \, = \, \frac{(2\pi)^2 \hbar}{m} \int d\theta' |\theta-\theta'| P_{\updownarrow} (|\theta-\theta'|, |\theta_\pm - \theta'_\pm |) \rho (\theta) \rho (\theta') (\rho_{\rm s}(\theta_+)- \rho (\theta_+)) ( \rho_{\rm s}(\theta_+) - \rho(\theta'_+))
\end{equation}
and similar expressions for $\mathcal{I}^-_{\rm p}$, $\mathcal{I}^+_{\rm h}$ and $\mathcal{I}^-_{\rm h}$, where the index `p' and `h' refers to `particle' and `hole', see \citep{moller_extension_2021}. Here $\theta_\pm = \frac{1}{2}(\theta + \theta') + \frac{1}{2}(\theta - \theta') \sqrt{1 \pm 8 m \omega_\perp/(\hbar (\theta-\theta')^2)}$ are the momenta of the two atoms after getting excited to the transverse state in that model of two-body collisions.

The various approximations involved in the construction of the effective Boltzmann equation (\ref{eq:Boltzmann-Moller})-(\ref{eq:collisionIntegralMoeller}) are particularly meaningful in the ideal Bose gas regime. There, this phenomenological approach is expected to work well. Notice that in the ideal Bose gas regime, the GHD dynamics reduces to that of free Bosons, with the effective velocity appearing in Eq.~(\ref{eq:Boltzmann-Moller}) reducing to the bare velocity $v^{\rm eff} (\theta) = \theta/m$. In that regime the aforementioned assumption that all intra- and inter-component coupling strengths are equal is actually no longer necessary for the gas to be integrable. Beyond the ideal Bose gas regime, the accuracy of the description of the 1D-3D crossover by Eqs. (\ref{eq:Boltzmann-Moller})-(\ref{eq:collisionIntegralMoeller}) remains to be investigated.

Motivated by the recent experiment in the Newton Cradle setup of~\cite{li_relaxation_2020}, which observed thermalization attributed to excitations of the transverse modes, \cite{moller_extension_2021} simulated the Newton Cradle with Eqs.~(\ref{eq:Boltzmann-Moller})-(\ref{eq:collisionIntegralMoeller}) in the ideal Bose gas regime, and found that excitations to the transverse modes do indeed induce thermalization, see Fig.~\ref{fig:newton_cradle_Dcrossover}.

\subsubsection{Beyond the Euler scale: Navier-Stokes diffusive corrections}
\label{sec:diffusive_corrections}

As mentioned in the introduction, Euler-scale hydrodynamic equations are but the zeroth-order in a gradient expansion. More precisely, under the assumption of local relaxation, the expectation value of the currents $\left<j(x)\right>$ are functions of the charges $\left< q(x) \right>$ and of their derivatives. Schematically: $\left<j(x)\right> = F(\left< q(x) \right>, \partial_x \left< q(x) \right>, \partial_x^2 \left< q(x) \right>, \dots)$, which is then expanded as $\left<j(x)\right> = F(\left< q(x) \right>, 0, 0, \dots) + \frac{\partial F}{\partial (\partial_x \left< q \right>)}(\left< q(x) \right>, 0, 0, \dots) \, \partial_x \left< q(x) \right>  +  \dots$ The zeroth order gives the Euler-scale hydrodynamic equations, while the next order gives the hydrodynamic equation at the diffusive scale. These hydrodynamic equations include a diffusive, entropy-producing (or Navier-Stokes) term.

\begin{figure}[ht]
    \centering
    \includegraphics[width=0.7\textwidth]{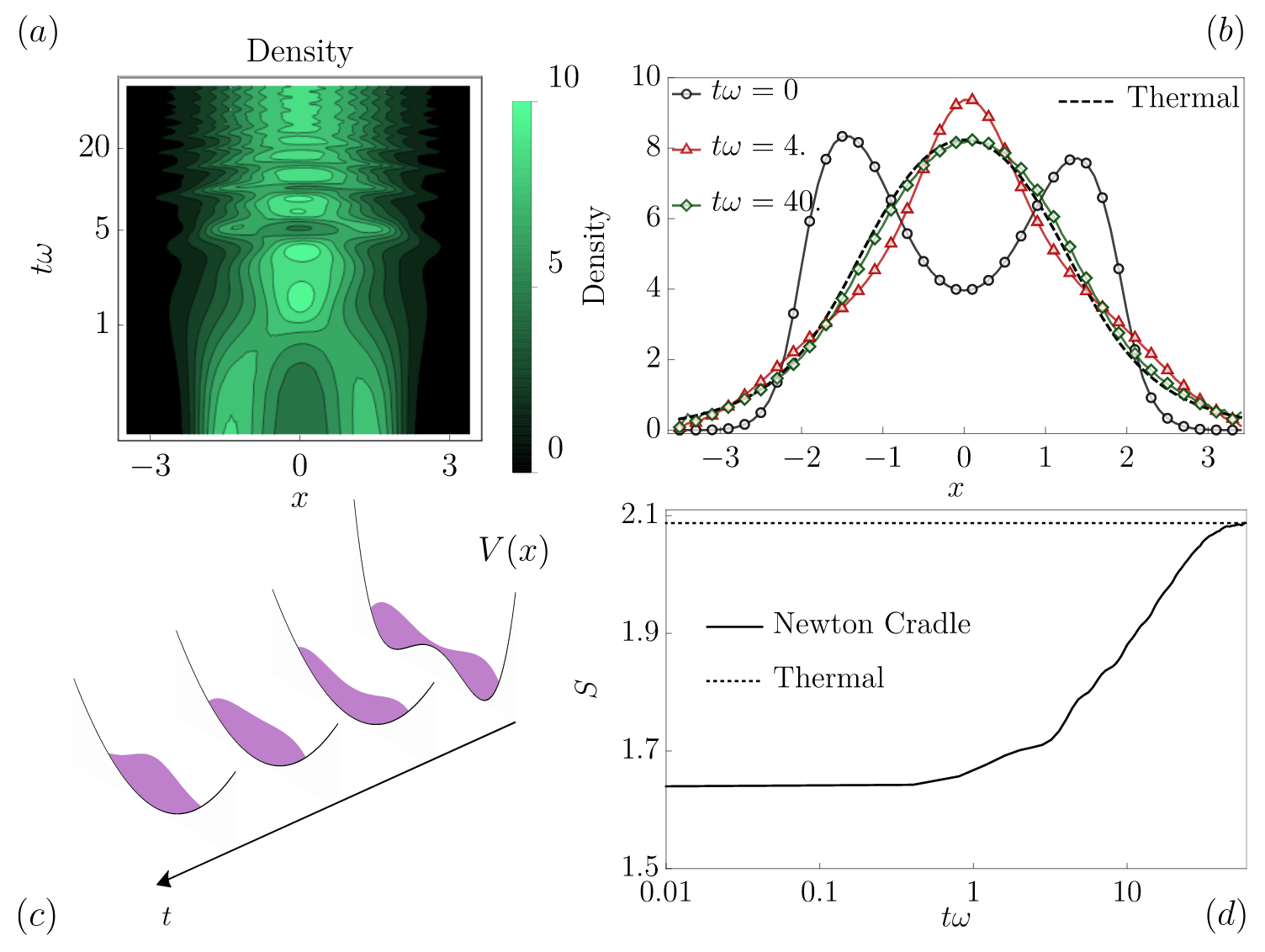}
    \caption{[From \citep{bastianello_thermalization_2020}] When the Navier-Stokes term (\ref{eq:NavierStokes}) is included in the GHD equation, it is found that the 1D Bose gas in the quantum Newton Cradle does reach thermal equilibrium. Bottom left: cartoon of the quench protocol (quench from double-well to harmonic potential). Top: evolution of the density profile $n(x,t)$ under GHD-Navier-Stokes evolution. In the top right plot, one sees that the density goes to the thermal equilibrium density at long times. Bottom right: the entropy invreases and ultimately reaches its thermal equilibrium value.}
    \label{fig:thermalization_diffusion}
\end{figure}

The `Navier-Stokes' GHD equation, with the diffusive term, was obtained in~\citep{de2018hydrodynamic}. The result reads
\begin{equation}
    \label{eq:NavierStokes}
    \partial_t \rho + \partial_x \left( v^{\rm eff} \rho \right) - (\partial_x V) \partial_\theta \rho  = \frac{1}{2} \partial_x \left( \mathfrak{D} \partial_x \rho \right) ,
\end{equation}
where $\mathfrak{D}$ is a kernel (i.e. $\mathfrak{D} f (\theta) = \int \mathfrak{D}(\theta,\theta') f (\theta') d\theta')$) describing a Markov process of random momentum exchanges via two-body collisions \citep{gopalakrishnan2018hydrodynamics,de2019diffusion}. It is defined by the relation
\begin{equation}
    [\mathfrak{D}(\theta,.) ]^{\rm dr} (\theta') \rho_{\rm s}(\theta') \, = \, [ \rho_s (.) \tilde{\mathfrak{D}} (., \theta') ]^{\rm dr} (\theta) ,
\end{equation}
with
\begin{eqnarray*}
    \rho_{\rm s} (\theta)^2 \tilde{\mathfrak{D}}(\theta, \theta') & = & \left(\int W(\alpha, \theta) d\alpha \right) \delta(\theta-\theta') - W(\theta, \theta') ,\\
    W(\theta,\alpha) &=& \rho(\theta) (1-\nu(\theta)) (\Delta^{\rm dr} (\theta - \alpha))^2 \left| v^{\rm eff}(\theta) - v^{\rm eff}(\alpha) \right| .
\end{eqnarray*}
For a detailed derivation of these equations, see \citep{de2019diffusion}.

As mentioned above, interestingly, taking into account the diffusive correction changes the conclusion about thermalization in the quantum Newton Cradle setup, see \citep{bastianello_thermalization_2020} and Fig.~\ref{fig:thermalization_diffusion}. With the inclusion of the Navier-Stokes term, the quantities (\ref{eq:Sf_conserved}) are no longer conserved, and it is found that the gas ultimately reaches thermal equilibrium, contrary to what had been observed previously in \citep{cao2018incomplete,caux2019hydrodynamics}.

For more studies of diffusion in the Lieb-Liniger gas, see e.g. \citep{panfil2019linearized,medenjak2020diffusion}. Diffusion has also been studied very extensively in spin chains; for this topic we refer to the review articles by Bulchandani, Ilievski and Gopalakrishnan and by Doyon, de Nardis, Medenjak, Panfil in this Volume.

\subsubsection{Quantum fluctuating hydrodynamics}

As emphasized above, Euler-scale hydrodynamic equations are based on separation of scales (Fig.~\ref{fig:sos}), which allows for a description of the gas as a continuum of independent fluid cells, each of which has relaxed to a stationary state.

In that hydrodynamic picture, a small perturbation of the system at spacetime position $(x,t)$ generates sound waves that propagate through the gas, so that the perturbation may be observable at a different position $(x',t')$ \citep{kadanoff1963hydrodynamic}. Such dynamical correlations have been studied in the framework of Generalized Hydrodynamics in \citep{doyon2018exact,moller2020euler}; see also the review by Doyon, De Nardis, Medenjak and Panfil in this Volume.

However, at equal time, the fluid cells at different positions $x$ and $x'$ are independent. Thus, all equal-time connected correlations vanish at the Euler scale: for any local observables $\mathcal{O}_j(x_j,t)$,
\begin{equation}
    \label{eq:correl_vanish}
    \left< \mathcal{O}_1(x_1, t) \mathcal{O}_2(x_2, t) \dots \mathcal{O}_n(x_n, t) \right>_{\rm conn.} \, = \, 0 .
\end{equation}
This is somewhat puzzling because, in a quantum system, equal-time correlations are typically non-zero: for instance, in the ground state of the Lieb-Liniger gas they are non-zero, see e.g. the reviews \citep{cazalilla2004bosonizing,cazalilla2011one}. However, this is not in contradiction with (\ref{eq:correl_vanish}): the reason is simply that such non-zero equal-time correlations are an effect occurring beyond the Euler scale.

A good illustration of this phenomenon is provided by the standard fluid described by the Euler equations (\ref{eq:euler}). At zero temperature, the third Euler equation (\ref{eq:euler}) is automatically satisfied as a consequence of the first two (see the discussion in Subsection~\ref{subsec:conventional_hydro}), so we are left with
\begin{equation}
    \label{eq:euler_zeroT}
    \left\{ \begin{array}{rcl}
        \partial_t n + \partial_x (n u) & = & 0 \\
        \partial_t u + u \partial_x u + \frac{1}{m n} \partial_x \mathcal{P} & = & - \frac{1}{m} \partial_x V .
    \end{array}  \right.
\end{equation}
For simplicity, we now consider the ground state of the spatially homogeneous system ($V(x)=0$), with $n (x,t) = n$ and $u(x,t) = 0$. Linearizing the system (\ref{eq:euler_zeroT}) around that solution leads to the equation of propagation of sound waves $(\delta n(x,t), \delta u(x,t))$
\begin{equation}
    \label{eq:sound_wave}
    \left\{ \begin{array}{rcl}
        \partial_t \delta n + n  \partial_x \delta u  & = & 0 \\
        \partial_t \delta u + \frac{1}{m n} \frac{\partial \mathcal{P}}{\partial n} \partial_x \delta n   & = & 0 ,
    \end{array}  \right.
\end{equation}
or equivalently
\begin{eqnarray}
    \label{eq:sound_wave2}
   \left( \frac{\partial}{\partial t}  -  \left( \begin{array}{cc}
        + v & 0 \\  0 & - v
    \end{array} \right) \frac{\partial}{\partial x} \right)  \left( \begin{array}{c}
       \pi \,  \delta n + K  \frac{m}{ \hbar} \, \delta  u  \\
         \pi \,  \delta n - K  \frac{m}{ \hbar} \, \delta  u
    \end{array} \right) \, = \, 0, \\
\nonumber  {\rm with} \qquad   v := \sqrt{\frac{1}{m}\frac{\partial \mathcal{P}}{\partial n}},  \qquad  K := \frac{\pi \hbar n}{m v} .
\end{eqnarray}
Here $v$ is the sound velocity in the gas, and $K$ is a dimensionless parameter (called Luttinger parameter, see below) normalized such that $K \rightarrow 1$ in the hard-core limit. We see from Eq.~(\ref{eq:sound_wave2}) that there are right- and left-moving sound waves, corresponding to specific linear combinations of $\delta n$ and $\delta u$ parameterized by $K$, traveling at velocity $\pm v$.

\begin{figure}
    \centering
    \includegraphics[width=0.98\textwidth]{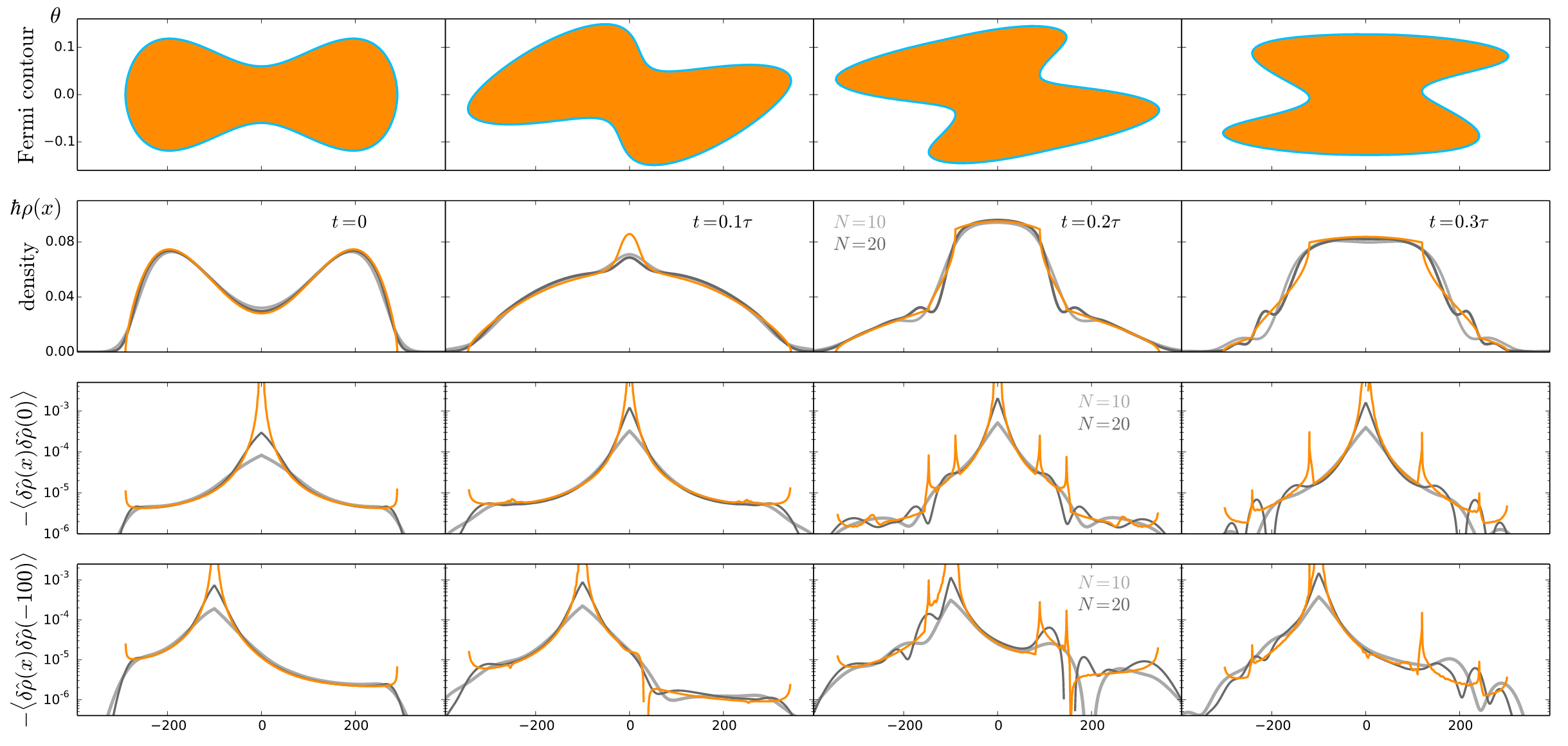}
    \caption{[From \citep{ruggiero2020quantum}] Quench from a double-well to harmonic potential (with period $\tau$) in a 1D Bose gas at zero temperature and $\gamma \simeq 1$. Top row: evolution of the contour $\Gamma_t$ according to the zero-entropy GHD equation (\ref{eq:zero_entropy_ghd}). Second row: the corresponding density profiles $n(x,t)$ (orange), compared to t-DMRG results for $N=10$ and $N=20$ particles. Third and fourth rows: the equal-time density-density correlation function $\left< \delta n(x_1) \delta n(x_2) \right>_{\rm conn.}$ obtained by quantizing the sound waves around zero-entropy GHD.}
    \label{fig:qghd}
\end{figure}

The sound waves can be used as the basic ingredient in a quantized theory of the fluid described by the Euler equations~(\ref{eq:euler_zeroT}). The basic idea is to look at $\delta n(x)$ and $\delta u(y)$ as operators in a quantum theory, and to impose the canonical commutation relations of a one-component fluid \citep{landau1941theory},
\begin{equation}
    \label{eq:landau_canonical}
    \left[ \delta u (x) , \delta n(y) \right] \, = \,  \frac{\hbar}{i m} \delta'(x-y) ,
\end{equation}
and $\left[ \delta n (x) , \delta n(y) \right] = \left[ \delta u (x) , \delta u(y) \right] = 0$. Then, one should find a Hamiltonian $H$ such that the Heisenberg equations $\partial_t \delta n = \frac{i}{\hbar} [H, \delta n]$ and $\partial_t \delta u = \frac{i}{\hbar} [H, \delta u]$ coincide with the equations of motion (\ref{eq:sound_wave}). The simplest choice is:
\begin{equation}
    \label{eq:Luttinger_ham}
    H \, = \, \frac{\hbar v}{2} \int \left[ \frac{K}{\pi} \left( \frac{m}{\hbar} \delta u(x) \right)^2 + \frac{\pi}{K} (\delta n(x))^2 \right] dx ,
\end{equation}
which is nothing but Hamiltonian of a Luttinger liquid, see e.g.~\citep{giamarchi2003quantum,cazalilla2004bosonizing,tsvelik2007quantum} for  introductions. Thus, we see that quantizing the sound waves of a standard Euler fluid at zero temperature (\ref{eq:euler_zeroT}) directly leads to a Luttinger liquid. With that observation, one can get the leading behavior of equal-time correlations beyond the Euler scale. For instance, the equal-time density-density correlation in the ground state of the Hamiltonian~(\ref{eq:Luttinger_ham}) is ~\citep{giamarchi2003quantum,cazalilla2004bosonizing,tsvelik2007quantum}:
\begin{equation}
    \label{eq:rhorho}
    \left< \delta n(x_1) \delta n(x_2) \right>_{\rm conn.} \, = \, \frac{-K}{2 \pi^2 (x_1-x_2)^2} .
\end{equation}
Such correlation functions can then be propagated in time using the sound wave equation (\ref{eq:sound_wave2}). For instance, propagating Eq.~(\ref{eq:rhorho}) in time leads to a combination of two terms, one coming form the right-moving sound wave, the other from the left-moving one:
\begin{equation}
    \left< \delta n(x_1,t_1) \delta n(x_2,t_2) \right>_{\rm conn.} \, = \, \frac{1}{4 \pi^2} \left( \frac{-K}{[(x_1 -v t_1) - (x_2 - v t_2) ]^2} + \frac{-K}{[(x_1 + v t_1) - (x_2+ v t_2) ]^2} \right).
\end{equation}
Other time-dependent correlations can be obtained in a similar way within the framework of Luttinger liquid theory, see e.g.~\citep{giamarchi2003quantum,cazalilla2004bosonizing,tsvelik2007quantum}. These dynamical correlation functions obtained from the propagation of linear sound waves are valid on time scales that are not too long, before non-linear effects kick in. For a review on such non-linear effects, see \citep{imambekov2012one}.

It is natural to ask whether such a program can be implemented, replacing the standard Euler equations at zero temperature~(\ref{eq:euler_zeroT}) by the ones of Generalized Hydrodynamics at zero temperature (\ref{eq:zero_entropy_ghd}). This was achieved, to some extent, by~\cite{ruggiero2020quantum}. In this work, the equation for linear sound waves around zero-entropy GHD are derived, and quantized by imposing canonical commutation relations. This procedure results in a time-dependent, spatially inhomogeneous, multi-component Luttinger liquid.

In the ground state of the gas, the Hamiltonian obtained by~\cite{ruggiero2020quantum} is the one of an inhomogeneous Luttinger liquid, see e.g.~ \citep{cazalilla2004bosonizing,dubail2017conformal,brun2018inhomogeneous,bastianello2020entanglement}. At later times it follows the evolution of the system under the zero-entropy GHD equation (\ref{eq:zero_entropy_ghd}), encoding the propagation of the quantum fluctuations from time $t=0$ to times $t>0$, see Fig.~\ref{fig:qghd}. For more details we refer to~\citep{ruggiero2019conformal,ruggiero2020quantum,collura2020domain}, and to the review by Alba, Bertini, Fagotti, Piroli and Ruggiero in this Volume.

We also mention that there is, in principle, another approach to quantum fluctuations in GHD \citep{fagotti2017higher,fagotti2020locally}, which consists in viewing the GHD equation (\ref{eq:ghd}) as the zeroth order in the evolution equation for the Wigner quasiprobability distribution~\citep{moyal1949quantum,bettelheim2006orthogonality,bettelheim2011universal,bettelheim2012quantum,protopopov2013dynamics,doyon2017large,dean2018wigner,ruggiero2019conformal,dean2019nonequilibrium,fagotti2020locally}. The evolution equation of the Wigner function may be expanded in powers of  $\hbar$~\citep{moyal1949quantum}, with higher order terms that give the corrections to the Euler-scale GHD equation. In this approach, it is not only the quantum fluctuations at time zero that are propagated in time as above. Corrections also appear dynamically because of the modified evolution equation. So far, this approach has been limited to the hard-core limit $g \rightarrow + \infty$, or more generally to spin chains that map to non-interacting fermions. To our knowledge, the extension of this approach to the interacting case is an open problem. We refer to the aforementioned review by Alba, Bertini, Fagotti, Piroli and Ruggiero in this Volume for a thorough discussion of this topic.

\newpage

\section{The 1D Bose gas in cold atom experiments}
\label{sec:experiments_beforeGHD}
Cold atom setups are well adapted to the study of model systems of
many-body physics. First, because of the small energy scales and large
inter-particle distances, the interactions in cold atom systems can be
modeled by simple terms. In particular, in many cases, the two-body interactions are
well represented by a contact interaction characterized
by its scattering length. Second, cold atom gases are extremely well isolated from their environment. Third, the different parameters controlling the physics, such as the interaction strength $g$, or the external potential $V(x)$, are controllable with great flexibility. For  these reasons, cold atom setups are ideal to simulate many-body Hamiltonians, and in particular, they can be used to investigate the physics of 1D Bose gases with contact interactions. In this Section we briefly review the main ideas and results that have led to these experimental achievements.

\subsection{The 1D regime}

\subsubsection{1D geometries}

Confining potentials for cold atoms can be realized by different
means~(see for instance the book by \cite{pethick_boseeinstein_2008}). One can use laser fields: the interaction
between the induced atomic dipole and the laser field results in a force acting on the center-of-mass motion of the atom, which is conservative if the laser frequency is sufficiently far from
any atomic resonance and which is proportional to the laser intensity gradient.
One can also rely on a non-vanishing  magnetic moment of the atoms to 
realize potentials using spatially-varying magnetic fields. For large
enough magnetic fields, adiabatic
following of the orientation of the magnetic dipole of the atoms ensures the realization of conservative potentials.

\begin{figure}
    \centering
    \includegraphics[viewport=47 615 215 750,clip]{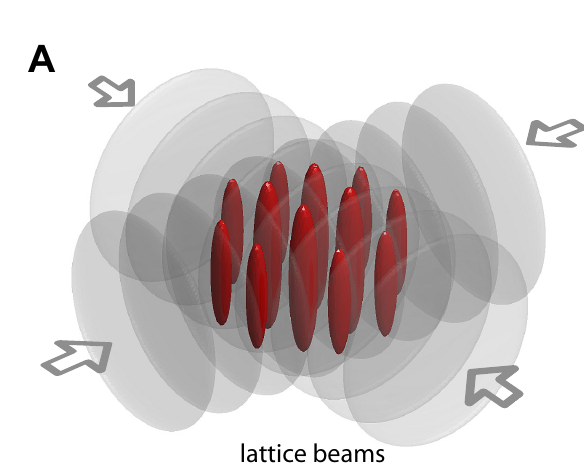}\\
    \includegraphics{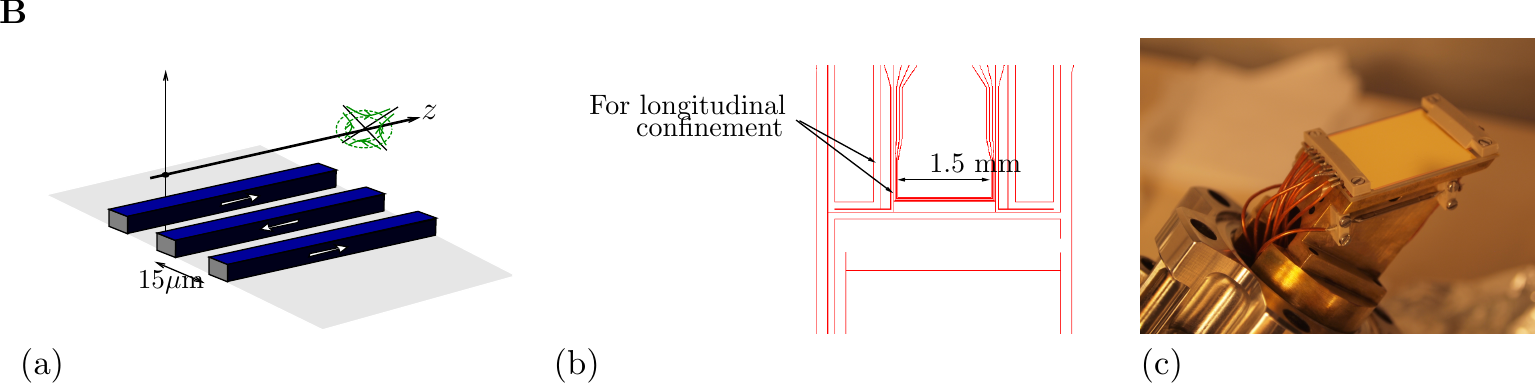}
    \caption{Creating 1D gases in cold atom experiments. {\bf A.} [From~\cite{haller_realization_2009}] : a 2D optical lattice produces an array of 1D tubes. {\bf B.} Guiding atoms along a 3-wire guide on an atom-chip [pictures from  the atom-chip setup in Palaiseau, France]. (a) Three parallel wires, running current in alternate directions, guide the atoms along a line above the central wire. (b) Chip layout (wires edges shown). In addition to the 3-wire guide, other wires are used for the longitudinal confinement and for preparation stages. (c) The atom-chip, covered with a gold mirror. }
    \label{fig:real1Dgasesexp}
\end{figure}

One-dimensional gases are realized in cold atom setups when the atoms are
confined in guides with a transverse confinement large enough so that the
energy gap between the transverse ground state and the first
excited state is much larger than the typical energy per atom. Then the
transverse degrees of freedom get  frozen and the resulting dynamics
in effectively one-dimensional.
The strong transverse confinement required to reach the 1D regime can be realized
using laser fields with large intensity gradients resulting from interference:
a 2D optical lattice
formed by lasers far-detuned from the atomic transitions realizes a 2D array
of 1D tubes for the atoms where the 1D regime can be accessed~\citep{kinoshita_local_2005,haller_realization_2009,fabbri_momentum-resolved_2011}, see Fig.~\ref{fig:real1Dgasesexp}.
Such setups present several advantages.
First, they offer the possibility to realize very strong transverse confinements, since 
the light intensity  is modulated on distances as small as  the laser wavelength. Second, cold atoms are loaded into such a lattice adiabatically from a 3D BEC, which permits to obtain very cold 1D gases. Finally, in this setup the magnetic field is a free parameter that can be tuned to approach Feshbach resonances and thus adjust the interaction strength between atoms~\citep{haller_realization_2009}.
On the other hand, optical lattice setups also have some limitations. The first one is that all measurements are ensemble measurements: the value of the measured observable is averaged over several 1D tubes, with parameters such as the number of atoms or the interaction strength varying from tube to tube, which complicates the data analysis. This averaging also prevents the investigation of fluctuations and correlations. Second, the 2D optical lattice has a finite length in the longitudinal direction -the direction along the tubes-, since it is produced using lasers with relatively small waists. This imposes restrictions on the spatial extension of the 1D gases: the clouds cannot be too long. However, this limitation does not prevent long enough cloud expansions allowing to access the rapidity distribution, see Section \ref{subsec:rapidity_dist_measurement}.

Another type of experiment uses magnetic trapping of atoms to reach the 1D regime.
Strong magnetic gradients are obtained approaching microwires
running an electrical current, in the so-called atom chip
setup first 
developed in Munich~\citep{reichel_atomic_1999} and 
in Vienna~\cite{folman_controlling_2000} (see
\citep{reichel2011atom} for a review). Going 
close enough to the microstructures, the 1D regime
can be reached~\citep{van_amerongen_yang-yang_2008,armijo_mapping_2011,jacqmin_sub-poissonian_2011}. The idea of the magnetic guiding is simple. 
Consider a current-carrying wire immersed in a homogeneous magnetic field perpendicular to it. The 
total magnetic field vanishes on a line parallel to the wire and atoms, when polarized in a low-field seeker magnetic state, will be guided along this line. 
The transverse homogeneous magnetic field can be replaced by the magnetic field produced by two wires placed on each side of the central wire, so that one has the configuration depicted in Fig.~\ref{fig:real1Dgasesexp}. Strong transverse confinement is obtained going close to the wires and, in some experiments, atoms are guided at a distance as small as 15$\mu$m from the wires~\citep{jacqmin_sub-poissonian_2011}. 
The advantage of atom chip setups is that a single 1D gas is realized and studied. This allows direct comparison with theoretical predictions. This also permits the study of fluctuations, and of their correlations. The guiding of atoms can be realized on very long distances, as the transverse confinement is invariant along the whole microwire. However,  atom chips also has some drawbacks. In particular, the 1D gases are prepared and cooled down directly in the 1D geometry, and it turns out that temperatures obtained in these setups are higher than those obtained using 2D optical lattices. The physical phenomena limiting the temperatures that can be reached are not completely elucidated, although the role of atom losses has recently been singled out~\citep{rauer_cooling_2016,schemmer_cooling_2018,bouchoule_asymptotic_2020}.

\subsubsection{Effective 1D interaction parameter}
\label{subsec:olshanii}
In experiments, although the transverse confinement is large enough to freeze the transverse degrees of freedom, it is typically small enough so
that the width of the transverse ground state wave-function is much larger than
the range of the 3D interaction potential. In that case, one can
model the 3D interaction by a contact potential characterized by the
scattering length $a_{{\rm 3D}}$. 
The relation between the 3D scattering properties and the 1D scattering properties
was worked out by~\cite{olshanii_atomic_1998}, and we briefly review the main steps of his argument.
A harmonic transverse confinement of frequency $\omega_\perp$
is assumed, and the scattering state of two atoms is explicitly constructed. Because of the harmonic nature of the transverse confinement, the center-of-mass motion decouples from  that of  
the relative coordinate. The wavefunction of the relative coordinate  obeys the Shr\"odinger equation for a particle of mass $m/2$ in a harmonic 
transverse confinement of frequency $\omega_\perp$ and with
the 3D contact potential at the origine. It can be expanded as a sum of transverse eigenstates, with $x$-dependent 
amplitudes. 
One considers a state whose energy is small enough so that,
apart from the transverse ground state, the amplitudes are exponentially decreasing functions of $x$. Then the 3D wavefunction reads, 
\begin{equation}
    \label{eq:expansion_harmonic_transv}
    \psi(x,r_\perp)=\psi_0(r_\perp)\cos(k|x|-\delta)+\sum_{n>0} a_n e^{-\kappa_n |x|}\psi_n(r_\perp)
\end{equation}
where the sum runs over the transverse excited states of vanishing angular momentum whose wavefunction are $\psi_n(r_\perp)$, where $r_\perp$ is 
the transverse distance to the origine. The parameters $\{a_n\}$ and $\delta$ are determined in the following way.
First, one imposes that $\partial \psi/\partial x$ is regular everywhere in the plane $x=0$ except at the 3D origin. Then, one  imposes
that the wavefunction fulfills the 3D boundary condition: for a zero-range 3D interacting potential of scattering length $a_{3D}$, the 
wavefunction diverges as $1/r-1/a_{3D}$ at short distance. This fixes the value of $\tan(\delta)$, which is found to scale as $1/k$ at small $k$~\citep{olshanii_atomic_1998}. For a pure 1D problem, with a contact potential 
$g_{{\rm 1D}}\delta(x)$, one would have $\tan(\delta)=-m g_{{\rm 1D}}/(2\hbar^2 k)$. Comparing with the small $k$ limit of the 3D problem, one finds that the effective 1D coupling constant is~\citep{olshanii_atomic_1998}
\begin{equation}
g_{{\rm 1D}}=\frac{2\hbar \omega_\perp a_{{\rm 3D}} }{1-{\cal C}a_{{\rm 3D}}/a_\perp} ,
\label{eq:g1D}
\end{equation}
where $a_\perp=\sqrt{2\hbar/(m\omega_\perp)}$ is the width of the transverse ground state and ${\cal C} \simeq 1.46$.

In most experimental realizations, $a_{{\rm 3D}}\ll a_\perp$, so that Olshanii's relation~(\ref{eq:g1D}) reduces to $g_{{\rm 1D}}=2\hbar \omega_\perp a_{{\rm 3D}}$. This expression can be recovered by a simple reasoning, valid
for a weakly interacting gas in the quasicondensate regime.
Since we assume $a_{\rm 3D}\ll a_\perp$,
interactions have a 3D nature and their effect is obtained simply by averaging over the transverse density profile. More precisely, the interaction energy per unit length in the longitudinal direction is
$e_{\rm int}=(1/2)\int d^{2} r_\perp \,
g_{{\rm 3D}} \, n_{\rm 1D}^2 |\psi_0(r_\perp)|^2$, where 
$g_{\rm 3D}=4\pi\hbar^2a_{\rm 3D}/m$
is the 3D interaction strength. 
On the other hand, considering the 1D problem one obtains 
$e_{\rm int}=(g_{\rm 1D}/2) n_{\rm 1D}^2$. 
Equating both expressions, and using the Gaussian shape of $\psi_0$, one recovers $g_{{\rm 1D}}=2\hbar\omega_\perp a_{\rm 3D}$.

Equation~\eqref{eq:g1D} shows a resonance behavior when ${\cal C} a_{{\rm 3D}}/a_\perp$ reaches one,
which can be interpreted as a Feshbach resonance involving a bound state
related to a transversely excited state~\citep{bergeman_atom-atom_2003}. 
Experimentally, approaching this resonance would require very large transverse confinements for standard 3D scattering lengths.
However, close to a 3D-scattering resonance, one can reach the regime where 
$a_{{\rm 3D}}$ becomes close or larger than $a_\perp$. This was used by~\cite{haller_realization_2009}
to reach the 1D hard-core regime, and even realize a metastable state
with attractive 1D interactions (i.e. $g_{{\rm 1D}}<0$).

\subsubsection{Validity of the assumption of separation of scales}
\label{subsubsec:validityLDA}
Generalized Hydrodynamics is a theory valid at large scales: it assumes that longitudinal
variations occur on length scales much larger than the microscopic
scale such that the gas can be described locally as a homogeneous Lieb-Liniger gas, see Fig.~\ref{fig:sos}.
In the static case, for equilibrium states, this assumption is the so-called local density approximation (LDA). 
In order to test numerically the LDA, one needs to compare 
results obtained within the LDA to exact results. This is possible for Bose gases at thermal equilibrium, and for moderate atom numbers,  using exact Monte-Carlo calculations, as done in~\citep{jacqmin_momentum_2012} and \citep{yao_tans_2018}. 
The LDA is found to be a very good approximation
for typical experimental parameters.  These numerical tests are confirmed by numerous experimental results which validate the LDA approach (see Section \ref{subsec:benchmarkingexp}).
For out-of-equilibrium situations, 
exact numerical calculations capable of 
capturing the long term dynamics are 
impossible as soon as the atom number is larger than about a dozen. Thus, the cold atoms experiments in this situation realize a quantum simulator that tests the validity of the large-scale approximation assumed by GHD.

The large scale approximation is expected to be valid  
for experiments based on atom chip setups: there, the inter-particle distance, as well 
as the length associated to the mean-field interaction energy, are both typically 
much smaller than the typical length scale 
of variation of the mean density.
In contrast, in some experiments performed with 2D arrays of 
1D tubes confined in an optical lattice, the 
atom number per tube is typically not very large number (it can be of order $\sim 10$), so the validity of the LDA is challenged. Nevertheless, 
experimental data are generally found to be in good agreement with LDA (see Section~\ref{subsec:testGHDWeiss}).

Very often, the atoms are confined in a  longitudinal potential that is harmonic, and thermal equilibrium in this situation 
has been discussed in several works. Here we 
present the final understanding of the situation.
In~\citep{ketterle_bose-einstein_1996}, the thermal equilibrium state of an ideal Bose gas is discussed. Although the 1D Bose gas does not undergo a Bose-Einstein condensation phenomenon, in a harmonic potential a sharp Bose-Einstein
condensation (BEC) due to
finite size effects was predicted. This phenomenon occurs
when the total atom number fulfills
$N \simeq k_{\rm B} T /(\hbar \omega_\parallel \ln(2k_BT/\hbar\omega_\parallel))$, where $\omega_\parallel$ is the frequency of the longitudinal confinement and $N$ is the total atom number, and it corresponds to the breakdown of the LDA.

By comparing the mean-field energy to the level spacing of the single-atom eigenstates in the trap, \cite{petrov_regimes_2000} noticed that
one expects this BEC phenomenon to 
be affected by interactions between atoms, 
in most experimental setups. 
The quantitative effect of interactions 
was investigated in~\citep{bouchoule_interaction-induced_2007} and the correct picture was established.
There, it was shown that the finite-size BEC phenomenon occurs only for extremely small interactions between the atoms, and the maximum interaction strength 
that allows the observation of the finite-size
BEC phenomena was computed. 
In most experimental cases, interactions are large enough so that Bose-Einstein condensation is not relevant. Instead, the LDA remains an excellent approximation, and 
crossovers between the different regimes of 1D Bose gases discussed in section~\ref{sec:regimes} are present. In particular, the crossover between the ideal Bose gas regime and the quasicondensate regime occurs, at the center of the trap, when the peak atomic density approaches the crossover density $n_{\rm cross.}= (m (k_{\rm B} T)^2 /(\hbar^2 g))^{1/3}$ (see Eq.~\eqref{eq:condition_IBG_equilibrium}).  In terms of the total atom number, it corresponds to $ N \simeq k_{\rm B} T /(\hbar \omega_\parallel \ln ( t^{1/3}))$, where 
$t=2\hbar^2k_{\rm B} T/(mg^2)^{1/3}$.
We recall that $n_{\rm cross.}$ is much 
larger than the degeneracy density $\sqrt{mk_{\rm B} T}/\hbar$ (see Subsection \ref{sec:regimes}), so that the gas could be highly degenerate but still in the ideal Bose gas regime. This is confirmed experimentally by the study of several observables. For instance, atom-number fluctuations that are well above the shot noise level~\citep{armijo_mapping_2011} and Lorentzian-like momentum distributions~\citep{jacqmin_momentum_2012}, both features being characteristic of degenerate gases, are observed while the gas lies
in the ideal Bose gas regime.

\subsection{Benchmarking experiments: results at equilibrium}
\label{subsec:benchmarkingexp}
The realization of the 1D Bose gas with contact repulsive interactions in
experimental cold atom setups has been established
by comparisons of experimental data with exact predictions from the Lieb-Liniger model. The theory predictions have used either 
the machinery of integrable systems reviewed in Section~\ref{sec:LiebLiniger}, or numerically exact Monte-Carlo calculations, valid for gases at thermal equilibrium. Here we present a few results which, in our view,
constitute landmarks in this field. All the results reviewed here are about gases of many atoms ($N \sim 10- 10^4$) and are in good agreement with the LDA analysis.

\subsubsection{Measurement of zero-distance correlation function}
\label{subsec:g20}
The zero-distance two-body correlation function can be obtained exactly using the Hellmann-Feynman theorem, see Eq.~\eqref{eq:HelmanFeynam}.
For the ground state it gives
\begin{equation}
n^2g^{(2)}(0)=2 \left( \partial e_0/\partial g \right)_n    ,
\end{equation}
where $e_0$ is the ground state energy density and the derivative is taken at constant linear density $n$. 
This quantity can be computed  numerically~\citep{lieb_exact_1963}.
As discussed in Subsection~\ref{sec:regimes}, the zero-distance correlation function characterizes the crossover between the quasicondensate regime  (where $g^{(2)}(0) \simeq 1$) and the hard core regime ($g^{(2)}(0) \simeq 0$).

\begin{figure}
    \centering
    \includegraphics[viewport=43 585 298 748,clip]{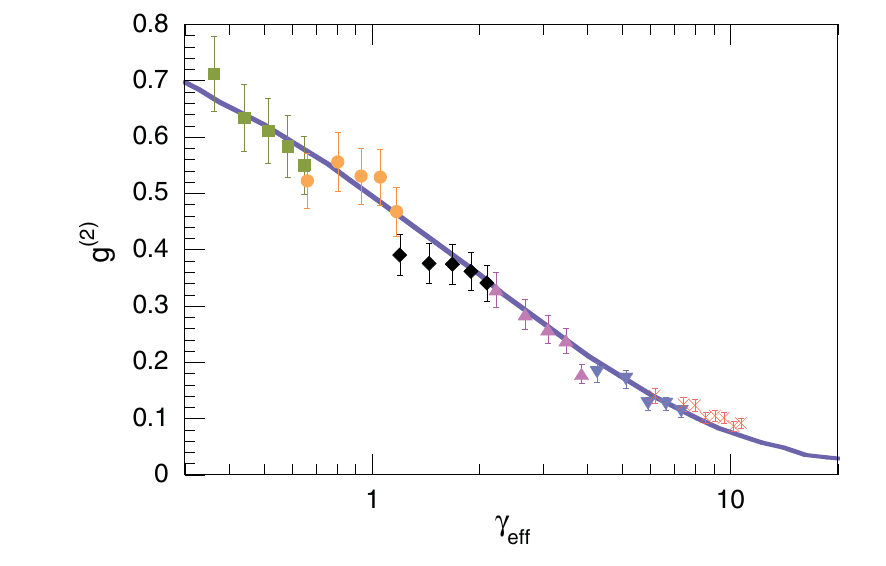}
    \caption{[From \cite{kinoshita_local_2005}] Measurement of the local two-body pair correlation, using photoassociation data and comparison with 
    exact prediction from Lieb-Liniger model for the ground state. $\gamma_{\rm eff}$ is the effective interaction parameter, that takes into account the inhomogeneity of linear densities across the cloud.}
    \label{fig:PairCorrelationWeiss}
\end{figure}

Experimentally, the zero-distance correlation
function can be measured by probing two-body losses
induced by photoassociation~\citep{kinoshita_local_2005}, see Fig.~\ref{fig:PairCorrelationWeiss}. The idea is to shine the cloud with
a laser that is sufficiently far from the atomic resonance to
leave isolated atoms at rest, but whose frequency is chosen to
 induce, for pairs of atoms that are very close to each other,
 a transition towards an excited diatomic molecule. The distance required to perform a photoassociation is much smaller than the typical
 length scale of variation of $g^{(2)}(x) := \left< (\Psi^\dagger(x))^2 (\Psi(0))^2 \right>/n^2$, such that 
 the rate of production of excited molecules is simply
 proportional to $n^2 g^{(2)}(0)$.
 When a photoassociated molecule decays, it typically produces atoms whose kinetic energy is much larger than the trap depth; these atoms thus leave the atomic cloud. In the end, one observes a decrease of the total atom number $N$, and the loss rate $dN/dt$ is proportional to $n^2 g^{(2)}(0)$.

The experiment of~\citep{kinoshita_local_2005} is realized in a 2D optical lattice, so that a 2D array of 1D
 inhomogeneous tubes is populated. More precisely, the atom density varies within each tube because of the harmonic longitudinal confinement $V(x)$, and the atom number $N$ per tube varies among the tubes. The data analysis uses the LDA. The local loss rate is proportional to the local value of $n^2 g^{(2)}(0)$, where $g^{(2)}(0)$ depends on $n$ via its dependence on the interaction parameter $\gamma=mg/(\hbar^2 n)$ :  $g^{(2)}(0)=g^{(2)}_0(\gamma)$.
One assumes that $g^{(2)}_0(\gamma)$
 is linear in $\log(\gamma)$, which is a good approximation for the  explored data range (see Fig.~\ref{fig:PairCorrelationWeiss}). 
 Then the loss rate is expected to be proportional to 
 $\bar{n}  g^{(2)}_0(\gamma_{\rm eff})$, where  $\bar{ n} $ is the mean linear density seen by atoms and
 $\gamma_{\rm eff}$ is such that 
 $\log(\gamma_{\rm eff})$ is the 
spatial average of $\log(\gamma)$, weighted by $n^2$.
 To evaluate $\gamma_{\rm eff}$ and $\bar{n}$,   the distribution of atoms
 among the tubes is assumed to be that corresponding to the initial  shape of the 3D BEC, and, within a 1D tube, the Lieb-Liniger equation of state is used.   
Finally, the value of $g^{(2)}_0(\gamma_{\rm eff})$ deduced from experimental data is shown in
 Fig.~\ref{fig:PairCorrelationWeiss}. It is in remarkable agreement
 with the prediction from the Lieb-Liniger model. The data show the crossover between the quasicondensate regime, where $g^{(2)}(0)\simeq 1$, as in a true BEC, to the hard-core regime, where $g^{(2)}(0) \ll 1$, as for a Fermi gas.

Shortly before that measurement of $g^{(2)}(0)$, the observation of 1D Bose gases in the hard core regime had been achieved  by~\cite{kinoshita_observation_2004}, who measured the total energy of an array of 1D gases by 1D expansion, as well as the length of the trapped atom clouds. The approach to the hard-core regime was also signaled by a measured reduction of the  three-body loss rate~\citep{tolra_observation_2004} (see Section \ref{sec:losses} for a discussion of three-body losses).

\subsubsection{Analysis of density profiles: Yang-Yang thermodynamics}
\label{subsec:analysisnx}
Remarkably, the results about $g^{(2)}(0)$ briefly reviewed in the previous Subsection are in good agreement
with theoretical prediction for the ground state of the 1D Bose. However, in many experimental situations, the gas is not in its ground state.
As reviewed in Subsection~\ref{subsec:yangyang}, exact theory results from \cite{yang1969thermodynamics} are available for thermodynamics quantities at thermal equilibrium.  
In several experimental works, the data were found to be in very good agreement with predictions from Yang-Yang thermodynamics.

\begin{figure}
    \centering
    \includegraphics{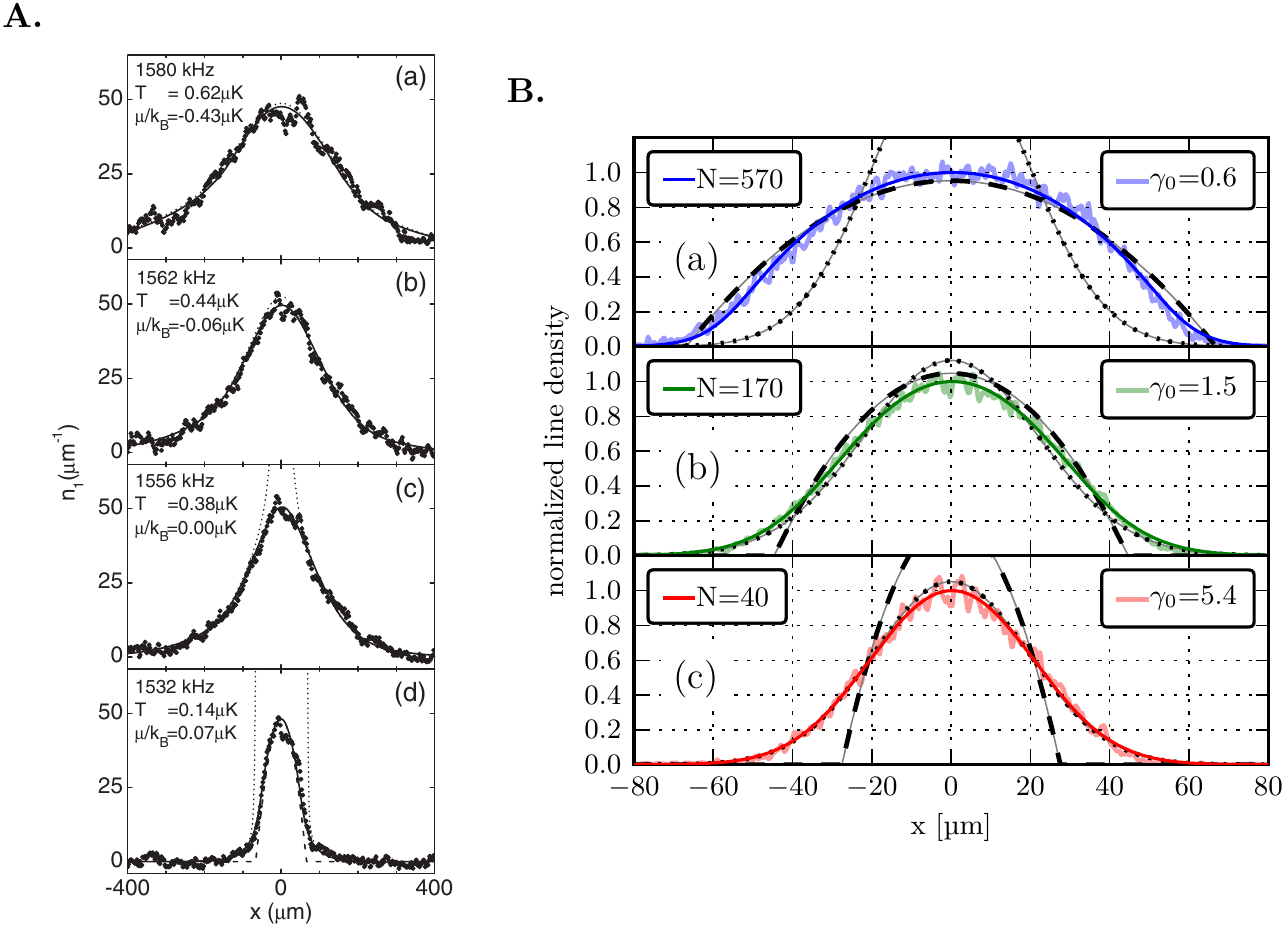}
    \caption{Density profiles of 1D gases, fitted with thermal equilibrium density profiles. The latter are computed, knowing the longitudinal potential, using the 
    equation of state of Yang-Yang~(\cite{yang1969thermodynamics}), and assuming a local density approximation.
    {\bf  A.} [From~\cite{van_amerongen_yang-yang_2008}] Measured density profiles (black circles), compared to Yang-Yang profiles (solid lines). Transversally excited states are taken into account as ideal Bose gases. {\bf B.} [From~\cite{vogler_thermodynamics_2013}]
    Density profiles for different 1D gases, located at different position in an array of 1D tubes.
    They are obtained from the raw data, which include a column integration of the signal, by an Abel's transformation. The Yang-Yang fit is shown in solid smooth line.}
    \label{fig:YangYangThermo}
\end{figure}

The first comparison between experiments and the finite temperature thermodynamics of Yang and Yang was done by~\cite{van_amerongen_yang-yang_2008}, who analyzed the density profiles of trapped 1D gases.
This work used an atom chip setup:
 atoms are confined in magnetic traps realized by microwires
 deposited on a chip. In contrast with optical trapping, a single 1D cloud is observed, which allows a more direct comparison
 with theoretical predictions, since no averaging over tubes is required.
 This permits to investigate the density profile of the 1D gas, confined in a longitudinal potential $V(x)$. As discussed in
 Subsection \ref{subsubsec:validityLDA}, the longitudinal trapping is weak enough so that 
 LDA is valid. For a gas at thermal equilibrium, the LDA means that
 the gas at position $x$ is 
 described by a gas at temperature $T$
 and chemical potential 
$\mu(x)=\mu_0 - V(x)$. Thus the density at position $x$ is 
 \begin{equation}
     n(x)=n_{{\rm YY}}(T,\mu_0-V(x)),
 \end{equation} 
 where $n_{{\rm YY}}(T,\mu)$ is the linear density of the homogeneous Lieb-Liniger gas at temperature $T$ and chemical potential $\mu$,  referred to as the Yang-Yang equation of state,.
 Thus, if $V(x)$ is known, the experimental density profile can be compared to the theory curve, with two adjustable parameters: the temperature $T$, and the chemical potential $\mu_0$. In the experiment of \cite{van_amerongen_yang-yang_2008}, however, the population of transversely excited states was not negligible. This was taken into account in the theoretical model by
 treating each transversely excited states as an ideal 1D Bose gas, at thermal
 equilibrium with the 1D gas in the transverse ground state.
 This treatment results in a modified Yang-Yang equation of state $n(T,\mu)$. Fig.~\ref{fig:YangYangThermo} reproduces the experimental density profiles of \citep{van_amerongen_yang-yang_2008} compared with the theory curves of the modified-Yang-Yang equation of state, with $\mu_0$ and $T$ as fitting parameters. The theory curves reproduce the measured profiles very well.

 Since these pioneering results, the analysis of density profiles of 1D gases using the Yang-Yang equation of state has been realized successfully in other experiments~\citep{armijo_mapping_2011,vogler_thermodynamics_2013}. In~\citep{vogler_thermodynamics_2013}, 
 the experimental setup uses an optical lattice and density profiles are acquired using an electron beam
 propagating perpendiculary to the longitudinal
 direction $x$.
 The resolution attained using electron-beam imaging is 
 sufficiently good to resolve individual 1D tubes.
However, the data analysis is complicated by the fact that column integrated density profiles are acquired, the integration being done over the direction of propagation of the electron beam. Thus raw data mix
 information about different 1D tubes
 that have different density profiles. The density profile of individual 1D tubes can however be extracted deconvoluting for the 
 effect of the column integration, if one assumes invariance by rotation of the tubes distribution.
 The deconvolution technique, that does not 
 require an {\it a priori} knowledge of the 
 tube distribution, uses the Abel transformation.
 The resulting density profiles of individual 1D gases, shown in Fig.~\ref{fig:YangYangThermo}, fit remarkably well with the Yang-Yang theory.

\subsubsection{Density fluctuations}
\label{subsubsec:densityfluctu}
A more stringent test of Yang-Yang thermodynamics 
can be made if one
investigates not
the density profile but its fluctuations. For this purpose
one can investigate the
atom number fluctuations in a pixel whose size $\Delta$ is much smaller
than the length of the cloud, and much larger than the correlation
length of the gas. We note $\delta N=N-\langle N\rangle$ the
atom number fluctuations in a pixel, where $N$ is the atom number in the pixel.

\begin{figure}
    \centering
    \includegraphics{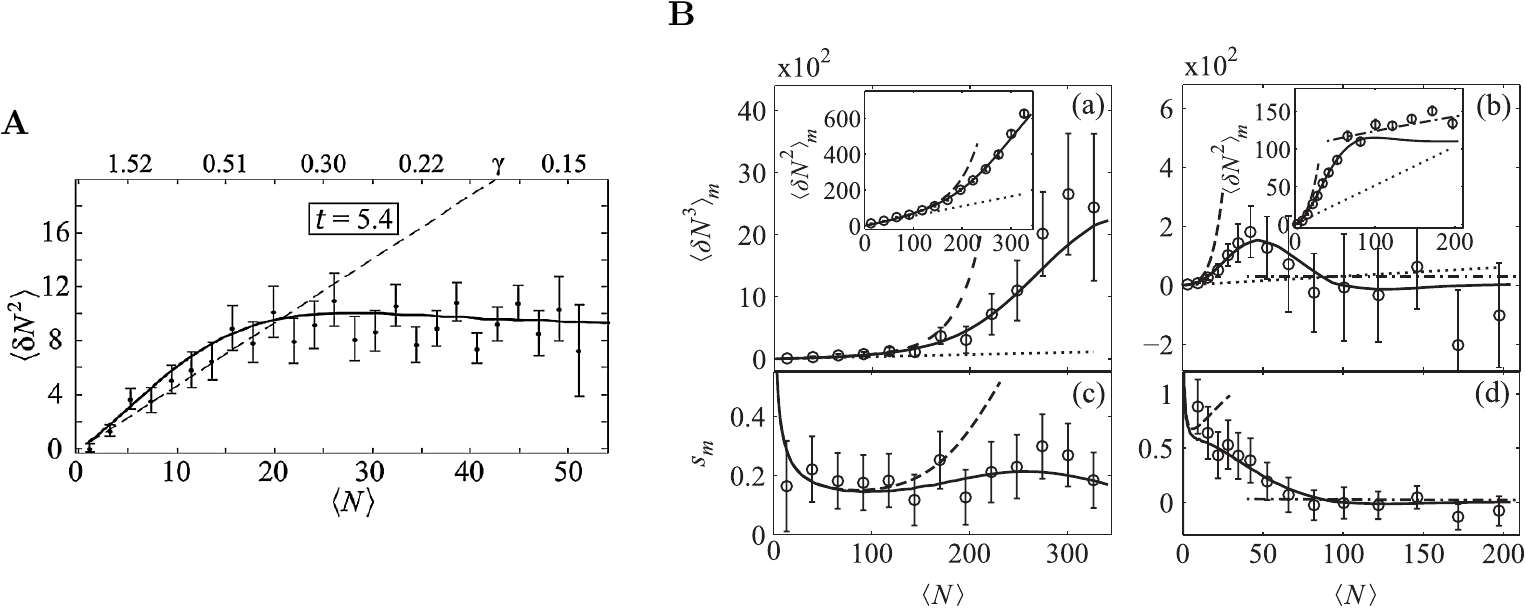}
    \caption{Density fluctuations in 1D Bose gases, compared to Yang-Yang prediction (see Eq.\eqref{eq.fluctudiss}). {\bf A}. [From \cite{jacqmin_sub-poissonian_2011}] Dots : measured density fluctuations. Solid line :  Yang-Yang predictions. The reduced temperature $t$ is $t=2\hbar^2 k_B T /(mg^2)$. The dashed line shows the shot noise limit, whose amplitude is reduced because 
    of the smearing produced by the finite optical resolution. The fact that gas never shows super-Poissonian fluctuations, that would be associated to bosonic bunching, signals the entrance into the strongly interacting regime. $\langle N \rangle = n\Delta$ where $n$ is the linear density and $\Delta$ the pixel size, equal to 4.5$\mu$m.  
    {\bf B}. [From \cite{armijo_probing_2010}] Third moment of the density fluctuations. Solid lines show the Yang-Yang prediction, with transversely excited states taken into account as ideal Bose gases. Dotted line gives the shot-noise level, corrected for the effect of finite resolution. The dashed lines shows the predictions for an ideal Bose gas. The graphs (c) and (d) shows the skewness of the atom number distribution, defined as $S_m= \langle \delta N^3\rangle/\langle \delta N^2\rangle^{2/3}$. 
    In the inset of Fig.(b), the deviation of the measured $\langle \delta N^2\rangle$ to the Yang-Yang prediction (solid line), is due to the swelling of the transverse wave function, which occurs because the condition $\mu \ll \omega_\perp$ is not fulfilled. This 3D effect is taken into account into the quasi-condensate predictions shown as dotted-dashed lines. }
    \label{fig:densityfluctu}
\end{figure}
Then, if one assumes thermal equilibrium,
the gas contained in the pixel can be described by a Gibbs ensemble, the rest of the cloud acting as a reservoir of energy and particles.
Then atom number fluctuations fulfill, for any integer $K$,  
\begin{equation}
  \langle \delta N^K\rangle = (k_{\rm B} T)^{K-1} \Delta \left( \frac{\partial^{K-1} n}{\partial^{K-1} \mu}\right )_{T},
\label{eq.fluctudiss}
\end{equation}
where $n(\mu,T)$ is the equation of state of the gas.
Thus, the measured density fluctuations can be compared to expectation values
derived from the Yang-Yang equation of state. Notice that, in contrast with the experiments presented above, no precise knowledge of the longitudinal potential $V(x)$ is
required. The longitudinal potential is irrelevant in the data
analysis: it is simply useful to sample different densities at a given temperature. 

In order to measure density fluctuations, ensemble
measurements, 
such as those realized in setups using 
2D optical lattices, are prohibited. Thus 
such measurements have been realized only in 
an atom chip setup.
To extract density fluctuations, one records an ensemble of density profiles, all taken with identical 
experimental parameters, from which a statistical analysis is performed.  
Measurements of density fluctuations in 1D gases
were first performed in~\citep{esteve_observations_2006}. 
Comparison with Yang-Yang predictions were performed in~\citep{armijo_mapping_2011,armijo_probing_2010,jacqmin_sub-poissonian_2011}. In \citep{armijo_mapping_2011}, the dimensional crossover from 3D to 1D was investigated, and it was shown that it is well accounted for by the aforementioned modified Yang-Yang equation of state, which includes the effect of transversely excited states treated as ideal Bose gases. 
Fig.~\ref{fig:densityfluctu} displays a selection of results from the two latter references.

\subsubsection{Techniques for measuring the momentum distribution}
\label{subsubsec:analysisnp}
 While all the above results concern observables that involve only the real-space density, information is also contained in other observables, in particular the momentum distribution. 
 The momentum distribution has been measured in several experiments using different techniques which we briefly review now. 

The first technique that was used 
was the so-called Bragg technique. The idea is to shine two laser fields with wavevector difference $q$ and frequency difference $\omega$ onto the atomic cloud. 
The action on the cloud is described by the addition of a new term to the Hamiltonian, which reads 
\begin{equation}
    V_{\rm Bragg} = V_0\sum_k \left ( \Psi_{k+q}^\dagger \Psi_k e^{i\omega t}
    +\Psi_{k+q}\Psi_k^\dagger e^{-i\omega t}
    \right ),
\end{equation}
where each term correspond to the
 absorption of a photon in one laser beam and stimulated emission in the other, a process called a Bragg process. 
 For counterpropagating lasers, 
 $q$ is usually very large compared to the typical momentum width of the cloud, and $\hbar^2q^2/m$ is typically much larger than the energy per atom. In this case, called the Doppler limit, the 
 atom promoted to the momentum state $k+q$ by the Bragg process can be considered as effectively removed from the system, and the
 final energy is about 
 $\hbar^2(k+q)^2/(2m)\simeq E_{\rm rec} + \hbar^2 kq/m$, where $E_{\rm rec}=\hbar^2q^2/m$ is the recoil energy, and the second term is the Doppler shift.
Then the Fermi golden rule leads to a rate of Bragg transfer $\Gamma$ which obeys   
\begin{equation}
    \Gamma \propto \langle \Psi_k^\dagger \Psi_k\rangle ,
\end{equation}
where conservation of energy imposes
$k=m (\hbar \omega-E_{\rm rec})/(\hbar^2 q)$.
Thus, measuring the loss rate of the atomic cloud as a function of the Bragg detuning $\omega$ gives access to the momentum distribution of the atoms. The energy deposited in the system per unit time is nothing but $\Gamma \hbar\omega$, such that one can 
equivalently deduce the momentum distribution from the measurement of  the energy increase rate versus $\omega$.
Note that, by choosing $\omega$
close to $N^2E_{\rm rec}$, where $N$ is an integer, one can induce N-th order Bragg processes: the momentum transfer is then $Nq$, which allows to be deeper into the Doppler limit. This technique was first used in~\citep{stenger_bragg_1999} in a 3D Bose-Einstein Condensate, where it confirmed the presence of long-range order in the cloud. Applied to a very elongated BEC, it was used to demonstrate the presence of longitudinal thermally excited phase fluctuations~\citep{richard_momentum_2003}. More recently, it was implemented in 1D gases realized in a 2D optical lattice~\citep{fabbri_momentum-resolved_2011}. 
Note that, going beyond the Doppler limit of large momentum transfer, the Bragg techniques allow the measurement of the dynamic structure factor. Such measurements have been compared to results based on the Lieb-Liniger model in \citep{fabbri_dynamical_2015} and \citep{meinert_probing_2015}.


The Bragg momentum spectroscopy suffers from  small signal, since it is based on a perturbative analysis, and, most importantly, it does not allow to record the whole momentum distribution in a single measurement. This impedes the measurement of correlations in momentum space. We now discuss another technique that enables the recording of the whole momentum distribution of 1D gases in a single measurement. 
The key point is to be able to 
switch off the interactions 
almost instantaneously with respect to the 
longitudinal motion. 
This can be achieved, close to Feshbach resonances, by manipulating the interaction strength with a magnetic field~\citep{stewart_verification_2010}. However, in the special case of 1D gases, this switch-off can be achieved very easily by removing the transverse confinement: after the switch-off of the transverse potential, the transverse wave function expands in a typical time of the order of $1/\omega_\perp$, much shorter than typical times associated with the longitudinal motion, and this amounts to an almost instantaneous  vanishing of the effective 1D interaction strength with respect to the longitudinal motion.

Once interactions have been effectively switched off, one is left with the task of measuring the momentum distribution of an ideal gas. 
This could be done by a simple ballistic expansion, following the sudden switch-off of the longitudinal potential: after an expansion time
long enough so that the final cloud size is much larger than its initial size, the density distribution becomes homothetic to the momentum distribution. This technique, usually referred to as the time-of-flight technique, is used for instance in~\citep{fabbri_momentum-resolved_2011} and \citep{wilson_observation_2020,malvania2020generalized}.
However, in experiments using very long initial clouds and/or gases lying deep in the quasicondensate regime, such as atom chip experiments, this technique typically requires unrealistic expansion times. To overcome this difficulty, one can use the so-called 
focusing method. This method, first implemented in~\citep{shvarchuck_bose-einstein_2002}, consists in applying a short pulse of a strong longitudinal harmonic potential. Atoms do not have time to move during this pulse but they acquire a momentum kick $\delta p =-A x$ proportional to their distance $x$ from the center. The cloud then undergoes free evolution. After a time equal to the focusing time $t_f=m/A$, the density distribution is homothetic to the initial momentum distribution.
This technique effectively erases the information on the initial cloud longitudinal spread.

\subsubsection{Results in momentum space}
All measurements reviewed in Subsections \ref{subsec:g20},\ref{subsec:analysisnx} and \ref{subsubsec:densityfluctu} 
probe 
thermodynamic quantities. On the theory side, those are 
easily accessible numerically using Yang-Yang thermodynamics for thermal equilibrium states.
In contrast, the momentum distribution is not a thermodynamic quantity. 

The momentum distribution of 1D gases lying in the quasicondensate regime was measured by Bragg spectroscopy by ~\cite{fabbri_momentum-resolved_2011}. 
The authors show that the measured momentum distribution is close to a Lorentzian. A Lorentzian shape of full width at half-maximum $m k_{\rm B} T/(\hbar^2 n)$
is expected for homogeneous gases in this regime, for wavevectors lying in the phononic regime~\citep{jacqmin_momentum_2012}.  Within the local density approximation, the total momentum distribution $n(p)$ is obtained by summing the contributions of each small fluid cell. For a harmonically confined quasicondensate, one finds a momentum distribution that stays very close to a Lorentzian~\citep{jacqmin_momentum_2012}.

The focusing method was used for 1D gases trapped on an atom chip in~\citep{davis_yang-yang_2012}. The measured momentum distribution $n(p)$ was compared to 
improved classical field calculations, expected to be valid in the quasi-condensate and ideal Bose gas regimes. The transversely excited states cannot be neglected in those experiments and 
they were taken into account assuming they behave as ideal Bose gases.
In fact, in this paper, the authors propose a thermometry method. More precisely, they evaluate the
total kinetic energy $E_K=\int  n(p) p^2/(2m) dp$
from 
the measured momentum distribution $n(p)$.  $E_K$, is  a thermodynamic quantity that can be calculated with  Yang-Yang thermodynamics, and that can be fitted to extract the temperature. 
The fact that $E_K$ is a thermodynamic quantity 
follows from the following argument. 
$E_K$ can be computed using the local density approximation as the integral of the 
kinetic energy density $e_K(x)$ over the cloud. The latter is 
$e_K(x)=e(n(x),T) - e_{\rm int}(n(x),T)$
where $e(n,T)$ is the energy density of the Lieb-Liniger gas at density $n$ and temperature $T$, and $e_{\rm int}(n,T)$ is its interaction energy. 
Both are thermodynamic quantities: the interaction energy can be obtained from the Hellmann-Feynamn theorem, see Eq.~\eqref{eq:HelmanFeynam}. For data analysis, one adds the contribution of the transversally excited,  accounted for as ideal Bose gases.


The first comparison between the measured momentum distribution of a 1D Bose gas and exact calculations was done in~\citep{jacqmin_momentum_2012}. The first order correlation function at thermal equilibrium was computed with a Monte-Carlo method, which gave very good agreement with the measured momentum distribution. The measured in-situ density fluctuations (Subsec.~ \ref{subsubsec:densityfluctu}) also fitted very well with the temperature 
fitted from the Monte-Carlo simulations.   
The momentum distribution $n(p)$ was measured across the smooth transition between the ideal Bose gas regime and the quasicondensate regime. No striking modification of $n(p)$ was observed across the transition; in particular it conserves a Lorentzian-like shape. [We recall that the crossover occurs for a degenerate gas whose momentum distribution, in the ideal gas regime, has a Lorentzian central shape.].

\begin{figure}
    \centering
    \includegraphics[width=0.9\textwidth,viewport=49 404 550 750,clip]{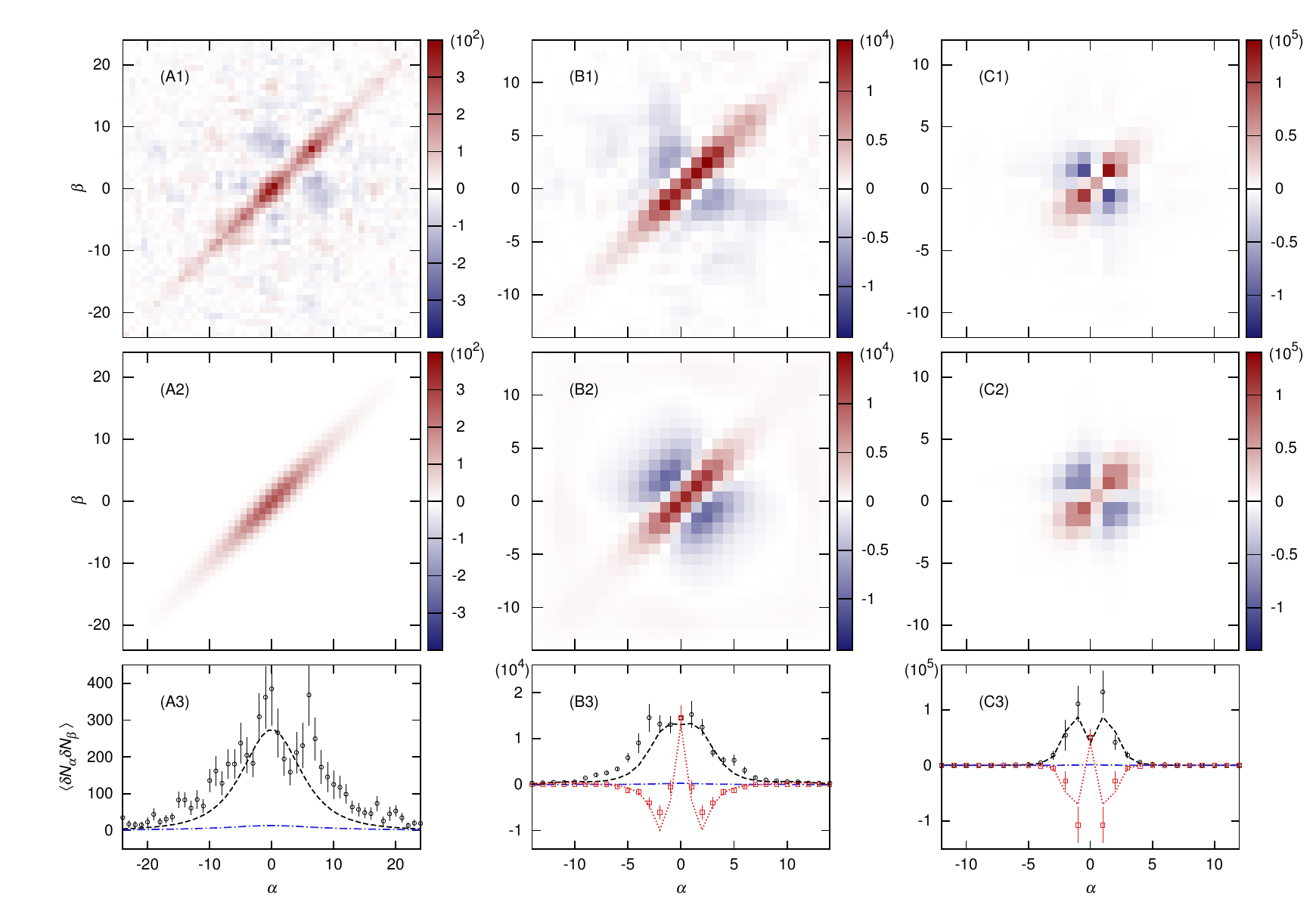}
    \caption{[From \citep{fang_momentum-space_2016}] Correlations in momentum space.
    (A1,A2,B1,B2,C1,C2) show the correlation  $\langle \delta n_\alpha \delta n_\beta\rangle$, where $\alpha$ and $\beta$ denote the index of the pixel in momentum space (equal to $\hbar \times 0.15\mu$m$^{-1}$). Column A corresponds to a  in the ideal Bose gas regime; column C to gas in the quasicondensate regime; column B to a gas at the crossover between both regimes. Experimental data (first line) are compared to Monte-Carlo calculations (second line). The third line shows cuts along the diagonal and the anti-diagonal.}
    \label{fig:npnpp}
\end{figure}

Using the focusing technique, the full momentum distribution can be recorded at a time in a sample. Thus, it is possible not only to extract the mean momentum distribution $n(p)$, but also its fluctuations $\delta n_p$. In~\citep{fang_momentum-space_2016}, the correlations
$\langle \delta n_p \delta n_{p'}\rangle$ are deduced from a statistical analysis of hundreds of images taken in the same experimental conditions. The results are in very good agreement to numerically exact results obtained, for a gas at thermal equilibrium, using a quantum Monte-Carlo algorithm, as seen in Fig.~\ref{fig:npnpp}.
The calculation uses a discretised model and a worm algorithm.
The crossover between the ideal-Bose gas
and the quasicondensate regime has a clear signature in momentum space correlations. The ideal Bose gas regime is characterized by a bunching phenomenon for equal momenta. In the quasi-condensate regime, on top of the bunching seen on the diagonal, anti-correlations appear for different momenta. Those anti-correlations ensure
small density fluctuations, the latter being inhibited in the quasicondensate regime because of the large interaction energy they would require. The presence of those anti-correlations in the quasicondensate regime was predicted using Bogoliubov theory in~\citep{bouchoule_two-body_2012}.

\subsubsection{Other higher-order correlation functions}
The two-body correlations in momentum space presented in Fig.~\ref{fig:npnpp} is related to the four-point correlation function of the atomic creation/annihilation operator. 
The four-point correlation function can also be probed by investigating density fluctuations resulting from a short free evolution time. The idea is to switch off the interactions between atoms abruptly, for instance by removing the transverse confinement (Subsec.~\ref{subsubsec:analysisnp}), and then to remove the longitudinal confinement and let the gas evolve freely for a short time $t_{{\rm free}}$. 
In contrast with the time-of-flight method used to measure momentum distribution (Subsec.~\ref{subsubsec:analysisnp}), here the free evolution time is assumed to be short, such that the mean density profile of the cloud barely changes.
Thus, no information is gained from the mean profile. Instead, one investigates the density fluctuations, also called density ripples. They are characterized by their spectral density $\langle |\rho(q)|^2\rangle$. 
For wavelengths much shorter than the size of the mean density profile, the latter is
related to the four-point correlation function
of the atomic field, 
\begin{equation}
    \langle |\rho (q)|^2\rangle = \int \int d\alpha dX e^{iqX}\langle \Psi^\dagger (\alpha) \Psi(\alpha+q t_{{\rm free}}) \Psi^\dagger (\alpha+X+q t_{{\rm free}}) \Psi(\alpha +X)\rangle ,
    \label{eq:densityripples}
\end{equation}
where expectation values are taken in the 1D gas before the expansion.
For very large $q$, since there is no long-range order in the 1D gas, the atomic field at positions separated by $q t_{{\rm free}}$ are uncorrelated and the above expression
vanishes.  On the other hand, for very small $q$ one recovers the 
spectral density of density fluctuations that were present in the initial gas.
This measurement turns out to be 
particularly relevant in the quasicondensate regime. There, the initial density fluctuations are negligible and  $\langle |\rho(q)|^2\rangle$ results from the 
transformation of phase fluctuations into density fluctuations during the free evolution.
Density ripples were first 
investigated in \citep{dettmer_observation_2001} in a quasi-1D setup. It was then used for thermometry in \citep{manz_two-point_2010} in a 1D gas realized in an atom-chip setup.
For $q$ small enough so that $\hbar q t_{{\rm free}}/m$ is very small compared to the correlation length of the phase fluctuations, one finds from Eq.~(\ref{eq:densityripples})
that the power spectrum of density fluctuations after free evolution during $t_{{\rm free}}$ is simply proportional to the power spectrum of the initial phase fluctuations at the same wave-vector. Thus, the analysis of density ripples allows to access each Bogoliubov mode individually.
This feature was used in~\citep{schemmer_monitoring_2018}
to monitor the dynamics produced by an interaction quench.
Finally, let us stress that the measurement of density ripples is clearly very different from the measurement of the momentum distribution, as the momentum distribution $n(p)$ mixes all  Bogoliubov modes.

\subsubsection{Validity of the thermodynamic equilibrium}
In the experiments reviewed so far, the 1D gases were shown to present a behavior in very
good agreement with that of a gas at thermal equilibrium.
However, as discussed in Section \ref{subsec:GGE}, the 1D Bose gas is integrable so there is no reason to assume that isolated 1D gases should be described by thermal equilibrium states. Instead,
it is expected that the local state of the gas is a Generalized Gibbs Ensemble (GGE) characterized by its rapidity distribution. The rapidity
distribution should depend on the preparation scheme of the gas. In particular, as discussed below in Section
\ref{sec:losses}, losses are expected to bring the system in a state that is not a thermal state. Since losses are usually present, at least in the preparation stage, one expects the state to lie in a non-thermal state.

The reason why the above data are in good agreement with thermal equilibrium
predictions is unclear. It may be do to the integrability breaking produced by  populated
transversely excited states~\citep{li_relaxation_2020,moller_extension_2021}.
The population of transversely excited states decreases exponentially with the 
ratio between the mean energy per atom and $\hbar \omega_\perp$, the gap between the transverse ground state and the transversely excited states. This population can be totally 
negligible for gases deep enough into the 1D regime.
However, three-body processes involving a virtual excitation of transversely excited states are still expected to lead to integrability breaking~\citep{mazets_breakdown_2008,mazets_dynamics_2011,tan_relaxation_2010}.
On long time scales, the longitudinal
potential is also expected to 
break the integrability and 
to bring the cloud towards a thermal state. This effect was first pointed out in~\citep{mazets_integrability_2011}, within a semi-classical approach taking into account the Wigner time delay associated to the two-body collision.  Using an improved GHD approach that includes diffusive terms, the relaxation towards a thermal equilibrium in presence of an external potential was recently computed in ~\citep{bastianello_thermalization_2020}, see Fig.~\ref{fig:thermalization_diffusion}.

In contrast with the results presented so far,
there do exist experimental data that show the presence of long-lived non-thermal states. 
In~\citep{langen_experimental_2015}, a 1D gas in the quasicondensate regime is cut into two 1D gases by transverse splitting, and the evolution of the relative phase between the two clouds $\theta_1(x,t)-\theta_2(x,t)$ is monitored. 
Within a Bogoliubov analysis, the antisymmetric degrees of freedom are decoupled from the symmetric ones and their Hamiltonian is a simple integrable model which reduces to a collection of independent modes. Within Bogoliubov theory, the GGE is simply parameterized by the population of each Bogoliubov mode, or equivalently by its temperature. 
Only large scale variations are probed 
in~\citep{langen_experimental_2015}, which belong to the phononic regime. For a rapid splitting process, all antisymmetric phononic modes are expected to share the same temperature~\citep{gring_relaxation_2012}, so one should recover a thermal ensemble. However, for another splitting procedure, the experimental results show that all 
Bogoliubov modes do not share a common temperature, so that one has a true GGE. 
The GGE realized in this experiment is a GGE relative to the Bogoliubov model, it is not a GGE of the Lieb-Liniger gas. 
On longer time scales, coupling between the Bogoliubov modes is expected to lead to the relaxation of this GGE towards a thermal ensemble for the phononic modes~\citep{mazets_dephasing_2009}.

In \citep{johnson_long-lived_2017}, a single 1D gas lying in the quasicondensate regime is investigated. Long-lived non-thermal states are reported with a mode occupation of the phononic modes corresponding to some temperature. This temperature is shown to be incompatible with the population of the short wave-length collective modes. It is proposed that such a non-thermal state emerges from the effect of atom losses. Such non-thermal states are robust with respect to the trivial Bogoliubov dynamics, however the Bogoliubov modes are not the real infinite lifetime quasiparticles of the Lieb-Liniger model. There is no one-to-one correspondence between the population of the Bogoliubov modes and the rapidity distribution. 
Yet, the short-wavelength Bogoliubov modes should correspond to  
high rapidities, while the phonons should be related to small deformations of the rapidity distribution near its zero-temperature edges. Thus, one expects that the non-thermal states reported in~\citep{johnson_long-lived_2017} are really robust against the Lieb-Liniger Hamiltonian, which means that they correspond to non-thermal rapidity distributions.

Finally, let us also mention that the most striking long-lived non-thermal state realized in a 1D Bose gas was reported as early as 2006 in the Newton Cradle experiment~\citep{kinoshita2006quantum}. This work is reviewed in detail in Section~\ref{subsection:NewtonCradle}. 

Although long-lived non-thermal states have been reported, integrability breaking mechanisms, such as effect of transverse
excited states in virtual processes~\citep{mazets_integrability_2011} or effect of 
longitudinal potential~\citep{bastianello_thermalization_2020}, are expected to bring the system  
towards a thermal equilibrium state described by the 
Gibbs Ensemble  at very long time. Note, however that loss mechanisms, whose 
effects are discussed in Sec.~\ref{sec:losses}, might prevent the 
observation of thermalization.

\begin{figure}
    \centering
    \includegraphics[viewport=134 480 475 593,clip]{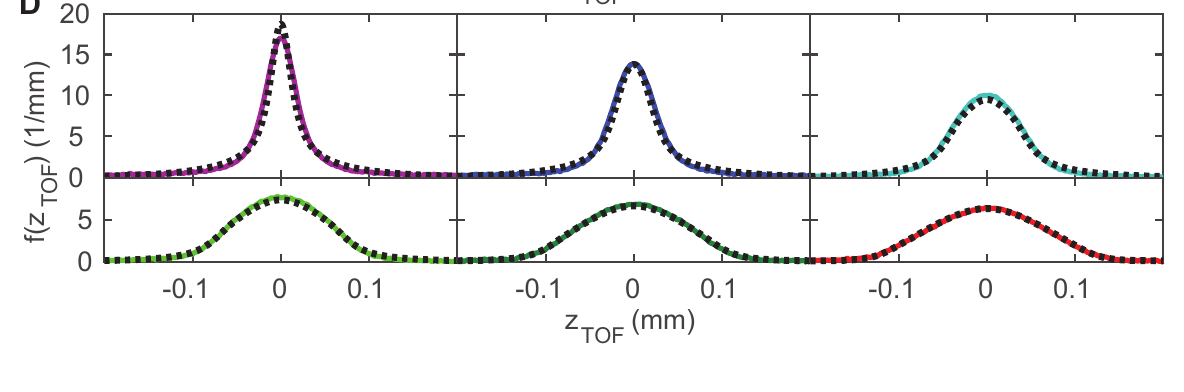}
    \caption{[From \cite{wilson_observation_2020}] Measurement of the rapidity distribution.
    Density profiles recorded after a 1D expansion during a time $t_{\rm{exp}}$, followed by a ballistic (free) expansion during $t_{{\rm free}}$.
    $t_{\rm exp}$ equals to (from left to right and top to bottom) 0,1,3,6,9,12 ms and $t_{\rm free}=70$ms$-t_{\rm exp}$. 
    Experimental data (solid lines) are in remarkable agreement with theoretical predictions for hard-core Bosons (dashed lines). For long 1D expansion times, the momentum distribution converges towards the rapidity distribution. }
    \label{fig:measurement_rapidity_distribution}
\end{figure}

\subsection{Measurement of the rapidity distribution}
\label{subsec:rapidity_dist_measurement}
As explained in sec.~\ref{subsec:asymptotic_momenta}, rapidities are the asymptotic momenta of the atoms after a 1D expansion. This 
characterization can be viewed as the definition of the rapidities. 
It also shows that the rapidity ditribution is an observable.
Owing to the important role of the rapidity distribution, the experimental ability to  measure it is a key development for the study of 1D gases. 

The first experimental measurement of the rapidity distribution was done by~\cite{wilson_observation_2020}, for gases lying quite deep inside the hard core regime.
In this experiment the trapping potential is the sum 
of a 2D array of 1D tubes realized by a 2D optical lattice, and a slowly varying trapping potential, which 
provides a longitudinal confinement along the tubes. 
Removing the slowly varying potential, a
1D expansion is performed within each tube. After a sufficiently long 
expansion time, the momentum distribution converges 
towards the rapidity distribution. The momentum distribution is then measured using the time-of-flight technique, performed by suddenly turning off all confining potentials (including both the 1D longitudinal confinement and the 2D array of tubes). Interactions are effectively almost instantaneously turned off by the rapid transverse expansion of each tube. Thus the cloud performs a ballistic expansion  such that, 
at long time, the density profile reflects the momentum  distribution.  

The momentum distribution measured for different 1D expansion times is found to be in very good agreement with 
theoretical predictions, see Fig.~\ref{fig:measurement_rapidity_distribution} from ~\citep{wilson_observation_2020}. The calculation of the evolution during the 1D expansion assumes hard-core bosons and is done for a gas initially in the ground state. 
For long enough expansion times, the momentum distribution converges towards the rapidity distribution.

\newpage

\section{Experimental tests of Generalized Hydrodynamics}
\label{sec:GHDexperiment}
Generalized Hydrodynamics (GHD) is an effective theory which assumes separation of scales, i.e. long wavelength and slow dynamics, see Fig.~\ref{fig:sos}. 
It is an approximate theory, and testing it experimentally is highly desirable to establish its relevance for the description of 1D Bose gases. The results reviewed in Section \ref{sec:experiments_beforeGHD} show that the Lieb-Liniger model is realized experimentally 
in cold atoms experiments.
On intermediate time scales, small enough such that
unavoidable integrability breaking mechanisms,
such as effect of transversely excited states~\citep{mazets_integrability_2011}, 
have negligible effects, but long compared to the 
relaxation time of the Lieb-Liniger model, one expects to 
be able to perform experimental tests of GHD. This has been done in two cold atom experiments up to now. The first one investigates a weakly interacting 1D gas in an atom chip setup. The second one uses strongly interacting atoms and investigates arrays of 1D gases. 
In both cases, an out-of-equilibrium 
situation is produced by a quench of the longitudinal potential and the subsequent time evolution is recorded.

Before we present those experimental tests of GHD, 
we briefly discuss the Quantum Newton Cradle experiment~\citep{kinoshita2006quantum}, which was performed a decade before the advent of GHD, and served as major motivation for many theoretical developments that occured during that time. The questions raised by that pioneering experiment have driven the research on the out-of-equilibrium dynamics of integrable quantum systems, including the development of GHD.

\subsection{The Quantum Newton Cradle experiment.}
\label{subsection:NewtonCradle}
In their famous experiment on non-equilibrium dynamics in a 1D Bose gas, \cite{kinoshita2006quantum} use an array of 1D tubes, in a blue-detuned 2D optical lattice setup, with a longitudinal confinement, approximately harmonic, realized by an additional smooth dipole trap. The  
interaction parameter $\gamma$ (see Eq.~\ref{eq:gamma}) at the center of the trap, averaged over the collection of 1D gases, ranges from 0.6 to 4, depending on the data set. 

\begin{figure}
    \centering
    \includegraphics[viewport=78 207 270 505,clip]{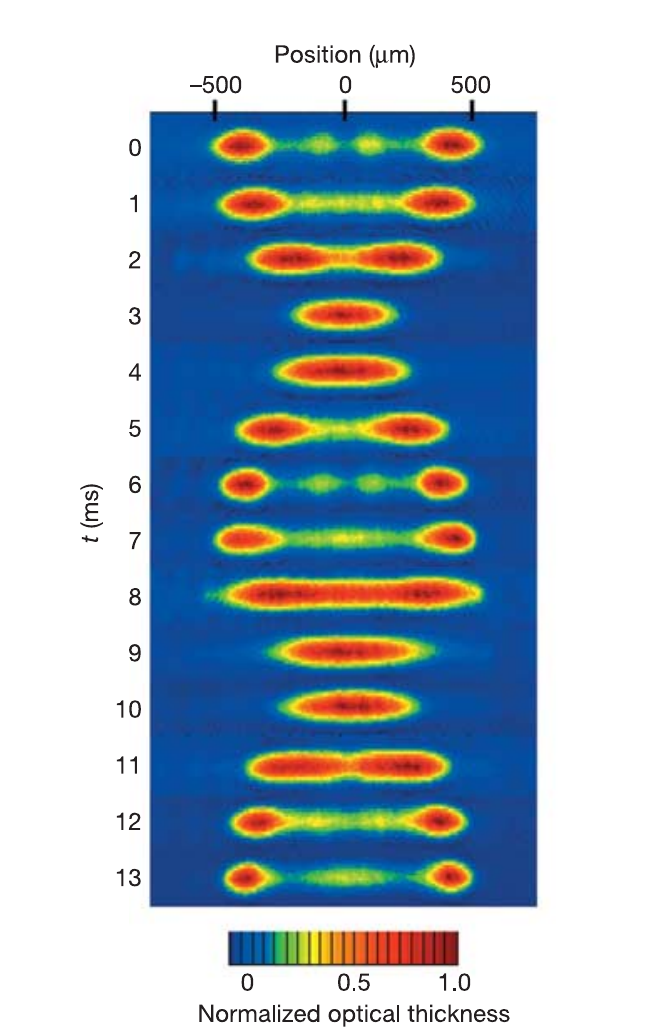}
    \raisebox{4cm}{
    \begin{tabular}{c}
    \includegraphics[viewport=325 354 533 449,clip]{figures/NewtonCradleWeiss.pdf}\\
    \includegraphics[viewport=325 156 533 176,clip]{figures/NewtonCradleWeiss.pdf}
    \end{tabular}}
    \caption{Quantum Newton Cradle experiment [taken from \cite{kinoshita2006quantum}]. Two "clouds" at different mean momenta are prepared by Bragg pulses. Left picture: absorption images of the cloud after an evolution time $t$ in the longitudinal potential, followed by a 1D expansion during a given time $t_{\rm exp}$. The period of the dipole motion in the longitudinal harmonic trap is spanned: 13 ms$=\tau=(2\pi)/\omega_z$. Right image: longitudinal density profiles. Green : profile averaged over the first oscillating period. For times $ t>10\tau$, the longitudinal profile barely changes during an oscillation period, as expected because of dephasing induced by the spread of $\omega_z$ among the 1D tubes. Bleu and red : profiles at $t=15\tau$ and $t=30\tau$. One observes a  decrease of the total atom number, due to 3-body recombination effect. The shape of the distribution however barely changes and does not converges towards the shape of a thermal equilibrium state. }
    \label{fig:NewtonCradleWeiss}
\end{figure}

An out-of-equilibrium initial situation is realized by applying two Bragg pulses on the cloud, such that the zero-momentum state is mostly transferred to a superposition of the two momentum states $p=\pm 2\hbar k$. [Because of the interactions between atoms, that momentum transfer is not perfect; for a theoretical study of the state generated by the Bragg pulse, see~\citep{van2016separation}.] 
The cloud then evolves freely inside the trap up to time $t$.
At time $t$, the longitudinal potential is turned off and atoms
undergo a 1D expansion during an expansion time $t_{\rm exp}$, large enough so that the final size of the atomic cloud is much larger than its initial size. After the 1D expansion,  an image is recorded. The image performs a column integration along a direction perpendicular to the longitudinal direction. Fig.~\ref{fig:NewtonCradleWeiss} shows such images, for different evolution times $t$ that span an oscillation period of the longitudinal potential: $\tau=(2\pi)/\omega_\parallel$,
where $\omega_\parallel$ is the frequency of the longitudinal potential.
At $t=0$, one observes two well separated clouds, corresponding to the two components of different momenta created by the Bragg pulses. Then, one sees a dynamics which resembles the one that would be found if the atoms were non-interacting. In particular, at time $\tau/2$, one recovers a situation close to the initial one: roughly speaking, the `cloud' of initial momentum $2\hbar k$
has performed half an oscillation in the harmonic trap and its momentum is now $-2\hbar k$.

After an evolution of about $10\tau$ the images show small variations on an oscillation period. This is consistent with the dephasing effect 
due to the spreading of $\omega_\parallel$ among the 1D tubes. A slow time evolution of the longitudinal profile is observed, see Fig.~\ref{fig:NewtonCradleWeiss}. 
This slow evolution is attributed to atom losses and to a small heating rate. 
Importantly, the measured longitudinal distribution does not evolve towards a thermal equilibrium distribution, at least not on the time scale probed in the experiment. 

Theoretical modeling of this experiment is easy only in two asymptotic regimes: the ideal Bose gas regime, and the hard core regime. In both cases, the dynamics is that of a non-interacting gas, see Subsection~\ref{sec:regimes}. In those limits, one expects to observe undamped oscillations going on forever, for a single 1D tube and a purely harmonic trap. The data of~\cite{kinoshita2006quantum} are compatible with this interpretation: at short times, the evolution is close to that of an ideal gas, and at longer times the observed damping can be attributed to dephasing between 1D gases. At very long times the evolution can be attributed to atom losses. 
The observed behavior is thus qualitatively similar to the one expected for hard core bosons, and this is because $\gamma$ is sufficiently large so the 1D gases are quite well inside the hard core regime.
However, a quantitatively accurate modeling of those results, properly taking into account the finite value of $\gamma$, was completely out of reach at the time when \citep{kinoshita2006quantum} was published. In the decade that followed the experiment, no theory was capable of simulating it, taking into account the finite value of the interaction strength.

GHD is the first, and, until now, only theory capable of obtaining quantitive predictions for such an experiment, valid for any initial situation (see section \ref{susec:modelingNC}). This illustrates the power of that theory. The physical picture, detailed in section \ref{susec:modelingNC}, is summarized below. The time-evolution of the rapidity density $\rho(x,\theta)$ develops sharp structures due to the trap anharmonicity~\citep{caux2019hydrodynamics}. At some point those structures will be so narrow that the large-scale approximation of Euler-scale GHD will fail. One then expects the fine structures to disappear and the rapidity distribution to tend to a rapidity distribution that is a stationary solution of the GHD equations and that is non-thermal~\citep{caux2019hydrodynamics,cao2018incomplete}. On even longer time scales, under the
combined effect of the integrability breaking produced by the longitudinal potential and diffusive terms, which are beyond Euler-scale GHD, the system will eventually drift towards a thermal equilibrium state~\citep{bastianello_thermalization_2020}.   
Relaxation towards a thermal 
equilibrium state has also been observed in classical field numerical simulations of the Newton Cradle setup~\citep{thomas_thermalization_2021}. 
The classical field, which is expected to describe weakly interacting gases with large mode population (see section \ref{subsubsec:classicalfield}), is not restricted to the description of long-wavelengths behavior: it is beyond Euler-scale GHD and this explains why it can lead to thermalization in the presence of an external potential.

In the Quantum Newton Cradle experiment, effects that are beyond the pure 1D physics may also play a role.  The population of transversely excited states is negligible since $\hbar \omega_\perp$ (i.e. the energy gap between the transverse ground state and the first excited state) greatly exceeds the typical longitudinal energy per atom. In particular, thanks to the use of blue-detuned lasers for the realization of the 2D lattice,  
the potential depth, limited by longitudinal trapping, is smaller than $\hbar \omega_\perp$, which ensures that the longitudinal energy stays smaller than $\hbar \omega_\perp$. However, even when they are not populated, transversely excited states can contribute as virtual states in three-body processes~\citep{mazets_breakdown_2008}. This phenomenon introduces an effective three-body interaction that breaks the integrability of the Lieb-Liniger model and it might contribute significantly to the relaxation towards a thermal equilibrium.

Similar Quantum Newton Cradle experiments have been reproduced in \citep{tang_thermalization_2018,schemmer2019generalized,li_relaxation_2020}. \cite{li_relaxation_2020} studied the effect of integrability breaking due to the presence of atoms in transversely excited states, while \cite{tang_thermalization_2018} investigated the effects  of dipolar interactions.

\subsection{Test of GHD in an atom chip setup}
A first experimental demonstration 
of the validity and relevance of GHD was carried out  by~\cite{schemmer2019generalized}.
In this experiment, a 1D gas is realized on an atom chip. 
The initial cloud is at equilibrium
in a regime close to the quasicondensate regime.
Dynamics is initiated by a quench of the longitudinal potential, and the time evolution of the density profiles is recorded. For all situations considered, the results are in very good agreement with the predictions of the GHD theory.
\begin{figure}[ht]
    \centering
    \includegraphics[viewport=306 578 567 745,clip,width=0.7\textwidth]{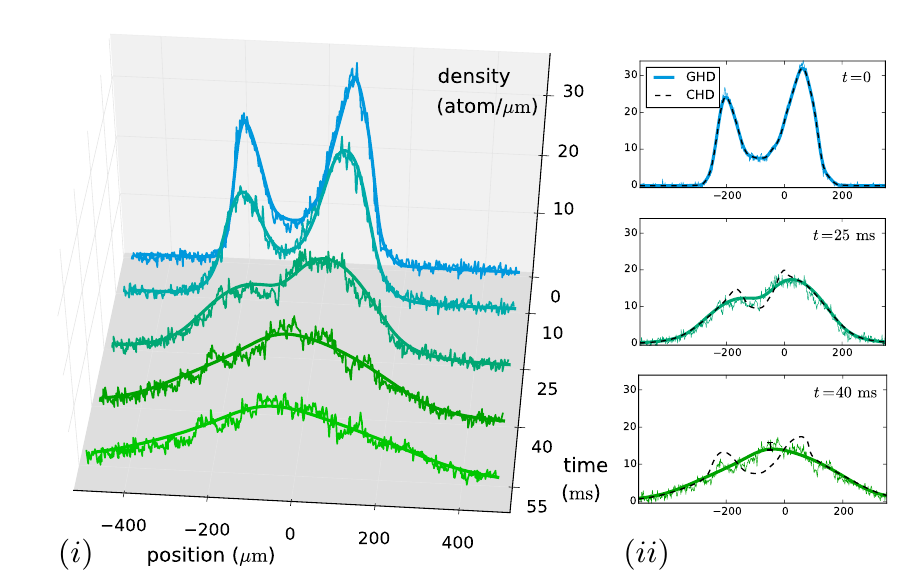} \\ \vspace{0.4cm}
    \includegraphics[width=\textwidth,viewport=34 406 568 730,clip,width=0.7\textwidth]{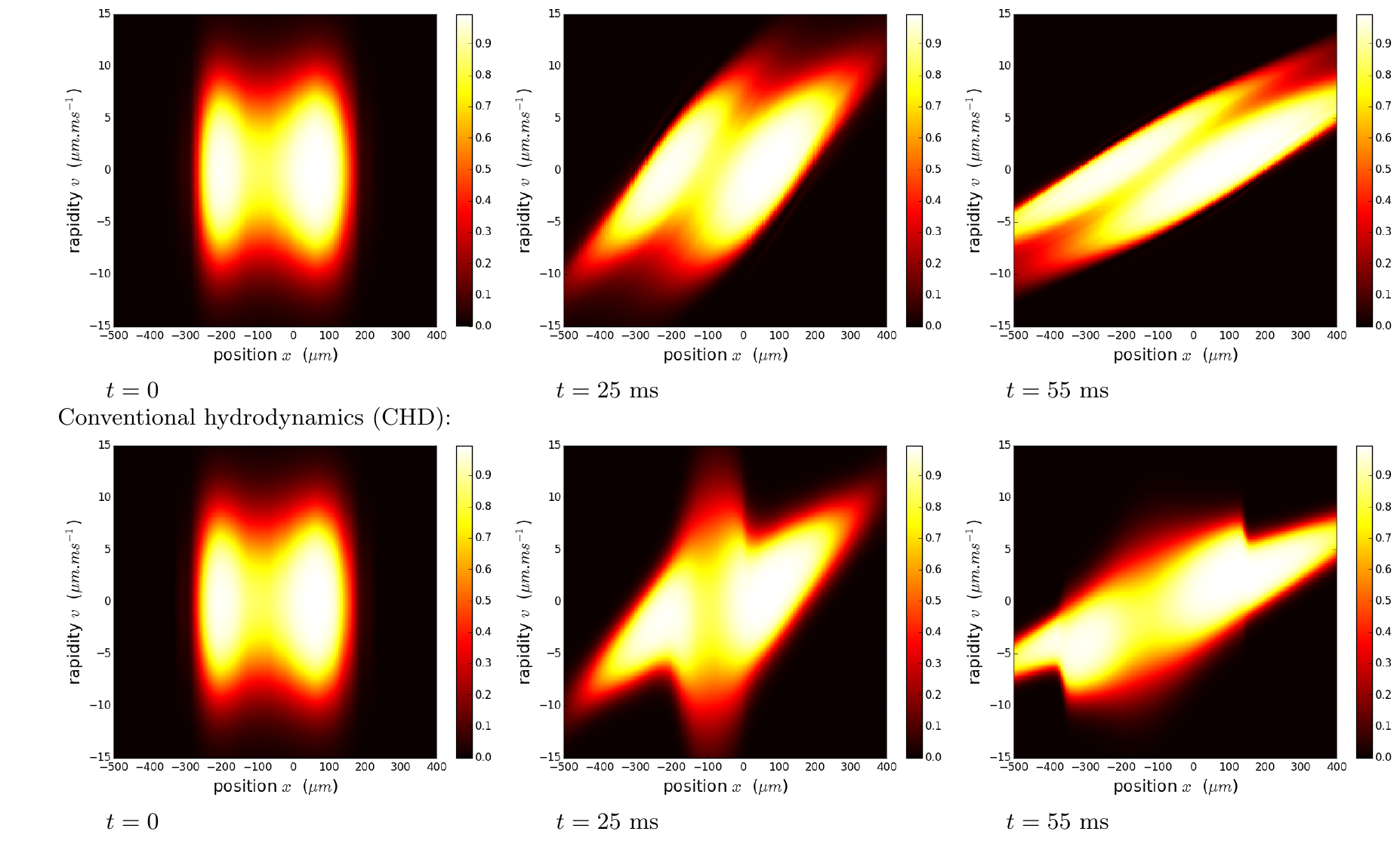}
    \caption{
    Top. [From  ~\cite{schemmer2019generalized}] Experimental test of GHD theory. Shown are density profiles after a quench from a double well potential to a flat potential. The experimental date (noisy curves) are in very good agreement with GHD predictions (smooth solid lines). The standart hydrodynamics (dashed lines on the right figures) fails to capture the physics. \\
    Bottom. [From Supplemental Material of \citep{schemmer2019generalized}] Calculations of the expected rapidity distributions after a release from a double-well potential. The figures show the evolution of the Fermi occupation ratio $\nu(x,\theta)$, which is in one-to-one correspondence with the rapidity distribution $\rho(x,\theta)$ (see Subsection~\ref{subsec:thermodynamic_limit}). The top line shows prediction from GHD theory, and the bottom line shows prediction from standard Euler hydrodynamics. The GHD predicts the appearance a of double peaked rapidity distribution around the center of the cloud (clearly visible on the plot at $t=55$ms). Standard Euler hydrodynamics, on the other hand, assumes that at each point $x$ the rapidity distribution $\rho(x,\theta)$ is the one at thermal equilibrium, which is a single peaked function. This theory is thus unable to capture the correct physics. 
    }
    \label{fig:testGHDSchemmer}
\end{figure}

The data are also compared to predictions of standard hydrodynamics, given by the Euler equations (\ref{eq:euler}) of the introduction  (see also Subsection~\ref{subsec:conventional_hydro}).
In contrast with GHD, this standard hydrodynamic approach assumes that the gas is locally
at thermal equilibrium. The numerical solution of the Euler equations involve the numerically tabulated pressure $\mathcal{P}(n,e)$ which is obtained from the Yang-Yang 
equation at thermal equilibrium, see Subsection~\ref{subsec:yangyang}.
A clear failure of standard
hydrodynamics is found when the 1D gas is prepared at equilibrium in a double-well potential, and the double-well is suddenly switched off and the gas is let to expand freely in 1D. 
In that case, the predictions of standard hydrodynamics differ strongly from those of GHD. Fig.~\ref{fig:testGHDSchemmer} shows the experimental data, together with 
GHD calculations and standard hydrodynamics calculations. 
 The experiment clearly discriminates between both theories and the data are found to be in agreement with GHD, but not with standard hydrodynamics. 

The origin of the failure of standard hydrodynamics in the above scenario is revealed by the following simple picture (Fig.~\ref{fig:testGHDSchemmer}, bottom). 
During the time evolution, the two 
clouds that were initially in each of the potential wells spread,  the negative rapidities moving to the left and the positive ones to the right. 
After some expansion time, at the central position, the positive rapidities from the left cloud meet negative rapidities coming from the right cloud. The resulting rapidity distribution is then double peaked.
 The standard hydrodynamic theory cannot capture this feature, because it assumes local thermal equilibrium and the rapidity distribution of thermal states are single-peaked. 
 This striking difference between GHD and conventional Euler hydrodynamics is clearly visible in Fig.~\ref{fig:testGHDSchemmer}, where the calculated Fermi occupation ratio $\nu(x,\theta)$ is displayed for both theories.

\begin{figure}[ht]
    \centering
    \includegraphics[width=0.7\textwidth,viewport=309 625 565 750,clip]{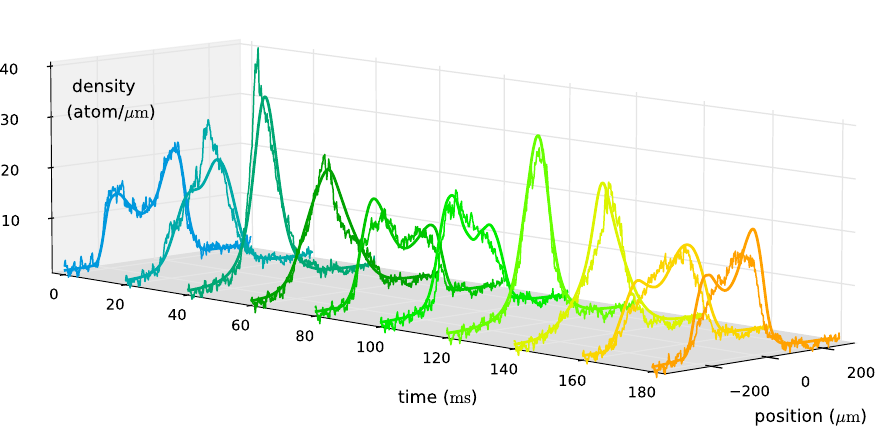}
    \caption{[From \citep{schemmer2019generalized}] Dynamics induced by a quench from a double well potential to a harmonic potential, which realizes a situation similat to the Quantum Newton Cradle. Experimental data (noisy lines) are compared to GHD calculations (smooth solid lines). Because of atom losses, the number of atoms drops by approx. $15\%$ from $t=0$ to $t=180$ms. This is not taken into account in the GHD theory.}
    \label{fig.SchemmerNC}
\end{figure}

 The Newton Cradle scenario has also been reproduced in~\citep{schemmer2019generalized}, 
 by quenching the longitudinal potential from a double-well to a harmonic potential, see Fig.~\ref{fig.SchemmerNC}. One then initiates a dynamics similar to that studied in \citep{kinoshita2006quantum}, with two clouds that oscillate and collide in a harmonic potential.
 The experimental results compare well with prediction from GHD.
 For this scenario, conventional Euler hydrodynamics completely fails: it predicts the formation of a very sharp structure in the denisty distribution,
  corresponding to large grandiant of the density, that eventually lead to a shock, at times as small as about 30 ms. The agreement between experimental data and GHD is less good that for the scenario of Fig.\ref{fig:testGHDSchemmer}. One of the reason might be the effect of three-body losses. In the Newton-Cradle scenario, large peak densities are attained when both cloud superpose, and three-body recombination process occurs, leading to atom losses. We find experimentally that the total atom number decreases by about 15\% on the time evolution shown in Fig.\ref{fig.SchemmerNC}.

\subsection{Test of GHD in strongly interacting gases}
\label{subsec:testGHDWeiss}

\begin{figure}
    \centering
    \includegraphics[viewport=43 339 313 762,clip,width=0.7\textwidth]{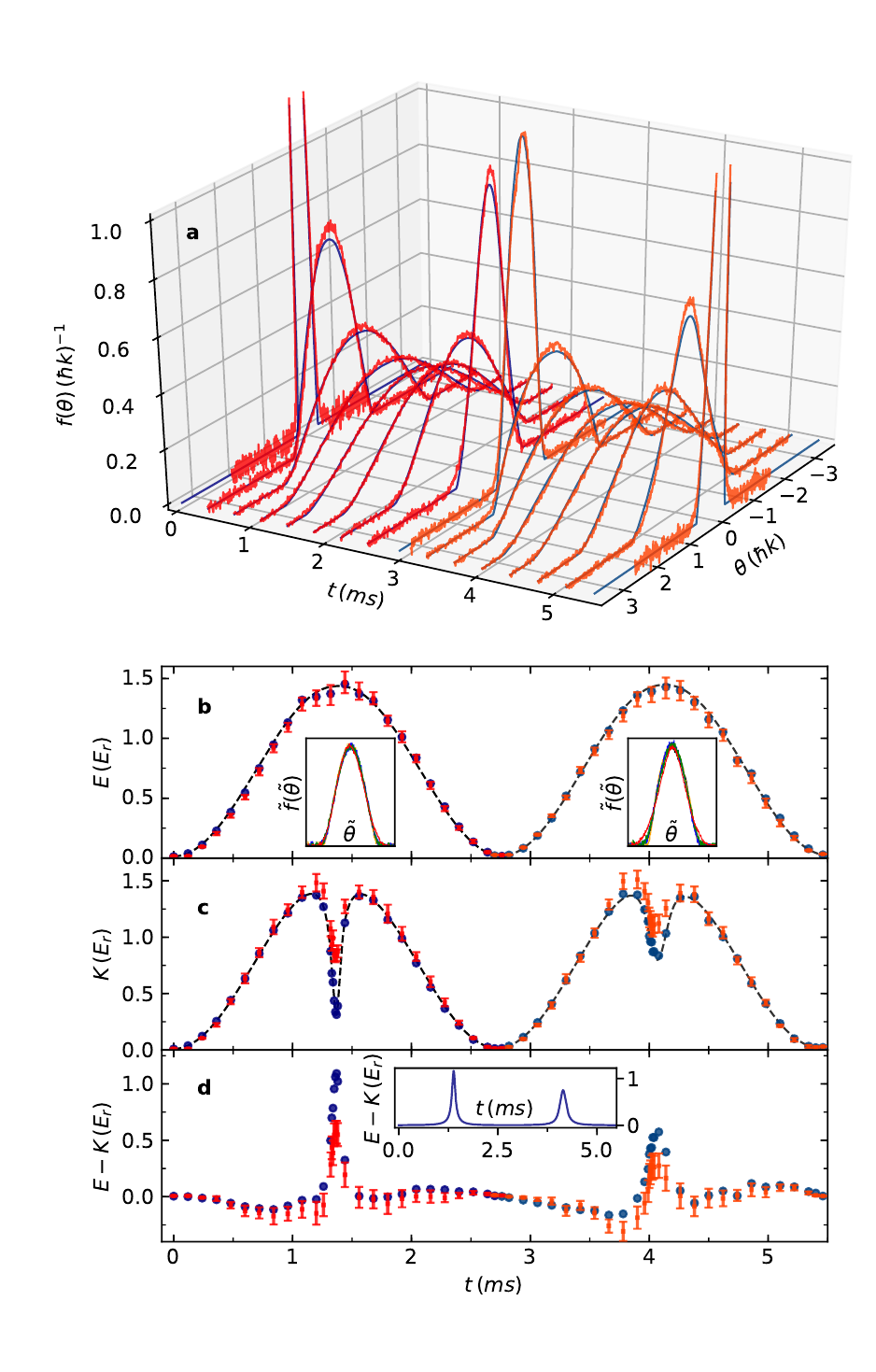}
    \caption{Test of GHD using strongly interacting gases [from \citep{malvania2020generalized}]. The dynamics is generated by a sudden increase of the depth of the Gaussian longitudinal confinement by a factor 100. The first two compression cycles are shown. Experimental data (red curves in (a) and red dots in (b-d)) are compared to  GHD predictions for a gas initially in the ground state (blue lines in (a)). The blue dots in (b-c) are the GHD calculation using the measured atom number at each time, while the dashed line in (b-c) is the theory using the average atom number. {\bf a} Measured rapidity distribution $f(\theta)$, integrated over positions $x$ and averaged over all 1D tubes, compared to the GHD prediction. {\bf b} Rapidity energy $E$, {\it i.e.} $\int d\theta f(\theta) (\theta^2/(2m))$, in units of the recoil energy $E_{\rm r}$. {\bf c} Evolution of the kinetic energy, obtained from the measured momentum distribution $w(p)$: $K = \int dp w(p) (p^2/(2m))$. 
    {\bf d} Interaction energy $E-K$. (To prevent confusion with notations used in sec.\ref{sec:losses}, $K$ is noted $E_K$ in the main text)}
    \label{fig:GHDWeiss}
\end{figure}

In the experiment of~\cite{schemmer2019generalized}, GHD is tested in a very large atom cloud that contains thousands of atoms, and whose longitudinal size, of the order of 100$\mu$m, is very large compared to microscopic scales. In these conditions, GHD is clearly expected to be valid. In a more recent experiment by~\cite{malvania2020generalized}, GHD is tested in a setup that uses a 2D lattice of 1D gases. In this experiment, the typical atom number $N$ is as low as 10 to 20 per 1D gas. Moreover, the (quasi)harmonic longitudinal potential $V(x)$ can be quenched very dramatically: at time $t=0$, the amplitude of the trap can be increased by a factor as large as 100, so the atom cloud is very strongly compressed. The validity of GHD is severely challenged in this situation. Yet, the experimental data, which are averaged over all the 1D tubes, are still correctly described by GHD over the first few oscillation cycles. We now discuss the results of~\cite{malvania2020generalized} in more detail.

The 1D clouds are prepared from a 3D Bose-Einstein condensate by adiabatically increasing the depth of the 2D lattice. When the 2D lattice depth is large enough, the gas decouples into independent 1D tubes. The temperature is extremely low, so that each 1D gas is close to the ground state of the Lieb-Liniger Hamiltonian in each tube. The dynamics is generated by suddenly increasing the longitudinal potential $V(x)$ felt by the atoms in each tube. The longitudinal potential $V(x)$, realized using an optical beam, has a Gaussian shape and its depth is increased by a large factor, 10 or 100 depending on the experimental data set. In sharp contrast with~\citep{schemmer2019generalized}, the initial cloud lies in the hard-core regime, with mean $\gamma$, averaged over the distribution of linear densities, as large as 9. \cite{malvania2020generalized} measure the time evolution of the rapidity distribution (Fig.~\ref{fig:GHDWeiss}), using the technique of~\citep{wilson_observation_2020} reviewed in Subsection~\ref{subsec:rapidity_dist_measurement}. This measurement is a global measurement: the rapidity distribution is integrated over positions $x$ within each tube, and averaged over all the tubes.

\begin{figure}
    \centering
    \includegraphics[width=0.7\textwidth]{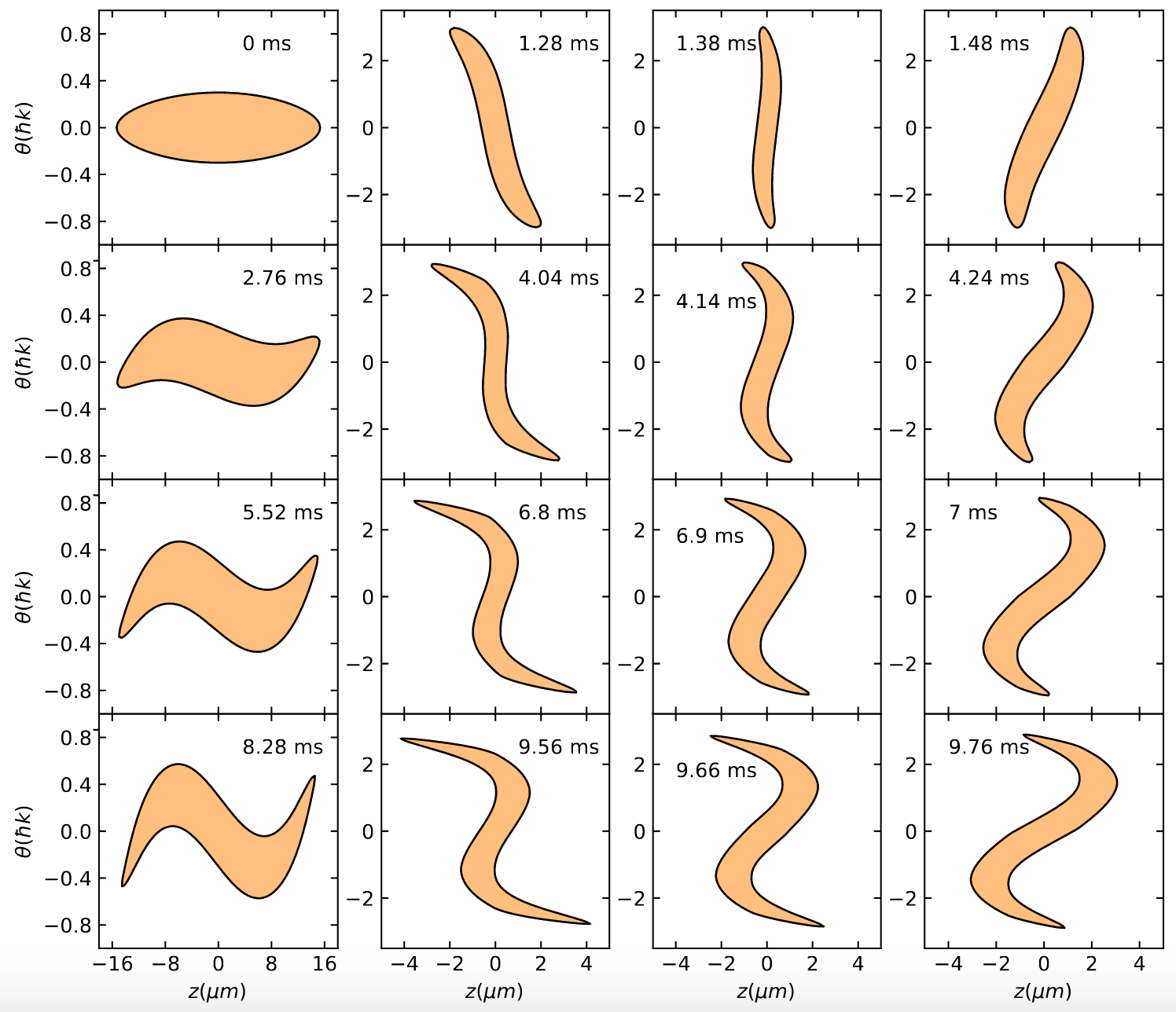}
    \caption{[From \citep{malvania2020generalized}] `Zero-entropy' GHD simulation of a sudden increase of the trap depth of the quasi-harmonic (Gaussian) potential $V(x)$: at $t=0$, the depth is increased by a factor 100, corresponding to the experimental situation of Fig.~\ref{fig:GHDWeiss}. Because the gas is initially at zero temperature, the Fermi occupation $\nu(x,\theta,t)$ is either one (in the orange area) or zero (white area) at any time. The contour $\Gamma_t$ (black curve) that separates the two regions evolves according to Eq.~(\ref{eq:zero_entropy_ghd}). Here the dimensionless repulsion strength $\gamma$ is initially of order $1$ in the center of the cloud, and it drops to a value of order $0.1$ at the maximum of the compression. One clearly sees that the contour $\Gamma_t$ gets deformed, as a combined effect of the interactions and of the small trap anharmonicity. Additionally, one observes the appearance of multiple Fermi seas: when a vertical line passing through a point $x$ can intersect the contour $\Gamma_t$ at more than two points, the gas is locally in state known as a `split Fermi sea' \citep{fokkema2014split,eliens2016general,eliens2017quantum}. For instance, a double Fermi sea appears near the first compression point (1.38ms panel). At this point, the conventional hydrodynamic approaches of Subsection~\ref{subsec:conventional_hydro} typically fail because of the appearance of shocks, and GHD is necessary to describe the full evolution. 
    }
    \label{fig:fermi_contour}
\end{figure}

The results are shown in Fig.~\ref{fig:GHDWeiss}, for a quench of the 1D trap amplitude by a factor 100. They are in excellent agreement with GHD predictions, which use the Lieb-Liniger ground state within the local density approximation as the initial state and the zero-entropy GHD equation~(\ref{eq:zero_entropy_ghd}) discussed in Subsection~\ref{subsec:other_setups} for the time evolution (Fig.~\ref{fig:fermi_contour}). In the theory calculation, it is found that, for a an increase of the trap depth of the quasi-harmonic (Gaussian) potential by a factor 100, the dimensionless repulsion strength $\gamma$, taken to be initially of order $1$ in the center of the cloud, drops to a value of order $0.1$ at the maximum of the compression: the variation of $\gamma$ as a function of position and time over one compression cycle is very large. In this setup, the contour $\Gamma_t$ (see discussion in Subsec.~\ref{subsec:other_setups} and the caption of Fig.~ \ref{fig:fermi_contour}) gets deformed very quickly. Already in the first compression cycle, there are positions in the cloud where the rapidity distribution is no longer that of a thermal equilibrium state. Instead, the local state of the gas consists of a split Fermi sea. The appearance of such exotic rapidity distributions originates both from the non-trivial effect of interactions between atoms - which are stronger at the maximum of the compression, where the atom density is high and the cloud is far from the hard-core regime - and from the anharmonicity of the longitudinal potential.
The presence of multiple Fermi seas rules out the possibility of describing the dynamics by standard hydrodynamic approaches (Subsec.~\ref{subsec:conventional_hydro}).

The measurement of the rapidity distribution, integrated over all atoms, allows to monitor the evolution of the rapidity energy. This is defined as $E= \int \rho(x,\theta) \theta^2/(2m) dx d\theta$ for a single 1D cloud, and it is averaged over all clouds in the measurement. The rapidity energy is the total energy, which is conserved, minus the potential energy associated to the confining potential. Since the potential energy oscillates as the cloud breathes, so does the rapidity energy $E$, as seen in Fig.~\ref{fig:GHDWeiss}.b.
Moreover, besides the rapidity distribution, \cite{malvania2020generalized} also measure the momentum distribution using a time-of-flight technique (see Subsection~\ref{subsubsec:analysisnp}), from which the kinetic energy $E_K$ can be extracted. The time-evolution of the kinetic energy is  shown in Fig.\ref{fig:GHDWeiss}.c 
(Notice that $E_K$ is simply noted $K$ in the figure).
The difference between the rapidity energy $E$ and the kinetic energy $E_K$ is the interaction energy. If the gas remained in the hard-core regime at all times, then one would always have $E_K=E$: the atoms would never be at the same position, so the contact interaction energy would be zero and the rapidity energy would be entirely in the form of kinetic energy. The narrow dip of $E_K$ at the time when the cloud is most compressed clearly demonstrates that, at this time, the cloud leaves the hard core regime. [Notice that it is important that it is the second moment of $w(p)$, i.e. the kinetic energy, that is used as a diagnostic for the departure from the hard-core regime. If, instead, one studied the half-width of $w(p)$, then its evolution would show pronounced dips at the time when the cloud is the most compressed, even if the gas stayed in the hard-core regime throughout the evolution~\citep{atas_collective_2017}.]

To conclude this section, we stress that the good agreement found by~\citep{malvania2020generalized} between experimental data with small atom number per tube and the predictions of GHD is a topic that deserves further investigation. As discussed in the Supplemental Material of \citep{malvania2020generalized}, the averaging over 1D tubes results in a smoothing of small atom number effects, which may explain the good agreement with the hydrodynamic theory, to some extent. It would be interesting to investigate the question of the accuracy of GHD for single 1D clouds of a few atoms more systematically. On the theory side, this could be done by performing numerical simulations of a single 1D gas for very small atom numbers (similarly to the numerical results shown in Fig.~\ref{fig:qghd}, but for longer times).

\newpage

\section{Atom losses}
\label{sec:losses}
When one describes experimental 1D Bose gases by the Lieb-Liniger Hamiltonian (\ref{eq:hamLL}), one assumes that they are perfectly isolated. However, even the cold atom experiments that realize the best isolated quantum many-body systems are never completely decoupled from their environment. Very often, the main coupling to the environment comes from loss processes in the gas.
The purpose of this Section is to give an introduction to recent progress on the effect of atom losses in the 1D Bose gas.

We stress that losses break the integrability of the model, so that they may be viewed as one special case of an integrability breaking mechanism ---a particularly relevant one, from an experimental viewpoint---. For a review of integrability breaking mechanisms in relation with Generalized Hydrodynamics, we refer to the article by Bastianello, de Luca and Vasseur in this Volume.

\subsection{Loss mechanisms in experiments}
Cold atom gases always suffer from losses. Different mechanisms for losses can be present, 
which are distinguished by the number of atoms $K$ ($K=1,2,3,\dots$) involved in each loss event.

\begin{itemize}
\item One-body losses ($K=1$) can occur due to collisions with hot atoms from the residual gas in the vacuum chamber: this typically imparts a kinetic energy to the (cold) atoms that is sufficiently large so that they leave the trap.  
One-body losses can also result from de-excitation for atoms lying in a metastable state, spin-flips to an untrapped magnetic state in the case of magnetically trapped atoms~\citep{burrows_nonadiabatic_2017},
collisions with energetic electrons~\citep{labouvie_bistability_2016}, or 
coupling to an untrapped state~\citep{rauer_cooling_2016,bouchoule_asymptotic_2020}.

\item Two-body losses ($K=2$) occur for instance when atoms are not in their internal ground state and exothermic two-body collisions that change the internal state of the atoms are present~\citep{traverso_inelastic_2009,yamaguchi_inelastic_2008}: the collision residues leave the trap because their internal state is not trapped, or because their kinetic energy exceeds the trap depth. In the presence of a laser, one could also have, starting from two nearby atoms, photoassociation towards excited molecules, which, after de-excitation, produce two very energetic atoms that leave the trap~\citep{kinoshita_local_2005}.

\item Importantly, cold atom experiments always suffer from three-body losses ($K=3$). This is due to three-body recombination, where a 
deeply bound molecule is formed: the binding energy, typically very large, is released in the form of kinetic energy, and the collision residues leave the trap~\citep{soding_three-body_1999,tolra_observation_2004}. 

\item In principle, loss processes involving more than three atoms also exist. In particular, losses involving $K=4$ atoms have been reported in ~\citep{gurian_observation_2012,ferlaino_evidence_2009}.
\end{itemize}
In the following we consider a general $K$-body loss process, for a fixed positive integer $K$. We now explain why the natural theoretical framework to model such a $K$-body loss process is the Lindblad equation~(\ref{eq:Lindblad}) below.
\vspace{0.5cm}

Because the lost atoms can be viewed as escaping towards a reservoir of particles whose state is not being monitored, the atoms remaining in the cloud no longer follow a unitary dynamics. 
Instead, the density matrix $\hat\rho$ of the remaining atoms in the gas (not to be confused with the rapidity distribution $\rho(\theta)$) evolves according to a Lindblad equation (see Eq.~(\ref{eq:Lindblad}) below). More precisely, the evolution of the gas under losses is described by a Lindblad equation if one assumes that the dynamics remains Markovian. This assumption holds if 
the energy-width of the reservoir, $E_{\rm res}$, which is the energy width spanned by the states of the continuum that are coupled to the trapped atoms by the loss process, is much larger than 
the energy width involved in the dynamics of the gas.
In temporal terms, it corresponds to the fact that  
the intrinsic duration of the loss event, equal to $\hbar/E_{\rm res}$, is much smaller than all other evolution time scales of the gas.

In experiments involving atoms in their internal ground state, the range of the interaction between atoms is typically much smaller than the typical distance between them. Losses can then be modeled by purely local processes: a loss event can occur only when $K$ atoms are found at the same position.

Under these assumptions, the evolution of the density matrix $\hat{\rho}$ of a uniform Lieb-Liniger gas of length $L$ under losses is
\begin{equation}
 \frac{d \hat{\rho}}{dt}   =  - i [H, \hat{\rho}]  	 + G \int_0^L \left(  \Psi^{K}(x)  \hat{\rho} \Psi^{\dagger K}(x)  - \frac{1}{2} \{ \Psi^{\dagger K}(x) \Psi^K(x) , \hat{\rho} \} \right) dx ,
 \label{eq:Lindblad}
\end{equation}
where $H$ is the Lieb-Liniger Hamiltonian (\ref{eq:hamLL}). Here $\Psi(x)^K$ is the operator that destroys $K$ bosons at position $x$, and $G$ is a constant characterizing the loss process, with dimension of ${\rm length}^{K-1}. {\rm time}^{-1}$.

\subsection{Theory of adiabatic losses}
Calculating the evolution of the density matrix $\hat \rho$ directly from the Lindblad equation~\eqref{eq:Lindblad} for more than a few atoms is, in general, an intractable task. Even numerically, the size of the matrix quickly becomes prohibitively large. Therefore, the role of Eq.~\eqref{eq:Lindblad} is merely to give a formal definition of the theory problem one would like to solve. To make progress, further assumptions are needed in order to simplify the description.

In the context of this review on hydrodynamics, where one focuses on effective hydrodynamic descriptions valid at large scales assuming local relaxation, a natural assumption is to consider the limit of adiabatic losses. When the parameter $G$ is small enough so that the dynamics induced by losses is much slower than the relaxation time $\tau_{\rm relax}$ of the system, the gas always remains in a relaxed state on long time scales. Its local properties are entirely described by the rapidity distribution $\rho(\theta)$, and the problem then boils down to computing the time evolution of $\rho(\theta)$. To lowest order in the small parameter $\tau_{\rm relax}G n^{K-1}$ (where $n=N/L = \int \rho(\theta) d \theta$ is the atom density), the evolution of the rapidity distribution must be of the form
\begin{equation}
    \label{eq:drho_dt_loss}
    \frac{d}{dt}\rho(\theta)= -G n^{K-1} F[\rho](\theta),
\end{equation}
where $F[\rho](\theta)$ is some functional of $\rho$ at time $t$, which needs to be determined from Eq.~(\ref{eq:Lindblad}). The functional $F[\rho]$ has been studied in~\citep{bouchoule_effect_2020}, which we briefly review now.

\vspace{0.5cm}

The idea is to consider the adiabatic evolution of the conserved local charges $Q[f]$, parameterized by some functions $f$ (see Eq.~\ref{eq:chargeQf}). To lighten the notation we simply write `$Q$' for such a generic charge. Under Lindblad evolution (\ref{eq:Lindblad}), the expectation value of the charge $\left< Q \right> := {\rm tr}( \hat{\rho} ~Q) $ changes as $\frac{d}{dt} \left< Q \right> \, = \, G \int \left( \frac{1}{2} \left< \Psi^{\dagger K} [Q, \Psi^K] \right> + \frac{1}{2} \left< [\Psi^{\dagger K} , Q \Psi^K \right> \right) dx$.
Because $Q$ is the integral of a local charge density, $Q= \int q(x) dx$, and because $\Psi^K(x)$ and $\Psi^{\dagger K}(x)$ are local operators, the two operators between the brackets are local. Then we know that, after the relaxation time $\tau_{\rm relax}$, their expectation value relaxes to their value in a Generalized Gibbs Ensemble, see Subsection~\ref{subsec:GGE}. This Generalized Gibbs Ensemble is a diagonal density matrix which can be characterized by its distribution of rapidities $\rho(\theta)$, see Subsection~\ref{subsec:yangyang}. Using the fact that $[\hat{\rho}_{\rm GGE} , Q] = 0$, and writing $\left< \mathcal{O} \right>_{[\rho]} = {\rm tr}\left( \hat{\rho}_{\rm GGE} \mathcal{O} \right)$ for the expectation value of an observable $\mathcal{O}$ w.r.t the GGE density matrix parameterized by the rapidity density $\rho$, we get the evolution equation for the expectation values of the charges
\begin{eqnarray}
    \label{eq:dqd_dt_loss}
\nonumber    \frac{d}{dt} \left< Q \right>_{[\rho]} & = & G \int_0^L  \left< \Psi^{\dagger K}(x) [Q, \Psi^K(x)] \right>_{[\rho]} dx \\
    & = & L G \left< \Psi^{\dagger K}(0) [Q, \Psi^K(0)] \right>_{[\rho]} .
\end{eqnarray}
We have used translation invariance, which implies that the integrand is independent of $x$, to go from the first to the second line.

Equation (\ref{eq:dqd_dt_loss}) determines the time evolution of the expectation values of all charges $Q$ under adiabatic losses. We see that the problem of modeling losses boils down to computing the r.h.s of (\ref{eq:dqd_dt_loss}), namely the  expectation value in a GGE of an operator of the form $\Psi^{\dagger K}(0) [Q, \Psi^K(0)]$, where $Q$ is a generic conserved charge, and $\Psi^K(0)$ is the operator that removes $K$ bosons at the same position.

\vspace{0.5cm}

To connect this simple general result to the evolution of the rapidity distribution (\ref{eq:drho_dt_loss}), we specialize $Q$ to an operator which measures the rapidity distribution. To elaborate, recall that the charges $Q[f]$ (\ref{eq:chargeQf}) are diagonal in the eigenbasis and that their expectation value in a Bethe state $\left| \{ \theta_a\}_{1\leq a \leq N} \right>$ is $\sum_{a=1}^N f(\theta_a)$. Formally, we can consider the charge $Q[f]$ corresponding to the choice $f(\alpha) = \delta(\theta-\alpha)$, which directly measures the distribution of rapidities $\rho(\theta)$. However, to ensure that the charge $Q[f]$ has good locality properties, it is safer to work with a regularization of the Dirac delta function, $\delta_\sigma$, of typical width $\sigma$ and of total weight $\int \delta_\sigma(\theta) d\theta = 1$, which remains a smooth function of $\theta$ for any $\sigma>0$ (for example a Gaussian of width $\sigma$). Then the choice $f(\alpha) = \delta_\sigma (\alpha-\theta)$ defines a charge $Q[f] = Q[\delta_\sigma(.-\theta)]$ which remains sufficiently local so that (\ref{eq:dqd_dt_loss}) applies. Thus, we see that the functional $F[\rho]$ entering the evolution equation (\ref{eq:drho_dt_loss}) must be given by
\begin{equation}
    \label{eq:F_limsigma}
    F[\rho](\theta) \, = \, \lim_{\sigma \rightarrow 0} \left( - n^{1-K} \left< \Psi^{\dagger K}(0) [Q[\delta_\sigma (.-\theta)], \Psi^K(0)] \right>_{[\rho]} \right) .
\end{equation}
It is essentially equivalent to calculate the r.h.s of (\ref{eq:dqd_dt_loss}) for arbitrary charges $Q$, or to compute the functional $F[\rho](\theta)$ defined by (\ref{eq:F_limsigma}). Either way, the difficulty lies in computing the expectation value of a specific local operator in a GGE.
\begin{figure}[tb]
    \centering
    \includegraphics[width=0.8\textwidth]{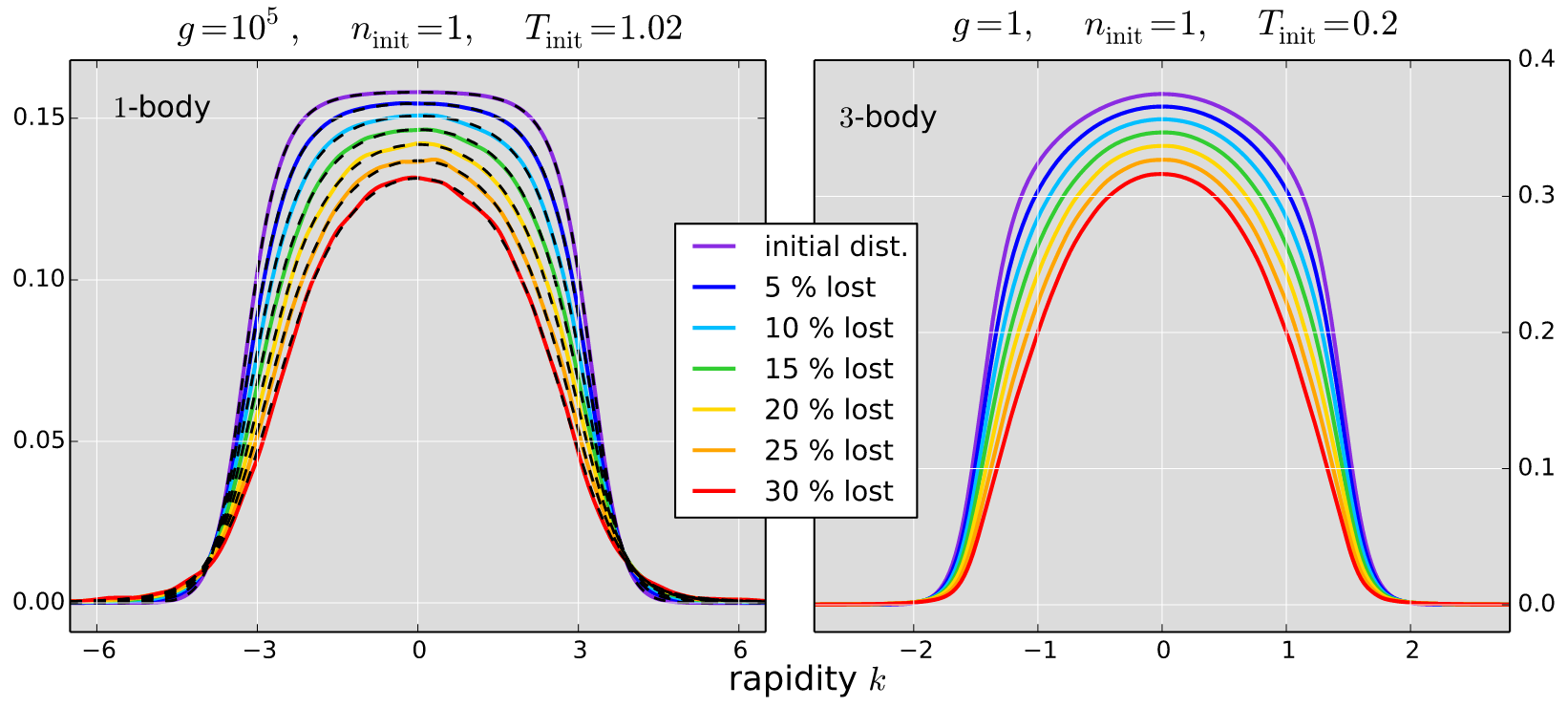}
    \caption{[From~\cite{bouchoule_effect_2020}] Rapidity distributions of the homogeneous 1D Bose gas, initially at thermal equilibrium at temperature $T_{\rm init}$, after a fraction of the atoms have escaped the system because of $K$-body loss processes. Left: results for one-body losses in the hard core limit. The colored lines are obtained by evaluating the functional $F[\rho](\theta)$ (Eq.~(\ref{eq:drho_dt_loss})) numerically, with a Monte-Carlo summation over eigenstates. The black dashed line is the analytical result available for the hard core limit. Right: numerical results for three-body losses at finite repulsion strength ($\gamma = 1$ in the initial state).}
    \label{fig:losses}
\end{figure}

\vspace{0.5cm}
Up to now, the following results have been obtained in connection with this problem.
\begin{itemize}
    \item The functional $F[\rho]$ can been evaluated numerically for a given $\rho$ by performing a Monte Carlo summation over Bethe states~\citep{bouchoule_effect_2020}, using exact formulas for the matrix elements of $\psi^K(0)$ between Bethe states, see~\citep{pozsgay2011mean,piroli2015exact}. The differential equation (\ref{eq:drho_dt_loss}) can then be integrated numerically, see the example shown in Fig.~\ref{fig:losses}. However this procedure is numerically heavy: it typically takes a few hours to compute $F[\rho]$ for a given rapidity distribution $\rho$ on a single core (but the procedure can be trivially parallelized). 
    
    \item The functional $F[\rho]$ is known analytically in the ideal Bose gas regime,
    \begin{equation}
        ({\rm ideal \; Bose \; gas}) \qquad  F[\rho](\theta) = K  \, K!  \, \rho(\theta) ,
    \end{equation}
    and also in the hard core regime,
    \begin{equation}
        ({\rm hard \; core}) \quad  F[\rho](\theta) = \left\{     \begin{array}{lll}
                \rho(\theta) - 2\pi [ \rho(\theta)^2 - (\mathcal{H} \rho(\theta))^2  ] + 2 n (\mathcal{H} \rho)'(\theta) & {\rm if} & K=1 , \\
                0 &{\rm if}& K \geq 2 ,       
            \end{array}
        \right.
    \end{equation}
    where $\mathcal{H} \rho(\theta) := \frac{1}{\pi} {\rm PV} \int \frac{\rho(\alpha) d\alpha}{\theta - \alpha}$ is the Hilbert transform of the rapidity distribution.
    
    The result for the ideal Bose gas is very simple, reflecting the fact that the rapidities are the momenta of the non-interacting bosons in that regime. The combinatorial factor $K \, K!$ comes from the local $K$-body correlation  $g^{(K)}(0) := \left< \Psi^{\dagger K}(0)\Psi^K(0) \right>/n^K = K!$ (a consequence of Wick's theorem), and the additional factor $K$ simply comes from the fact that there are $K$ atoms lost in each event.
    
    In contrast, the result for the hard core regime for $K=1$ is much more complex, even though it is also related to an underlying model of non-interacting particles, see Subsection~\ref{sec:regimes}. In particular, one sees that $F[\rho]$ is both non-linear in $\rho(\theta)$, and non-local in rapidity space (because the Hilbert transform $\mathcal{H} \rho(\theta)$ depends on $\rho$ at all values of the rapidity, not just at $\theta$). These properties are thus expected to hold generically for finite repulsion strength.
    
    In the hard core regime, $F[\rho](\theta)$ vanishes for $K\geq 2$ because two atoms (or more) can never be found at the same position; thus local $K$-body processes with $K \geq 2$ are suppressed.
    
    \item Instead of aiming directly at the time variation of the full rapidity distribution $\rho(\theta)$, another possibility consists in studying the variation of particular conserved charges. For instance, specifying $f(\theta) = 1$ in (\ref{eq:dqd_dt_loss}) leads to the evolution of the density of particles,
    \begin{equation}
        \frac{dn}{dt} \, = \, -G K \left< \Psi^{\dagger K}(0) \Psi^K(0) \right> \, = \, -G K n^K g^{(K)}(0) .
    \end{equation}
    Similarly, specifying $f(\theta) = \theta^2/2$ gives to the evolution of the energy density,
    \begin{equation}
        \label{eq:de_dt_loss}
        \frac{de}{dt} \, = \, G \left< \Psi^{\dagger K}(0) [H, \Psi^K(0) ] \right> ,
    \end{equation}
    and so on. General expressions are available for the local $K$-body correlation $g^{(K)}(0)$ as a functional of the distribution of rapidities $\rho(\theta)$, so that it is possible to compute $dn/dt$ efficiently~\citep{pozsgay2011mean,bastianello2018exact,bastianello2018sinh} (the topic of the evaluation of $g^{(K)}(0)$ in the Lieb-Liniger gas has a long history, see e.g.~\citep{kheruntsyan2003pair,gangardt2003local,cheianov2006three,cheianov2006exact,kormos2009expectation,kormos2011exact}). 
    On the experimental side, the 
    measurement of $dn/dt$ was used by \cite{tolra_observation_2004} as a first demonstration that the inferred zero-distance correlation $g^{(K)}(0)$ can take a value substantially below 1 in strongly interacting 1D gases. More recently, a strong dependence of the effective loss constant $K_{\rm eff}=Kg^{(K)}(0)$ on the energy 
    of colliding clouds has been reported for strongly interacting gases in \citep{zundel_energy-dependent_2019}. Such a behavior is compatible with the expected strong dependence of $g^{(K)}(0)$ with the spread in rapidity space \citep{gangardt2003stability}.This behavior is also recovered by an analysis of the three-body problem~\citep{mehta_three-body_2007}.

    Until very recently not much was known about the evolution of other charge densities. \cite{hutsalyuk2020integrability} focused on the energy density and managed to compute the r.h.s of (\ref{eq:de_dt_loss}) as an explicit functional of the rapidity distribution $\rho(\theta)$. It is likely that their method can be generalized to some other charge densities, and this could perhaps ultimately lead to the variation of the full distribution of rapidities. At the moment, nothing is known beyond the atom density and the energy density though, and finding expressions similar to the ones of \citep{pozsgay2011mean,bastianello2018exact,hutsalyuk2020integrability} for the variation of other charges remains an open problem.
\end{itemize}
Again, we also refer to the review of Bastianello, de Luca and Vasseur in this Volume for a thorough discussion of related recent results.

\subsection{$1/\theta^4$ tails in the  rapidity distribution}
One striking effect of losses is found when investigating the evolution of the high-rapidity tails of the rapidity distribution $\rho(\theta)$. 
It was shown in~\citep{bouchoule_breakdown_2020} that, under losses, $\rho(\theta)$ develops algebraically decaying tails,  $\rho(\theta) \sim 1/\theta^4$ when $|\theta|\rightarrow \infty$. This is in contrast with more standard cases of rapidity distributions, e.g. the ones in thermal equilibrium states, which typically decay exponentially or even as Gaussians.  

The physical origin of the development of those $1/\theta^4$ tails in the rapidity distribution lies in the cusp singularity of the wave function when two atoms are at the same position, see Eq.~(\ref{eq:singularity}):
\begin{equation}
    \label{eq:cusp_condition}
    (\partial_{x_i} \psi )_{|_{x_i\rightarrow x_j^-}}-( \partial_{x_i} \psi) _{|_{x_i\rightarrow x_j^+}}
= \frac{mg}{\hbar^2} \,\psi(\dots,x_i=x_j,\dots). 
\end{equation}
Consider the case of one-body losses ($K=1$), for simplicity. At a time $t_l$, the $l^{\rm th}$ atom is suddenly removed from the system. Immediately after the loss event, the wavefunction of the remaining atoms is $\psi_{t=t_l^+}(\dots, x_i,  \dots, x_l, \dots)$, where $x_l$ is the position of the lost atom. This wavefunction, viewed as a function of $x_i$, still has a cusp singularity at $x_i = x_l$, even though there is no longer a particle at $x_l$. On the other hand, the  eigenstates of the Lieb-Liniger Hamiltonian (\ref{eq:hamLL}) for $N-1$ atoms are smooth functions of $z_i$ around $z_i=z_l$ (for fixed values of $x_j \neq x_l$ for $j\neq i$). Expands the wavefunction $\psi_{t=t_l^+}(\dots, x_i, \dots, x_l, \dots)$ over the eigenstates for $N-1$ particles, one finds that the coefficients of the Bethe states $\left| \{ \theta_a \}_{1 \leq a \leq N-1} \right>$ decay as $\sim \left( {\rm max}_{ 1 \leq a \leq N-1 } \, | \theta_a |  \right)^{-2}$. The rapidity distribution is obtained by averaging w.r.t to the squared amplitudes of these coefficients, so it must decay as $1/\theta^4$.

Rapidity distributions decaying as $1/|\theta|^4$ are not very common, however there is at least one other known physical scenario where they appear: a sudden quench of the repulsion strength $g$. In~\citep{de2014solution}, the rapidity distribution $\rho(\theta)$ after a quench from $g=0$ at zero temperature (Bose-Einstein condensate) to $g>0$ is computed exactly, and it is found that $\rho(\theta) \sim 1/ \theta^4$ for large rapidities. Notice that this is consistent with the discussion above based on cusps of the wavefunction: when the repulsion strength $g$ is suddenly changed, the wavefunction immediately after the quench  violates the cusp condition (\ref{eq:cusp_condition}),  resulting in an expansion over Bethe states with coefficients decaying slowly at large rapidities, as in the case of losses.


\vspace{0.5cm}

One important physical consequence of the presence of the $1/\theta^4$ rapidity tails, worked out in \citep{bouchoule_breakdown_2020}, is the breakdown of a famous exact relation between the tails of the momentum distribution $w(p)$ in the gas and the `contact'~\citep{minguzzi2002high,olshanii2003short},
\begin{equation}
    \label{eq:Tan}
    w(p) \underset{|p| \rightarrow \infty}{=} \left(  \frac{m^2}{2\pi \hbar} g^2 \, n^2 \, g^{(2)}(0) \right) \frac{1}{p^4} .
\end{equation}
Here the momentum distribution $w(p)$ is normalized such that $\int w(p) dp = n$. The relation (\ref{eq:Tan}) has been extended to higher dimensions and to fermionic gases or general Bose-Fermi mixtures with contact interaction~ \citep{tan2008energetics,tan2008large,tan2008generalized}. It is known as `Tan's adiabatic theorem' or simply `Tan's relation', and it has been studied extensively in the past fifteen years, both theoretically ---see e.g.~\citep{minguzzi2002high,olshanii2003short,tan2008energetics,tan2008large,tan2008generalized,braaten2008exact,barth2011tan,werner2012generalI,werner2012generalII,yao_tans_2018}--- and experimentally~\citep{kuhnle2010universal,stewart_verification_2010,wild2012measurements}.

However, in \citep{bouchoule_breakdown_2020}, it is argued that the relation (\ref{eq:Tan}) breaks down when the rapidity 
distribution  decays as  $1/\theta^4$ at large $\theta$. 
In that case, the rapidity tail adds to the one caused by the two-body contact term. More precisely, introducing $C_r=\lim_{\theta\rightarrow\infty} \rho(\theta)\theta^4$ the amplitude of the $1/\theta^4$ rapidity tails, the relation (\ref{eq:Tan}) is superseded by the relation 
\begin{equation}
    \label{eq:CrCc}
    w(p) \underset{|p| \rightarrow \infty}{=} \left( C_{\rm r} + \frac{m^2}{2\pi \hbar} g^2 \, n^2 \, g^{(2)}(0) \right) \frac{1}{p^4} .
\end{equation}
In most known stationary states of the gas, in particular in thermal equilibrium states, $C_{\rm r}=0$, so that Tan's relation (\ref{eq:Tan}) holds. However, in a gas subject to losses, or after a quench of the interaction strength $g$, $C_{\rm r} > 0$ and Tan's relation is violated. Moreover, in~\citep{bouchoule_breakdown_2020}, the amplitude of the  term $C_{\rm r}$ is found to be potentially much larger than $\frac{m^2}{2\pi \hbar} g^2 \, n^2 \, g^{(2)}(0)$. For one-body losses ($K=1$) the ratio $C_{\rm r}/ [ \frac{m^2}{2\pi \hbar} g^2 \, n^2 \, g^{(2)}(0) ]$ increases exponentially in time under lossy evolution, while it grows  logarithmically for $K=2$ and remains bounded (but not necessarily close to $1$) for $K\geq 3$.

\subsection{Results in the quasicondensate regime}

In the asymptotic regime of quasicondensate, the 
effect of losses
can be investigated using the Bogoliubov description of the gas. 
The Bogoliubov approach is an approximate description of the gas, that constitutes a trivial integrable model since it resumes to independent bosonic modes: the integral of motion are nothing else than the population in each mode.
Effect of losses within the Bogoliubov approach has been 
first investigated in~\cite{rauer_cooling_2016,grisins_degenerate_2016}, where the emphasis was put on the 
long wave length modes, the so-called phononic modes.
These studies were then extended to all Bogoliubov modes, and also to $K$-body processes~\cite{johnson_long-lived_2017,bouchoule_cooling_2018}.
 The  population of the phononic modes is expected to reach a value corresponding to a temperature $T_p$ that fulfills
\begin{equation}
    k_B T_{\rm p} = \alpha \mu
    \label{eq:Tphonons}
\end{equation}
where $\mu\simeq gn $ is the chemical potential of the gas
and $\alpha$ is a numerical factor, of order 1, which depends on $K$. 
This prediction is in agreement with experimental studies made for three-body~\citep{schemmer_cooling_2018} and one-body losses~\citep{bouchoule_asymptotic_2020}. 
Bogoliubov modes of shorter wavelength are on the other hand expected to reach a higher temperature~\citep{grisins_degenerate_2016,johnson_long-lived_2017}. This peculiar state, with a temperature of the short wavelength modes larger than the temperature of the phonons, might be at the origin of the experimental observation of long-lived non thermal states~\citep{johnson_long-lived_2017}. 

The integral of motion of the Bogoliubov model are however not the real integrals of motion, which are given by the rapidity distribution. The precise link between the Bogolubov modes and the rapidities is still an open question. For short wavelength modes, however, one can 
identify the population of the Bogoliubov modes to rapidities, as done already by~\cite{lieb_exact_1963} in a seminal contribution. When doing this identification, the results of the Bogoliubov theory for the effect of losses on short wavelength modes coincide with the expected behavior for the tails of the 
rapidity distribution~\citep{bouchoule_breakdown_2020}.
For the phonons on the other hand, there is no one-to-one correspondence with rapidities. Neither the validity of Eq.~\eqref{eq:Tphonons} on long terms, nor 
its compatibility with the time evolution of the rapidity distribution has been established yet.

\subsection{Other related recent results}

In \citep{rossini2020strong}, a lattice Bose gas with onsite repulsion (Bose-Hubbard model) and onsite two-body losses is studied. In the limit of very fast losses, all configurations with more than one particle per site decay extremely quickly, so the slow dynamics occurs within the restricted subspace of configurations with at most one boson per site. In the effective dynamics of these emerging hard core lattice bosons, two-body losses are still present, however they occur on nearest-neighbor sites, and they are very slow~\citep{garcia2009dissipation}. Thus, \cite{rossini2020strong} effectively work with a lattice hard core boson model subject to adiabatic two-body losses. The corresponding (lattice) rapidity distribution is evaluated in a way that parallels the above discussion, see Eq.~(\ref{eq:drho_dt_loss}), and results analogous to the ones of~\citep{bouchoule_effect_2020} are obtained.

\vspace{0.5cm}

As mentioned above, losses in the Lieb-Liniger gas are but one particular example of a mechanism that breaks the integrability of the underlying model. Other mechanisms have been studied, for instance the coupling between tubes in an array of 1D gases~\citep{caux2019hydrodynamics}, or dephasing~\citep{bastianello2020generalized}. Let us also mention closely related works on integrable spin chains evolving under Lindblad evolution  \citep{lange2017pumping,lange2018time,lenarvcivc2018perturbative}, where an adiabatic limit analogous to the one discussed above is implemented using truncated Generalized Gibbs Ensembles, or the crossover from ballistic to diffusive transport induced by weak integrability breaking terms~\citep{friedman2020diffusive,vznidarivc2020weak}. These results, and more, are discussed in the review of Bastianello, de Luca and Vasseur in this volume.

\newpage

\section*{Conclusion and perspectives}

    Generalized Hydrodynamics theory has proven to be very efficient to describe non-equilibrium dynamics in 1D Bose gases. Its broad applicability domain is confirmed by the first experimentals tests.  
    Nevertheless, investigation of non-equilibrium dynamics using the GHD theory is still in its infancy. Many situations are still to be explored. On the experimental side, it would be interesting to investigate the so-called bi-partite quench protocols where two clouds with different rapidity distributions and separated by a barrier at $x=0$ are merged by a sudden removing of the barrier. This fundamental problem in the theory of gases and in hydrodynamics ---where it is usually known as a Riemman problem~\citep{riemann1860fortpflanzung}--- seemed completely out of reach in integrable spin chains and integrable gases before 2016, and it played a major role in the discovery of~\cite{bertini2016transport} and \cite{castro2016emergent}. 
    For an experimental study of this setup, a complete characterization of the system, the implementation of a local measurements in such non-equilibrium protocols would be a great progress. 
    
    At the heart of GHD there is the idea that the gas is locally described by its rapidity distribution. While the meaning of the rapidity distribution and its time evolution under GHD is quite transparent in the hard core regime, where it simply corresponds to the 
    momentum distribution of the equivalent ideal Fermi gas, it is less obvious in the weakly interacting case. On the other hand, in weakly interacting regimes, powerful techniques have been developed, such as the Bogoliubov techniques or the classical field approach. Making the link between those approximate techniques and GHD, including the  notion of rapidity distribution, would be a substantial progress. 
     The Bogoliubov model is a trivially integrable model, although its integrals of motion are not the true integrals of motion of the underlying Lieb-Liniger model. Some open questions are: what are the  Bogoliubov distributions which are stationary with respect to the Lieb-Liniger dynamics?  What is the Bogoliubov distribution of a given rapidity distribution ? Vice-versa, what is the rapidity distribution of a given Bogoliubov distribution ?
     At sufficiently high temperatures such that quantum fluctuations become negligible, one can describe the gas within the classical field framework. The link between GHD and classical field predictions also deserves more investigation.
    
    Many-body dynamics in 1D Bose gases is a wide research area, and the effects that can be described within the GHD framework are, presumably, only a small part of it. It is of great interest to study phenomena that are beyond the original GHD theory.
    Many questions have still to be elucidated and we propose here a non-exhaustive list of research directions.
    \begin{itemize}
        \item {\bf Beyond the 1D regime.}  In experiments, physics lies in the 3D space, and the 1D model is only an approximated description. Effects that are beyond the 1D physics still need investigation. In the case of a harmonic transverse trap, the effect of transversally excited states is not completly established yet. In~\citep{moller_extension_2021}, a first theoretical framework was proposed to take into account atoms populating a higher transverse excited state. 
        On the experimental side, the thermalization observed by~\cite{li_relaxation_2020} in the Quantum Newton Cradle setup is attributed to the presence of atoms in transversely excited states. In this experiment, relaxation occurs for a gas lying in the ideal Bose gas regime. Exploring other regimes would be highly desirable. 
        Even if the transverse degree of freedom is energetically frozen, transverse excited states may play a role as intermediate states in virtual 3-body processes~\citep{mazets_breakdown_2008}. Such an effect, which is of course beyond the GHD theory, is expected to lead to  thermalization of the gas. How this thermalization occurs is an open question. 
        
        Another situation that goes beyond the 1D model is the case of an array of 1D tubes coupled by a small tunnel effect. Again, the effect of such a coupling on the evolution of the rapidity distribution in not known. One expects that this coupling will permit thermalization. This situation allows the study of the dimensional  crossover between 1D and 2D or 3D physics. 
        \item {\bf Diffusion effect.} Generalized Hydrodynamics, which was initially formulated in the Euler limit, has been extended by the addition of a Navier-Stokes diffusive term. The experimental test of such a diffusive term would be an important achievement.  
        \item {\bf Breakdown of integrability due to a potential. } The combined effect of the diffusive term in GHD and of a spatially varying  external potential is expected to lead to a relaxation towards a thermal state, as shown in numerical  simulations reproducing the Newton Cradle setup. 
        From the theoretical viewpoint different situations could be considered. On the experimental side, such an effect has not been observed, probably because the potentials are usually varying on too large distances. 
        The experimental investigation of such an effect would certainly permit to increase the understanding of the phenomena.
    \end{itemize}

We would like to finish this conclusion by some considerations on the effects of losses. As seen in the recent studies presented above, the effect of losses is highly non trivial. Even one-body losses, whose effect is trivial for a non-correlated gas, have a non trivial effect in presence of interactions between atoms. This is also true in higher dimension. Higher dimensional gases, when they are interacting, are not integrable, such that the effect of adiabatic losses can be characterized by the time evolution of only two quantities: the particle density and the energy density. However, the computation of the latter still needs be done. Thus, amazingly, the effect of losses is now better understood in 1D gases, where it is {\it a priori} more complicated since one has to keep track of the whole rapidity distribution, than in higher dimensions.

\begin{acknowledgments}
We have greatly benefited from discussions  with many colleagues. Among them, we would like to thank especially our co-authors on the topics reviewed in this article: 
Julien Armijo, Maxim Arzamasov, Yasar Yilmaz Atas, Alvise Bastianello, Pasquale Calabrese, Jean-S\'ebastien Caux, Benjamin Doyon, Bess Fang, Dimitri Gangardt, Thibaut Jacqmin, Aisling Johnson, Karen Kheruntsyan, Robert Konik, Yuan Le, Neel Malvania, Marcos Rigol, Tommaso Roscilde, Paola Ruggiero, Max Schemmer, Gora Shlyapnikov, Jean-Marie St\'ephan, Stuart S. Szigeti, David Weiss, Takato Yoshimura, Yicheng Zhang. 
We are also grateful to David Weiss, Frederik M\o ller and J\"org Schmiedmayer for comments on the manuscript.


This  work  was  supported by the ANR Project QUADY - ANR-20-CE30-0017-01.

\end{acknowledgments}


\end{document}